\documentclass[acmlarge]{acmart}
\makeatletter
\newcommand{\myconfshort}{\acmConference@shortname}
\newcommand{\myconffull}{\acmConference@name}
\newcommand{\myconfdate}{\acmConference@date}
\newcommand{\myconfloc}{\acmConference@venue}
\AtBeginDocument{
  \fancypagestyle{firstpagestyle}{
    \fancyhead{}%
    \fancyfoot[C]{}%
  }
  \fancyhf{}
  \fancyhead[LO]{\@headfootfont\shorttitle}%
  \fancyhead[RE]{\@headfootfont\@shortauthors}%
  \fancyhead[LE]{\@headfootfont\footnotesize \myconfshort, \myconfdate, \myconfloc}%
  \fancyhead[RO]{\@headfootfont\footnotesize \myconfshort, \myconfdate, \myconfloc}%
  \fancyfoot[C]{}%
}
\makeatother
\acmBooktitle{\conffull\@ (\confshort), \confdate, \confloc}

\usepackage{placeins}
\usepackage{array}
\usepackage{ragged2e}
\usepackage{bbm}
\usepackage{multirow}
\usepackage{makecell}
\usepackage{enumitem}
\usepackage{adjustbox}
\usepackage{siunitx}  
\usepackage{bbding}
\usepackage{tablefootnote}
\usepackage{subcaption}
\usepackage{caption}
\usepackage{bbm}
\usepackage{longtable}
\AtBeginDocument{%
  }

\newcommand\newsubcap[1]{\phantomcaption%
       \caption*{\figurename~\thefigure\thesubfigure. #1}}

\copyrightyear{2026}
\acmYear{2026}
\setcopyright{cc}
\setcctype{by-nc-nd}
\acmConference[FAccT '26]{The 2026 ACM Conference on Fairness, Accountability, and Transparency}{June 25--28, 2026}{Montreal, QC, Canada}
\acmBooktitle{The 2026 ACM Conference on Fairness, Accountability, and Transparency (FAccT '26), June 25--28, 2026, Montreal, QC, Canada}
\acmDOI{10.1145/3805689.3812405}
\acmISBN{979-8-4007-2596-8/2026/06}

\begin{document}
\title[Scrutinizing Index-Based Risk Assessments]{Scrutinizing Index-Based Risk Assessments: A Case Study in NYC Decision-making for Heat Emergency Management}

\author{Jennah Gosciak}
\affiliation{%
  \institution{Cornell Tech}
  \city{New York City}
  \state{NY}
  \country{USA}
}
\email{jrg377@cornell.edu}

\author{Luke Boyce}
\affiliation{%
  \institution{New York City Emergency Management}
  \city{New York City}
  \state{NY}
  \country{USA}
}
\email{lboyce@oem.nyc.gov}

\author{Angelina Wang}
\affiliation{%
  \institution{Cornell Tech}
  \city{New York City}
  \state{NY}
  \country{USA}
}
\email{angelina.wang@cornell.edu}

\author{Allison Koenecke}
\affiliation{%
  \institution{Cornell Tech}
  \city{New York City}
  \state{NY}
  \country{USA}
}
\email{koenecke@cornell.edu}

\renewcommand{\shortauthors}{Gosciak et al.}

\begin{abstract}
    Cities are increasingly turning to large-scale data analysis and machine learning to make consequential decisions.
    While the algorithmic fairness community has focused on analyzing the risks and benefits associated with these complex methods, there has been much less scrutiny of the many simpler, but still widely used, data-driven tools that support government decision-making in a variety of settings.
    In this work, we study hand-crafted indices for geographic targeting and decision-making in emergency management --- a field responsible for coordinating preparedness and response efforts to hazards ranging from natural disasters to human threats. 
    Indices, which capture abstract principles and overarching priorities (e.g., reducing social vulnerability), are low-complexity models that statistically aggregate chosen variables. They are generally flexible and interpretable, but can also be sensitive to key design choices and require strong assumptions.
    Through a case study of decision-making for extreme heat emergencies in NYC, we examine the challenges that practitioners may face in selecting an index for preparedness and response actions. We map empirical findings from index-based simulations to concerns related to validity and reliability from the measurement literature and show via sensitivity analyses that different reasonable choices of input variables or spatial scale can result in substantive differences to index risk scores, thereby affecting downstream government decision-making.
    We contrast these challenges with considerations for developing predictive algorithms that more narrowly relate to concrete, measurable outcomes. Ultimately, we provide generalizable recommendations that practitioners and public-sector technologists can use for navigating the trade-offs between indices and predictive algorithms in other government settings.
\end{abstract}




\maketitle

\section{Introduction}
\label{sec:introduction}
Predictive algorithms have been the subject of significant controversy in the algorithmic fairness community, and raise a variety of new concerns for local governments related to 
fairness, transparency, and recourse.
However, hand-crafted indices --- the status quo in many public policy and government settings --- are often just as opaque and arbitrary.
Examples of indices include tools to measure poverty~\citep{alkire2021global}, social vulnerability~\citep{cutter2003social}, and flood risk~\citep{balica2012flood}. We define indices as simple, low-complexity models that summarize a set of observed characteristics for spatial units (countries, cities, or neighborhoods\footnote{There are also indices for individual- or household-level risk, often based on survey responses (e.g., \citep{orgcode2015vulnerability, maxwell2003coping}), but we do not consider these in this paper.}) into a single ranking~\citep{joint2008handbook}; for example, an index measuring social vulnerability might standardize and add together 16 different variables related to socioeconomic status, household characteristics, race and ethnicity, and housing type and transportation~\citep{svi_cdc_site}. 
The methodologies for developing indices can vary -- from hierarchical, deductive methods to principal component analysis~\citep{beccari2016comparative}. The resulting outputs, however, can all be described as a standardized score that is meant to facilitate comparisons between disparate units.
Though widely used, indices are also contentious in the literature; prior work has shown them to be sensitive, easy to manipulate, and ineffective~\citep{huynh2024mitigating, dobbie2013robustness, greco2019methodological, grupp2004indicators, grupp2010review, beccari2016comparative, kaiser2021should, razavi2020global, abbey2020global, aitken2020rethinking}. 

By focusing on indices, we build on calls for studying the ``bureaucratic counterfactual'' in algorithmic fairness research, i.e., considering how decisions would be made in the absence of an algorithm~\citep{johnson2022bureaucratic}.
Other examples of bureaucratic counterfactuals include threshold-based triggers (e.g., when the maximum daily heat index passes 95°F for two consecutive days~\citep{hazard_mitigation_heat}), discretion and expertise (e.g., a street-level bureaucrat responds to citizen complaints~\citep{lipsky2010street}), heuristics (e.g., taking the same action as neighboring jurisdictions~\citep{roberts2019decision}), or ``categorical prioritization'' (e.g., prioritizing individuals for housing vouchers based on pre-determined criteria~\citep{johnson2022bureaucratic}).

To concretize our analysis, we focus on the field of emergency management through a case study on decision-making for extreme heat emergencies in New York City (NYC). We chose this case study for several reasons. First, indices feature prominently in vulnerability assessment (e.g., heat~\citep{reid2009mapping} and flood vulnerability indices~\citep{balica2012flood}), in part inspired by seminal work on social vulnerability~\citep{cutter2003social}. Vulnerability assessments inform high-stakes decisions in emergency planning and response across a range of hazard types.
Second, 
a collaboration with 
a local government agency provided insight into current index tools and heat-related decision-making.
Third, relatively little scholarship in algorithmic fairness has focused on emergency management. Decisions in this setting primarily focus on geographic targeting as opposed to predictions about individuals, which are more commonly studied in the algorithmic fairness literature~\citep{wang2024against}.
Lastly, extreme heat specifically is understudied in both the literature on emergency management and machine learning. Prior research tends to focus on more visible disasters (e.g., coastal storms or wildfires). But, extreme heat is a serious concern:
2024 was the hottest year on record, and $19$ of the $20$ hottest years in recorded history have occurred since $2000$~\citep{nasa, zhao2021global}.

The core of our analysis centers on \textbf{NYC’s Heat Vulnerability Index (HVI)}, a tool widely cited in citywide planning efforts such as the \textit{NYC Urban Forest Agenda}~\citep{forestry_plan} or \textit{Cool Neighborhoods}~\citep{cool_neighborhoods}.
Governments may prefer to use indices like the NYC HVI, given its methodological simplicity.
However, as we show, modifications to the choice of inputs or spatial scale can lead to fluctuations in risk assessment rankings.
These fluctuations present challenges for policymakers and government officials who may need to select a tool in specific contexts to inform a downstream task (e.g., intervening on an at-risk population or conducting outreach).
Such challenges can be exacerbated with tasks that are removed from the original purpose for developing the index tool (e.g., anticipating where power outages may occur during a heat wave, as power outages are only indirectly related to heat).

Indices may support long-term planning processes and allow governments to make certain values or commitments more explicit by incorporating data inputs associated with those values. But they are also harder to validate, lack specificity, and bear limited connection to concrete downstream tasks.
In the context of extreme heat, we briefly highlight two examples illustrating this tension (we discuss several others in Section~\ref{sec:discussion}). The first is a setting where indices may be effective: a long-term \textbf{public health advertising campaign} to raise awareness about extreme heat, like New York City Emergency Management's (NYCEM) ``Beat the Heat'' campaign~\citep{extreme_heat_beat}. Such campaigns involve physical advertisements throughout the city, social media announcements, and outreach. 
Planning can occur months in advance and may reflect explicit values (e.g., reaching elderly populations) as opposed to a data-driven goal.
In contrast, an alternative setting where predictive algorithms may be preferable is \textbf{a text message notification warning New Yorkers about power outages}. In this setting, decisions are made quickly and often focus on narrow actions (e.g., reduce electricity usage based on power outage risk).

The algorithmic fairness community has typically been concerned with the potential for bias and harm resulting from predictive algorithms.
However, the investigation of the bureaucratic counterfactual in many settings might suggest that these concerns are more nuanced; predictive algorithms in some settings may offer an improvement over the status quo. Our argument is not to do away with existing tools, but rather to characterize the kinds of decision-making contexts where different methods are most useful.

In this paper, we test the sensitivity of the NYC HVI, along with several other publicly accessible index tools for extreme heat.
Sensitivity analyses are useful for policymakers, particularly when it is challenging to define ground truth. Highly correlated rankings from different indices may suggest overall alignment with high-level concepts while divergent rankings call for greater scrutiny, particularly when applying an index to a new task.
We connect our sensitivity analyses to concepts from measurement and validity theory~\citep{jacobs2021measurement, cronbach1955construct, messick1987validity}, providing a theoretical basis for understanding why indices may be inconsistent and arbitrary.

Based on reflections and observations from developing prototypical predictive algorithms for heat-related outcomes in collaboration with a government agency, we discuss the trade-offs associated with using algorithmic tools for decision-making (e.g., spatiotemporal models that predict measurable impacts related to extreme heat) as opposed to hand-crafted indices (e.g., the NYC HVI).
Ultimately, we enumerate a list of holistic considerations for practitioners: the scope of the problem, the relevance of clearly defined outcomes, the need to express explicit values, the time scale of decisions, the potential for evaluation and validation, the capacity of relevant stakeholders, and the intended audience.

Our work highlights the value in broadening the scope of algorithmic fairness research to include research on index design and usage. Indices are simple, easier to develop, and widely used in government decision-making, but have their own limitations that should be rigorously considered. To this end, we provide a public code repository that researchers can use to extend our analyses to other domains.\footnote{\href{https://github.com/jennahgosciak/ScrutinizingIndexRiskAssessments}\href{https://github.com/jennahgosciak/ScrutinizingIndexRiskAssessments}} 

\textbf{To summarize, we make the following contributions: (1) we provide a case study on decision-making in NYC for extreme heat emergencies; (2) we argue for the use of sensitivity analysis as a method to test validity and reliability concerns in algorithmic fairness settings where clear ground truth outcomes are challenging or hard to define; (3) we propose seven trade-offs for practitioners to consider between indices and predictive algorithms.}

The rest of the paper proceeds as follows. In Section~\ref{sec:related_work}, we
connect our paper to several lines of related work, such as the use of algorithms in government decision-making.
In Section~\ref{sec:case_study}, we provide background for our case study on extreme heat emergencies in NYC.
In Section~\ref{sec:tool_review}, we then empirically evaluate the validity and reliability of the NYC Heat Vulnerability Index (HVI). We compare the HVI to alternative specifications, three other index tools available to local governments for decision-making related to extreme heat, and measurable heat-related impacts. 
In Section~\ref{sec:discussion}, we contrast index tools more broadly with considerations for developing predictive algorithms, using examples from decision-making for extreme heat to guide our comparison. We comment on the generalizability of our study to other domains and discuss the limitations of our work. In Section~\ref{sec:conclusion}, we conclude with a summary of findings and overarching thoughts on impact.

\section{Related Work}
\label{sec:related_work}

\subsection{Algorithmic Fairness and Government Decision-Making}
In recent years, cities have grown concerned with the expansion of algorithmic tools used to make government decisions.
Algorithmic tools offer attractive benefits to policymakers: they can improve the enforcement of regulatory policies, better identify anomalous events, and automate routine tasks~\citep{engstrom2020government, pencheva2020big}.
At the same time, they are often opaque, whether because of the nature of the models themselves, or the available public information.
Assessing the validity of these tools or the potential for bias is challenging.
In response to these concerns, municipal laws aimed at promoting algorithmic transparency and accountability have become more prominent. For example, in 2020, Helsinki and Amsterdam became the first cities to pilot the idea of an AI register~\citep{floridi2020artificial}.
More recently, Local Law 35 in NYC has similarly mandated public reporting on any algorithmic tools used by city agencies~\citep{govtech}.
Yet even as more cities require transparent documentation, in many cases the information made available is insufficient to understand or challenge the design decisions made in the development of the tool.

\textbf{Numerous consequential decisions made at every level of government already involve some data manipulation or standardization. 
But many of these tools do not conventionally ``count'' as an algorithm and would not appear on a transparency register like the one mandated by Local Law 35.} As we discuss in this paper, \emph{indices} are one class of such tools. They are widely used, particularly for risk and vulnerability assessment. These indices are typically constructed once, based on prioritizing a set of data inputs, and incorporate some form of standardization and aggregation. They capture the abstract, multi-dimensional values of different stakeholders. However, they can be sensitive to the choice of inputs and methods and require strong assumptions. Though simple and transparent, they can still lead to inconsistent and arbitrary outcomes~\citep{huynh2024mitigating}.

\subsection{Connections to Algorithmic Fairness Literature}
There are several connections with ongoing work in the algorithmic fairness literature.
We draw on prior work on geographic targeting for eviction outreach \citep{mashiat2024beyond, mashiat2025pays}, environmental compliance~\citep{benami2021distributive, huynh2024mitigating, karasaki2024machine}, and opioid prevention~\citep{allen2024provident, allen2026ethical, heuton2025spatiotemporal}. 
In particular, \citet{benami2021distributive} highlight how algorithm design can help clarify some of the implicit assumptions in high-level policy decisions.
These papers also raise concerns related to spatial inequality, stakeholder capacity, and the potential to exacerbate rather than mitigate disparities. 

We build on \citet{johnson2022bureaucratic}'s notion of the ``bureaucratic counterfactual,'' which in many social policy allocation decisions is known as categorical prioritization. Categorical prioritization involves: (1) deciding which individual-level attributes to prioritize, (2) grouping individuals into categories, and (3) mapping categories to decisions. This definition bears similarity to the indices we discuss in this paper. For example, the Vulnerability Index-Service Prioritization Decision Assistance Tool (VI-SPDAT)~\citep{orgcode2015vulnerability} is a flexible, widely used index to prioritize unhoused individuals for assistance that aligns with Johnson and Zhang's criteria. 
However, in contrast, we focus on population-level tools used for geographic targeting. These types of indices typically involve standardization and aggregation to produce a unified score or ranking and may be repurposed in new settings, as they often reflect abstract values and highly general concepts.

Our work also relates to causal approaches to algorithmic fairness~\citep{kasy2021fairness, stevenson2024algorithmic} and prediction as intervention problems~\citep{liu2025bridging} -- both of which consider the effect of decision-making tools in the broader context in which they are deployed. 
Prior work on predictive optimization~\citep{barocas2023automated, wang2024against}, which questions whether the uses of individual-level predictions for decision-making are legitimate, is relevant as well. Importantly, in this paper, we do not consider individual-level predictions and the interventions proposed are assistive in nature. We start from the premise that simple, data-driven tools are already used in practice. Given this reality, we contrast these approaches with alternative methods (i.e., predictive algorithms) to understand the decisions better suited to each. 

Lastly, some work in algorithmic fairness has emphasized the importance of evaluating model robustness, uncertainty, and sensitivity in relation to fairness concerns~\citep{huynh2024mitigating, rosenblatt2024fairlyuncertain, benesse2024fairness, simson2024one, kuzucu2024uncertainty, fawkes2024fragility}. Drawing on the precedent for conducting such analyses in the context of designing indices~\citep{schmidtlein2008sensitivity, beccari2016comparative, saltelli2020five}, we similarly advocate for employing sensitivity analyses as part of a broader assessment of validity in algorithmic fairness~\citep{jacobs2021measurement, cronbach1955construct, messick1987validity}. These methods are particularly useful when ground-truth labels are challenging to define or when practitioners need to understand the trade-offs associated with adapting existing tools (e.g., vulnerability indices) to new tasks.

\subsection{Index Design in Policy Analysis}
Indices are an important tool in policy analysis. The OECD Joint Research Centre \citep{joint2008handbook} provides a comprehensive handbook on this topic. Prior work has explored the implications of different methodologies (e.g., standardization vs. aggregation)~\citep{dobbie2013robustness, huynh2024mitigating}. \citet{greco2019methodological} discuss the dramatic increase in the popularity of indices, and the historical controversy between ``aggregators'' vs. ``non-aggregators.'' For example, in \citep{grupp2004indicators, grupp2010review}, the authors note inconsistencies across composite indicators in science and technology policy and argue for preserving multi-dimensional representations rather than aggregating (e.g., spider charts). 
Most relevant to this paper, \citet{beccari2016comparative}'s work on indices for disaster risk finds limited use of sensitivity or uncertainty analysis~\citep{schmidtlein2008sensitivity}.

There are several approaches for conducting sensitivity analyses. Local sensitivity analyses test single changes in the index design (e.g., changing one of the inputs to an index)~\citep{tate2012social, tate2013uncertainty}. The resulting rankings of the indices can be compared using correlation or analysis of variance tests, similar to our approach in this paper. More complex, global sensitivity analyses use methods like Monte Carlo simulation to determine which key design decisions proportionally contribute to the uncertainty~\citep{tate2012social}. 

\section{Background: Extreme Heat Emergencies}
\label{sec:case_study}
We now turn to our case study on government decision-making for extreme heat emergencies in NYC. This section provides relevant background for understanding our recommendations in Section~\ref{sec:discussion} and details that may help with generalizing our findings to other settings (e.g., walkability indices in urban planning~\citep{frank2010development} or flood vulnerability~\citep{nyc_fvi, balica2012flood}).

In emergency management, risk and vulnerability indices are a standard tool for decision-making, particularly for hazard mitigation and preparedness. 
However, we also observe the potential for predictive algorithms to improve decision-making, 
from integrating with early warning systems to supporting post-disaster damage assessments~\citep{kyrkou2022machine}. Analogously, public health researchers may often leverage large datasets like electronic health records or syndromic surveillance data to anticipate disease outbreaks and better target interventions~\citep{lewis2002disease, henning2004syndromic}.
But, while academic research has prototyped predictive algorithms for a range of hazard types (e.g., flood early warning systems~\citep{nearing2024global,  tran2025ai} or earthquake monitoring~\citep{natgeo_article}), prediction in
emergency management has been relatively under-explored in practice even though city governments manage large (and often public) datasets that may be useful for this purpose. 
Potential reasons include: concerns of reliability~\citep{kyrkou2022machine}, data and resource limitations~\citep{apwa}, outdated procurement systems~\citep{johnson2025legacy}, and the time-sensitive nature of many decisions~\citep{lentz2022information, tufts_feinstein}.

\subsection{How Are Cities Preparing for Extreme Heat Emergencies?}
\label{sec:how_cities_prepare}

In the U.S., there are few consistent or standardized approaches for combating extreme heat. 
However, indices for heat, which we will describe in the next section, are often used as inputs to \textit{hazard mitigation plans} or \textit{heat action plans}~\citep{kimutis2024emergency}.
Hazard mitigation plans concern actions for long-term planning and risk minimization. These federally mandated plans include detailed information on risk assessments, general purpose information about the type of emergency, and recommendations for both individual and community preparedness.
Heat action plans govern the precise types of coordinated activities that different government agencies may take in the event of an emergency or disaster (usually in response to a temperature trigger or advisory from the National Weather Service).
Typical actions for heat emergencies are similar across cities. 
In some cities, heat action plans are not public, as they contain sensitive information. In contrast, hazard mitigation plans are designed for a public audience.

NYC's Hazard Mitigation Plan provides high-level information on extreme heat -- information such as the number of historical high-heat days, the potential for adverse health impacts and populations most at risk, the NYC neighborhoods that are \emph{heat vulnerable}, and actions to minimize indirect impacts like power outages~\citep{hazard_mitigation_heat}. NYC's heat action plan, which is not public, triggers a set of actions that government agencies must take once a heat emergency has been declared. 
Key components include: opening cooling centers, conducting outreach to people experiencing homelessness, increasing public messaging (particularly to at risk populations), and coordinating with utility and healthcare providers and community organizations~\citep{heataction_nyc, heat_response_plan}.
At present, none of the actions in the heat action plan appear to vary based on heat vulnerability.
We discuss additional details for New York City, Phoenix, and Los Angeles in Appendix~\ref{sec:comparison_other_cities}.

\subsection{The Creation of a Heat Vulnerability Index in NYC}
Neighborhood heat vulnerability is an important concept for extreme heat emergency planning, particularly as NYC is one of the few cities with a heat vulnerability index (HVI).
The NYC HVI is a simple and flexible tool that identifies areas throughout the city with the highest risk for heat-related mortality~\citep{int_hvi}.
Depicted in Figure~\ref{fig:hvi}(a), the NYC HVI was developed by the NYC Department of Health and Mental Hygiene (DOHMH). There are comparable HVIs for New York State and other municipalities.\footnote{An HVI was developed for New York State. It incorporates additional inputs related to heat and has a different methodology~\citep{nayak2018development}.} The NYC HVI is comprised of the following five inputs: daytime summer surface temperature, the percentage of households with air conditioning, the percent vegetative cover, median household income, and the percentage of residents who are non-Latinx Black~\citep{int_hvi, madrigano2015case}. 

NYC DOHMH selected these inputs based on a case-only\footnote{Case-only is a term from epidemiology that describes a study with only positive cases and no comparison group (e.g., \citet{madrigano2015case} study all non-external mortality that occurred during an extreme heat event)~\citep{hamajima1999detection, khoury1996nontraditional}.} study of heat-wave related mortality in NYC~\citep{madrigano2015case}.
Several of these inputs were statistically significant in a logistic regression model that predicted the likelihood of individual heat-wave related mortality~\citep{madrigano2015case}.\footnote{To clarify, the inputs were selected based on their relationship to heat-related mortality in a logistic regression model. None of the values in the NYC HVI directly come from a logistic regression model. The current NYC HVI from DOHMH also uses slightly different inputs compared to the original study from \citet{madrigano2015case} -- e.g., median household income as opposed to the percentage of households receiving public assistance. The original study also did not have information on air conditioning.}
The five inputs are then summed together (or subtracted) to produce a composite index (shown in the first row of Table~\ref{tab:hvi_specification_formulas}). Quintiles of the raw composite scores serve as the basis for a 5-category risk score (1 = low risk, 5 = high risk). To validate the NYC HVI, \citet{madrigano2015case} implemented a multinomial logistic regression model to test whether heat-related mortality predicts the risk scores of the NYC HVI and found a significant association for the highest risk scores (scores $4$ and $5$). These scores commonly denote high heat vulnerable neighborhoods~\citep{comptroller_report}.

The NYC HVI is widely used for heat-related decision-making in NYC. It appears in the \textit{NYC Urban Forest Agenda} to prioritize neighborhoods for tree canopy expansion~\citep{forestry_plan} and is included in citywide plans for cooling features (such as sprinklers and drinking fountains)~\citep{coolit}. According to the NYC hazard mitigation plan, the NYC HVI can influence: programming, outreach, green roof installation, and tree planting~\citep{cool_neighborhoods}. 

For the sensitivity analysis, which we discuss in the next section, we replicate the NYC HVI on existing public data at the neighborhood and census tract level.\footnote{Census tracts are the primary statistical geographic unit that the U.S. Census uses for data collection. They typically have a population between 1,200 and 8,000~\citep{census_definition}. Since neighborhoods can be subjective, but are politically and socially important in NYC, the Census created neighborhood tabulation areas (NTAs) that use census tracts to approximate NYC neighborhoods~\citep{nta_definition}. Henceforth, when we refer to neighborhoods, we are actually describing NTAs.} Appendix~\ref{sec:replication} provides a  more detailed explanation as to how we compared each of the indices. There are slight differences between our reconstruction of the NYC HVI and the official NYC HVI categories obtained from NYC DOHMH, even though we base our approach on available documentation~\citep{madrigano2015case}. We are able to replicate the categorical risk scores with 98\% accuracy.\footnote{There are four neighborhoods with different quintile-based risk scores in our replication out of 197 neighborhoods, even when we use the same data as the NYC HVI. All four neighborhoods have percentile rankings that are close to the numeric thresholds for defining quintiles.} 
For our analyses, we examine changes in the coarser quintile-based scores as well as changes in the underlying percentile rank.

\section{A Critical Review of the NYC HVI}
\label{sec:tool_review}

In this section, we evaluate the reliability and validity of the NYC HVI, drawing on prior research advocating for measurement theory in algorithmic fairness~\citep{jacobs2021measurement, messick1987validity, cronbach1955construct}.
First, we assess \textit{construct reliability}. Construct reliability evaluates the stability of a measure, and we operationalize this through the sensitivity of a model to changes in which features are index inputs, and the geospatial granularity for these inputs. We compare the sensitivity of the NYC HVI risk scores to alternative specifications that might reflect different decisionmaker values.
While the design of the NYC HVI originated from a study on heat wave-related mortality, the tool has a multi-purpose function and informs a variety of planning initiatives (e.g., the same index numbers are used for tree planting as well as outreach related to cooling centers) where other concerns may take priority (e.g., reducing average surface temperatures or prioritizing seniors).

We then assess \textit{convergent validity}, by comparing the NYC HVI to several other public tools that cities can realistically use for decision-making. 
Convergent validity assesses the correlation of measurements for the same construct.
Specifically, we consider the \textbf{National Risk Index (NRI)} developed by the Federal Emergency Management Agency (FEMA)~\citep{fema_nri} and the \textbf{Heat and Health Index (HHI)} from the Centers for Disease Control and Prevention (CDC)~\citep{cdc_tracker}.\footnote{We briefly discuss comparisons to the experimental HeatRisk product developed by the National Weather Service (NWS)~\citep{noaa_heatrisk} in Appendix~\ref{sec:heatrisk}.} To our knowledge, these are the most widely accessible tools for measuring heat-related vulnerability and risk.
All three also represent distinct classes of indices, as depicted in Table~\ref{tab:tool_comparison}. The NYC HVI is a \textit{simple additive index}. The formula (in Table~\ref{tab:hvi_specification_formulas}) only involves addition and subtraction. The FEMA NRI is a \textit{theory-based formula}, relating measurements for economic loss, social vulnerability, and community resilience according to a theoretical notion of risk. The CDC HHI is a \textit{hierarchical additive index} comprised of three evenly weighted sub-indices for sociodemographics, health outcomes, and the built environment.\footnote{This broad taxonomy aligns with prior efforts to review and classify indices~\citep{beccari2016comparative}. There are potentially other classes of indices to consider, such as indices that use principal component analysis.}

Lastly, we consider \textit{predictive validity}, whether indices predict theoretically relevant and measurable outcomes. We examine the relationship of different indices to reasonable correlates of heat vulnerability: power outages, heat-related emergency (EMS) calls, and  311 complaints for open hydrants.

Prior work has already shown that HVIs (in other locations) can be sensitive to the choice of inputs and spatial scale, and only weakly correlated with heat-related health outcomes~\citep{conlon2020mapping, niu2021systematic}. Researchers have cautioned against using HVI tools for critical decision-making for these reasons~\citep{conlon2020mapping}. 
In focusing on the NYC HVI, we examine whether changes to the index can improve our understanding of heat vulnerability in NYC.
Drawing on measurement and validity theory, we show how conflicting measurements of risk (obtained from comparisons to other specifications, other indices, and historical data on heat-related correlates) may frustrate attempts to meaningfully use the NYC HVI in decision-making.

\begin{table*}[t!]
    \centering
    \setlength{\extrarowheight}{2pt}
    \adjustbox{max width=\textwidth}{\begin{tabular}{|>{\raggedright\arraybackslash}p{0.6in} >{\raggedright\arraybackslash}p{0.6in} >{\raggedright\arraybackslash}p{1.2in} >{\raggedright\arraybackslash}p{1.8in} >{\raggedright\arraybackslash}p{0.8in} >{\raggedright\arraybackslash}p{0.8in} >{\raggedright\arraybackslash}p{0.8in} |}
        \hline
        \textbf{Index Type} & \textbf{Tool} & \textbf{Who Uses It} & \textbf{Input Variables} & \textbf{Outputs} & \textbf{Spatial Resolution} & \textbf{Last Updated} \\
        \Xhline{2pt}
        Additive & NYC HVI & NYC City Agencies (NYCEM, Parks Department, DOHMH) & Sociodemographics (race, income); environmental factors (temperature, vegetation); air conditioner access  & Vulnerability score (1-5) & Neighborhood, zipcode & 2024 (every 3-5 years) \\
        \hline
                Formula & NRI & FEMA, Local emergency management agencies for hazard mitigation & Historical hazard information, social vulnerability characteristics~\citep{svi_cdc_site}, community resilience~\citep{nri_resilience_data} & National percentile rankings; a qualitative risk score 
        & Census tract, county & 2025 (four major updates since 2020) \\
        \hline
        Hierarchical & HHI & CDC or the general public; no mention in emergency management plans & EMS calls; prior heat events; sociodemographics (e.g.,  age, poverty, unemployment, etc.); comorbidities; environmental factors (e.g., tree-canopy) & National percentile rankings and a qualitative risk score  
        & Zipcode, ZCTA & 2024 (updated ``periodically'')  \\
        \hline 
    \end{tabular}}
    \caption{\textbf{Indices for extreme heat.} All indices rely on static sociodemographic and environmental information. To our knowledge, these are the most widely accessible data tools that cities can use for measuring heat risk. 
    These indices are taxonomized by index type, discussed further in the main text, and stand in contrast to predictive algorithms (which include spatial and temporal models).
    }
    \vspace{-2em}
    \label{tab:tool_comparison}
\end{table*}

\begin{table*}[htbp]
    \centering
    \setlength{\extrarowheight}{2pt}
    \begin{tabular}{| > {\raggedright\arraybackslash}p{1.5in}| > {\raggedright\arraybackslash}p{4.5in}|}
        \hline 
        Composite Indicator & Formula \\
        \Xhline{2pt}
         Original & $(Z_{\text{Avg. Surface Temp.}} + Z_{\text{\% Black Pop.}}) - (Z_{\text{\% Greenspace}} + Z_{\text{\% HH with AC}} + Z_{\text{\% Median HH Income}})$ \\
         Alt. 1: Environmental & $Z_{\text{Avg. Surface Temp.}} - Z_{\text{\% Greenspace}}$ \\
        Alt. 2: Seniors and poverty & $(Z_{\text{Avg. Surface Temp.}} + Z_{\text{\% Over 65}}  + Z_{\text{\% In Poverty}}) - (Z_{\text{\% Greenspace}} + Z_{\text{\% HH with AC}})$ \\
         Alt. 3: Comorbidities & $\text{Original} + Z_{\text{\% High Blood Pressure}} + \ldots + Z_{\text{\% Stroke}}$ \\
         Alt. 4: All & $\text{Original} + Z_{\text{\% Over 65}}  + Z_{\text{\% In Poverty}} + Z_{\text{\% High Blood Pressure}} + \ldots + Z_{\text{\% Stroke}} $\\
         \hline
    \end{tabular}
    \caption{\textbf{Four different alternative specifications.} The original NYC HVI formula is in the first row. The NYC HVI is the result of summing or subtracting z-scores (denoted ``Z'') for each of the data inputs. The alternative specifications are: (1) environmental characteristics only, (2) the percentage of individuals over 65 and the percentage in poverty from the 2016-2020 American Community Survey~\citep{acs_data} (in place of median household income and race), (3) the percentage of individuals with comorbidities such as high blood pressure from the 2024 CDC Places Data~\citep{cdc_places_data}, and (4) combined inputs (2) and (3).}
    \label{tab:hvi_specification_formulas}
    \vspace{-1em}
\end{table*}

\subsection{The NYC HVI Is Highly Sensitive to the Choice of Specification }

Inspired by concerns related to construct reliability, we examine the sensitivity of the NYC HVI to alternative specifications and design choices, such as spatial resolution. 
There is no standard methodology for developing an HVI. Common approaches include simple standardization and aggregation (e.g., z-scores and summing data inputs) along with principal component analysis~\citep{conlon2020mapping}.

We focus on a sensitivity analysis, as opposed to other kinds of evaluation that rely on ground-truth labels (e.g., accuracy), because defining and measuring ground truth outcomes is challenging. The NYC HVI, for example, was developed based on individual-level mortality data. Heat mortality data is both sparse and undercounted. In NYC, there are on average only five heat-stress deaths per year, though NYC DOHMH estimates that heat-exacerbated deaths are on average over 500~\citep{dohmh_mortality_2025}.
Additionally, heat mortality may not be the only outcome that decisionmakers care about when using the NYC HVI to inform a task.
Public health officials may care about emergency calls or heat-related morbidity (e.g., how many individuals show up at the emergency department with a different, but potentially heat-related diagnosis?). In emergency management, decisionmakers may care about power outages and access to cooling interventions such as air-conditioned community spaces or outdoor spray showers. If a sensitivity analysis indicates that some areas are highly likely to fluctuate, it may be prudent for policymakers to examine more closely the alignment between the initial development of the tool and their task-specific goals or consider using multiple indices (similar to the recommendations in \citep{huynh2024mitigating}). 

\begin{figure*}[t!]
    \centering
        \includegraphics[width=\linewidth]{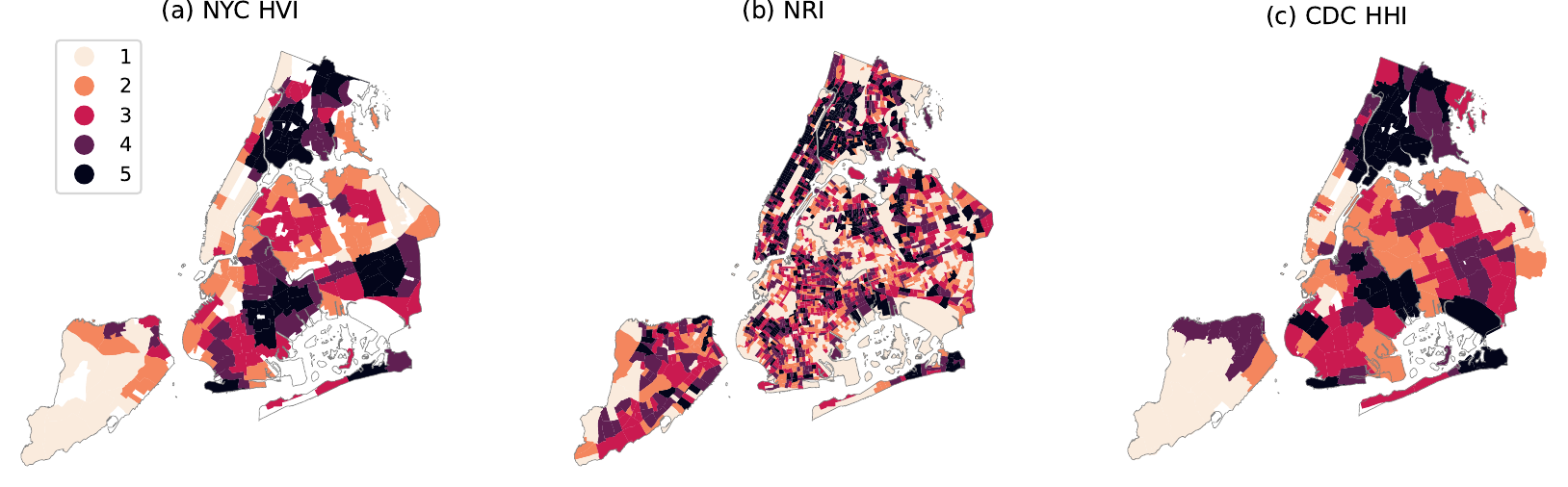}
        \newsubcap{\textbf{Comparison of different index tools.} We compare the NYC Heat Vulnerability Index (HVI), which combines and standardizes information on five sociodemographic and environmental characteristics related to extreme heat, to two other indices: the FEMA National Risk Index (NRI) and the CDC Heat and Health Index (HHI).
        The NYC HVI is at the neighborhood level (n=197), the NRI is at the census tract level (n=2,324), and the CDC HHI is at the 2010 zipcode tabulation level (n=183). We use quintiles to convert the underlying values for each index into a  5-point scale (1=lowest risk, 5=greatest risk).
        \vspace{-1em}
        }
        \label{fig:hvi}
        \Description[Three choropleth maps of NYC that illustrate different index tools.]{Three choropleth maps of NYC that illustrate different index tools: the NYC HVI, the NRI, and the CDC HHI. All maps use an orange and red colorscale that denotes five risk score categories ranging from 1 (low risk) to 5 (high risk).}
    \end{figure*}
    
    \begin{figure*}[htbp!]
    \centering
    \includegraphics[width=\linewidth]{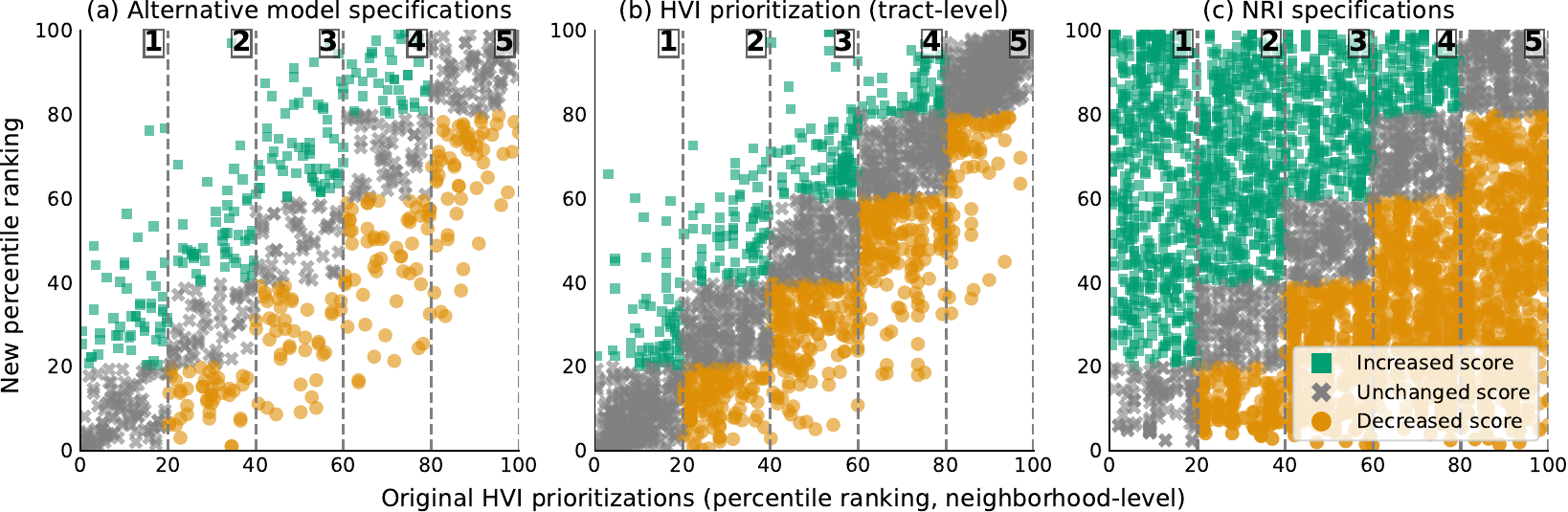}
        \caption{\textbf{Sensitivity of the NYC HVI.} 
        Each of the three graphs represents a different sensitivity analysis of the NYC HVI, corresponding to (a),(b) construct reliability, and (c) convergent validity.
        Substantial fluctuation is observed in each graph, as evidenced by the large number of green squares and gold dots, rather than points clustering in 5 gray boxes along the diagonal.
        Green, orange, and gray
        illustrate whether there is an increase, decrease, or no change to the original NYC HVI vulnerability score (ranging from 1-5).
        In (a), each point represents a combination of a neighborhood and specification. There are 197 neighborhoods and 4 alternative specifications (see Table~\ref{tab:hvi_specification_formulas}).
        In (b), each point is a census tract (n=2,240). 
        In (c), each point represents a combination of census tract (n=2,240) and NRI specification (n=2; Appendix~\ref{sec:appendix_nri}).
        }
        \Description[Three different scatter plots test the sensitivity of the NYC HVI.]{Three different scatter plots test the sensitivity of the NYC HVI. Figures (a) and (b) test construct reliability. Figure (c) tests convergent validity. The figure is described fully in the caption. In addition to the points that are green, gold, and gray, vertical lines denote the percentile rank thresholds that define the five point risk scores (from 1 to 5). The x and y axes both capture percentile rankings and range from 0 to 100.}
    \label{fig:hvi_alt}
\end{figure*}

To conduct the sensitivity analysis, we first evaluate the NYC HVI in relation to several alternative specifications (enumerated in Table~\ref{tab:hvi_specification_formulas}), and then evaluate its sensitivity to different spatial scales. For the alternative specifications, we measure changes in the HVI neighborhood percentile rankings with (1) environmental characteristics only, (2) additionally, information on age and poverty status instead of median household income and percent Black or African-American, (3) original inputs, as well as the percentage of individuals with different comorbidities,\footnote{We use the 2024 CDC Places Data to estimate comorbidities (see Appendix~\ref{sec:replication} for more details).} and (4) combined inputs from (2) and (3).
We choose these alternative specifications for the following reasons. Elderly individuals are a high-risk population for extreme heat~\citep{extreme_heat_beat}, and other HVI tools~\citep{reid2009mapping, nayak2018development} include inputs for both poverty and age. Individuals with comorbidities are another high-risk group~\citep{extreme_heat_beat}, but \citet{madrigano2015case}'s study, which influenced the HVI NYC, did not have access to comorbidity data at the time of their analysis.

Figure~\ref{fig:hvi_alt}(a) presents the effect of constructing the NYC HVI at the neighborhood level (n=197) under four alternative specifications (each of the 4 x 197 dots represents one alternative specification for each neighborhood, relative to the original specification). While there is a consistent upward trend (indicating overall correlation),\footnote{
We include a correlation analysis in Appendix~\ref{sec:correlation_alt_specifications}. Even moderate and strong correlations (in the range of 0.5 - 0.9) do not necessarily translate to high alignment across the coarser 1-5 category risk scores.} changing the inputs induces large shifts for some neighborhoods. 
For example, 9 neighborhoods with HVI scores of 1 or 2 would become high risk (HVI score = 4 or 5) under an alternative specification (represented in Figure~\ref{fig:hvi_alt}(a) by green squares with original HVI prioritization $\leq$ 40 and new percentile ranking > 60).
If changing the index specification did not have a robust effect on the output, we should expect to see only dots along a positive diagonal line (i.e., the gray boxes in each figure).
When considering all four alternative specifications, only 12\% of neighborhoods would not experience changes to their original NYC HVI score.
When considering any pairwise comparisons of alternative specifications, 48\% of neighborhoods would not experience changes, while 26\% would see an increase in HVI score.
Appendix Figures~\ref{fig:hvi_sens}-\ref{fig:hvi_sens_full} visualize these results spatially, illustrating how some specifications prioritize different clusters of neighborhoods throughout the city. 
 These fluctuations matter because higher HVI risk scores are typically used to prioritize neighborhoods for heat-related interventions such as opening new cooling centers or extending cooling center hours. For example, a 2022 NYC Comptroller report advocates for increasing cooling centers in underserved, heat vulnerable neighborhoods (HVI 4 or 5)~\citep{comptroller_report}. Similarly, the NYC °CoolRoofs Strategic Implementation Plan used the NYC HVI for neighborhood outreach~\citep{cool_neighborhoods}.
 
In Figure~\ref{fig:hvi_alt}(b), we test the sensitivity of the NYC HVI to spatial scale by re-creating the NYC HVI with census tracts, as opposed to neighborhoods. Let $i$ represent a census tract and $j$ represent a neighborhood. Census tracts are subsets of neighborhoods; each neighborhood $j$ contains many census tracts $i$. We first define the ``Original HVI prioritization'' as the neighborhood percentile ranking, which we assign to each corresponding census tract: $\forall i \in j : \text{HVI}^{\text{Original}}_{i} = \text{HVI}_j$. Then, we produce a new percentile ranking at the census tract level using the formula from Table~\ref{tab:hvi_specification_formulas}: $\text{HVI}^{\text{New}}_{i} = \text{RANK}\left( Z_{\text{Avg. Surface Temp.},i} + Z_{\text{\% Black Pop.},i} - (Z_{\text{\% Greenspace},i} + Z_{\text{\% HH with AC}} + Z_{\text{\% Median HH Income},i})\right)$.\footnote{The percentage of households with an air conditioning unit is not available at the census tract level. Additionally, as we discuss in Appendix~\ref{sec:replication}, there are potentially small numeric differences between our estimates and the original HVI due to differences in data sources.} As we observe in Figure~\ref{fig:hvi_alt}(b) and spatially in Appendix Figure~\ref{fig:hvi_tract}, changing the spatial scale can lead to large fluctuations for some tracts, even some originally categorized as low risk (e.g., two census tracts
fluctuate from HVI scores of 1 to 4). 
Known as the modifiable areal unit problem, this problem is well-studied in geography~\citep{openshaw1979million}.

To be clear, the sensitivity of an index to the choice of inputs and modeling specifications is not an undesirable property on its own. It may be useful for decision-makers to know that an index is sensitive to key inputs, particularly if the inputs are  meaningful to a policy or long-term goal.
Similarly, it is important to recognize what a tool is missing prior to deploying it in a new context. For example, individuals with comorbidities (such as heart conditions) are at greater risk of heat-related complications. By not including comorbidities in the index, emergency managers may miss high-risk neighborhoods where such individuals live -- particularly if comorbidities are not correlated with other inputs.

\subsection{The NYC HVI Diverges from Other Reasonable Indices}
Next, we assess the convergent validity of the NYC HVI relative to other indices for extreme heat (Figure~\ref{fig:hvi}). Low correlations between indices may be meaningful if the substantive goals of the indices differ but may raise concerns if the differences are unintentional. Though all of the indices in this section invoke notions of risk and vulnerability, as we discuss, the actual quantities they measure and the assumptions they make are more distinct.

\subsubsection{Comparison to the NRI}

We compare the NYC HVI scores to the FEMA National Risk Index (NRI)~\citep{fema_nri}. The NRI aims to provide a standardized measure of risk for 18 different hazard types. There is both an overall risk score and hazard-specific risk scores. We focus on risks associated with heat waves.

The NRI is not directly used in decision-making, but appears in several hazard mitigation plans (e.g., LA's hazard mitigation plan notes that LA County has the greatest climate risk according to the NRI~\citep{la_hmp}). 
It was developed over several rounds of expert review and measures the economic damage of natural disasters in dollars~\citep{nri_technical}. Risk is defined as the ``potential for negative impacts.'' It multiplies together the expected annual loss (EAL) of a hazard and a community's risk factor (social vulnerability divided by community resilience~\citep{nri_technical}).

Figure~\ref{fig:hvi_alt}(c) compares the NRI to the NYC HVI at the neighborhood level. We estimate the percentile rankings of census tracts within NYC based on two versions of the NRI: the EAL alone and the full NRI risk score (multiplying EAL and the community risk factor). We compare these percentile rankings to the percentile rankings obtained from the NYC HVI at the neighborhood level. Refer to Appendix~\ref{sec:appendix_nri} for further details.
The almost uniform distribution of the points suggests that the NRI and the NYC HVI are only weakly related.
As shown in Table~\ref{tab:correlation_matrix}, the Spearman correlation for the percentile rankings, though positive, is 0.07.

There are several key differences between the NYC HVI and NRI methodologies that likely contribute to these differences. Unlike the NYC HVI, which produces a composite index of relevant heat-related characteristics, the NRI considers historical losses related to heat incidents. For example, the EAL draws on the annualized frequency of extreme heat event-days that affect a census block. Two adjacent census tracts can have drastically different frequencies (see 
Appendix Figure~\ref{fig:nri_annualized_freq}) and these patterns do not bear any spatial resemblance to other temperature measures such as land surface temperature (shown in Appendix Figure~\ref{fig:lst_temperature}). The NRI also prioritizes quantifiable economic loss, such as building damage, agricultural loss, injury, and death. Person losses are converted to dollar amounts using the Value of Statistical Life formula. This approach standardizes the concept of loss, but this standardization may result in distorted comparisons -- especially if there is an important reason to prioritize one kind of loss over another. Lastly, even though the NRI produces relatively granular estimates at the census tract level, many of the NRI components rely on coarser estimates. For example, the ``Historic Loss Ratio,'' which estimates the rate of building, person, and agricultural loss, is a county-level estimate, as is community resilience. In major cities with large, diverse populations, county-level estimates obscure significant heterogeneity. 
Adapting the NRI with local data, such as the local NYC version of the index called the Urban Risk Index (URI)~\citep{uri}, can potentially address this problem. We discuss extensions of our analysis to the URI in Appendix~\ref{sec:uri}.

\subsubsection{Comparison to the CDC HHI}
\label{sec:cdc_hhi_comparison}

Another national-level index for extreme heat is the CDC's Heat and Health Index (HHI). The HHI 
takes a more explicit focus on health than the NYC HVI. Key components of the index include: (1) the number of extreme heat days and (2) heat-related illness~\citep{cdc_tracker}. 
The index is hierarchical, based on four sub-indices (using 23 data inputs overall) for  heat and health concerns, sociodemographics, community-level health outcomes (i.e., comorbidities), and the built environment. Percentile rankings for these sub-indices are averaged together using 2010 zipcode tabulation areas.\footnote{The Census uses zipcode tabulation areas (ZCTAs) to approximate zipcodes~\citep{census_definition}.} Mirroring Figure~\ref{fig:hvi_alt}, we find that the HHI similarly diverges from the HVI in Appendix Figure~\ref{fig:cdc_hhi_original_rankings}. A limitation in our implementation of the HHI is scope; we rely on national percentile rankings from the CDC and, as a result, our estimates may be less extreme compared to local tools like the NYC HVI. Spatially, we observe that the HHI does produce different risk assessment rankings: for example, the NYC HVI would prioritize areas in the eastern and southern parts of NYC, while the HHI would prioritize more northern areas (see Appendix Figures~\ref{fig:hvi_appendix}-\ref{fig:hhi}). Despite these differences, as shown in Table~\ref{tab:correlation_matrix}, the HHI is positively correlated with both the NYC HVI and the NRI. The correlations with the NYC HVI are stronger, which may reflect the shared focus on health.

\subsection{The NYC HVI Poorly Predicts Relevant Heat-Related Impacts}
\label{sec:heat-impacts}

We now compare the NYC HVI to several heat-related impacts, which are themselves correlates of heat vulnerability: power outages, heat-related EMS calls~\citep{ems_data}, and 311 hydrant complaints~\citep{311_data}.\footnote{311 is a non-emergency phone number and online web portal that allows New Yorkers to report on minor quality of life issues~\citep{311_website}.}
This approach supports an assessment of the \textit{predictive validity} of the NYC HVI.
Power outages frequently arise during extreme heat events due to stress on the electrical grid. Emergency managers often warn residents to reduce electricity consumption during heat waves~\citep{extreme_heat_beat}.
Prior work has also demonstrated links between power outages and emergency hospitalizations~\citep{do2025impact, deng2022independent, dominianni2018health}, and between EMS calls and extreme heat \citep{dolney2006relationship, seong2024spatio, calkins2016impacts, ke2023effects}. 311 data, while useful for tackling quality of life issues, has offered insight into residents' needs, especially after disasters like hurricanes~\citep{eugene2022using}.
Motivated by these examples and available public data, we compare these heat-related impacts to indices for heat. We connect these outcomes to relevant decisions in emergency management in Appendix~\ref{sec:heat_impacts_appendix}.

As shown in Table~\ref{tab:correlation_matrix_impacts}, correlations between index-based rankings and rankings based on heat-related impacts are much weaker compared to Table~\ref{tab:correlation_matrix}, and even negative. For example, the NYC HVI’s focus on heat-related mortality leads to low alignment with power outages ($\rho=0.163$). This comparison illustrates the problem with adapting an index from one domain to another without any critical reflection.

\begin{table*}[t!]
    \centering
\begin{minipage}[t]{0.44\textwidth}
    \vspace{0pt}
    \centering
    \adjustbox{max width=\textwidth}{\begin{tabular}[t]
    {l S[table-format=3.2] S[table-format=3.2] S[table-format=3.2] S[table-format=3.2]}
    \toprule
        & \parbox{0.8in}{\centering HVI} & \parbox{0.4in}{\centering HVI} & \parbox{0.4in}{\centering NRI}  \\
    & \parbox{0.8in}{\centering (Neighborhood)} & \parbox{0.4in}{\centering (Tract)} & \parbox{0.4in}{\centering (Tract)} \\
    \midrule
    HVI (Tract) & 0.89 & & \\
    NRI (Tract) & 0.07 & 0.106 & \\
    HHI (ZCTA) & 0.693 & 0.65 & 0.289 \\
    \bottomrule
    \end{tabular}}
    \caption{Spearman correlations at the census tract level (n=2,240) comparing the percentile rankings from the NYC HVI (at the neighborhood and census tract levels) to percentile rankings based on the NRI at the census tract level and the HHI at the zipcode tabulation area level. See Table~\ref{tab:correlation_matrix_kendalls} for Kendall's Tau. While all four indices are positively correlated, the strength of the association varies (e.g.,  the NRI is weakly correlated with the NYC HVI and the HHI).} 
    \label{tab:correlation_matrix}
    \end{minipage}
    \hfill
    \begin{minipage}[t]{0.54\textwidth}
      \vspace{0pt}
    \adjustbox{max width=\textwidth}{\begin{tabular}[t]  {l S[table-format=3.2] S[table-format=3.2] S[table-format=3.2] S[table-format=3.2]}
    
    \toprule
    & \parbox{0.8in}{\centering HVI} & \parbox{0.4in}{\centering HVI} & \parbox{0.4in}{\centering NRI} & \parbox{0.4in}{\centering HHI} \\
    & \parbox{0.8in}{\centering (Neighborhood)} & \parbox{0.4in}{\centering (Tract)} & \parbox{0.4in}{\centering (Tract)} & \parbox{0.4in}{\centering (ZCTA)} \\
\toprule
Power outage & 0.163 & 0.135 & -0.188 & 0.028 \\
EMS & 0.181 & 0.19 & 0.258 & 0.35  \\
Hydrant & 0.307 & 0.354 & 0.087 & 0.398 \\
\bottomrule
\end{tabular}}
    \caption{Spearman correlations comparing census tracts (n=2,240) ranked by indices for heat and heat-related impacts. See Table~\ref{tab:correlation_matrix_impacts_kendalls} for Kendall's Tau. The correlations here are much weaker compared to Table~\ref{tab:correlation_matrix}. We compare index-based percentile rankings to percentile rankings based on (1) average daily maximum power outage rates at the locality level, (2) total counts of heat-related EMS calls at the zipcode level~\citep{ems_data}, and (3) total counts of 311 hydrant complaints (per 1,000 residents) at the census tract level~\citep{311_data}. All heat impacts use data from May - September (2021-2025). }
    \label{tab:correlation_matrix_impacts}
    \end{minipage}
    \vspace{-1em}
\end{table*}

\subsection{Takeaways}

Capturing multi-dimensional concepts such as heat vulnerability is challenging, especially since defining ground truth outcomes is extremely difficult.
We advocate for evaluating the trade-offs associated with different tools through sensitivity analysis and concepts like construct reliability and validity. These methods are valuable both to a low-complexity index or more complex machine learning model.

Our evaluation of the NYC HVI's reliability and validity raises a number of concerns. First, the NYC HVI has low to moderate \textit{construct reliability}. Changes to input features and spatial scale can have unintended effects on index rankings. These changes may be desirable in some settings, but practitioners should be aware of them and consider whether the default inputs truly align with their goals. Differences between the NYC HVI and other indices like the NRI and HHI may reflect issues with \textit{convergent validity}. While these indices appear to measure the same concepts, the weak relationships we observe reflect deeper conflicts in values and priorities (e.g., prioritizing health or economic loss). Lastly, we highlight the NYC HVI's low \textit{predictive validity}, a consideration if relying on the NYC HVI for concrete, downstream tasks.

\begin{table*}[htbp!]
    \centering
    \adjustbox{max width=\textwidth}{%
    \begin{tabular}{|>{\raggedright\arraybackslash}p{0.6in}|> {\raggedright\arraybackslash}p{2.2in}|> {\raggedright\arraybackslash}p{2.2in}|> {\raggedright\arraybackslash}p{2.6in}|}
         \hline
         \textbf{Trade-off} & \textbf{Better-Suited: Index} & \textbf{Better-Suited: Algorithms} & \textbf{Recommendations} \\
        \hline
                Problem Formulation & 
                \textbullet~Abstract and high-level goals (e.g., vulnerability) \par
                \textbullet~When it is challenging to define a tractable data science task
                 & 
                 \textbullet~Narrow and measurable goals (e.g., reduce heat emergency calls) that can be translated into one or several discrete tasks
                 & Use indices for strategic planning and other abstract goals. Use predictive algorithms when there is an actionable goal that can be achieved via discrete tasks. \\
                \hline
                Outcome Selection  & \textbullet~Sparse outcomes (e.g., heat mortality) \par \textbullet~Outcomes are difficult to measure or subjective (e.g., well-being) \par \textbullet~Sensitive outcomes (e.g., health information) & \textbullet~Public or internal data (e.g., 311 data; power outages) \par \textbullet~Meaningfully triggers actions \par \textbullet~Reasonable to predict & Predictive algorithms rely on high-quality data to define outcomes (e.g., agency administrative data). If agency data is inaccessible or highly aggregated, combining public data sources into an index may be more feasible.
                \\
                \hline 
                Value Alignment & \textbullet~Explicit value choices (e.g., prioritize some groups based on selection of inputs) & \textbullet~Prevent unfairness (i.e., ensure equalized error rates, change decision thresholds) & When certain groups should be prioritized due to historical or social justifications, indices can make these value commitments more explicit. \\
                \hline
                Time Horizon & \textbullet~Long-term, and independent of recent or historical events \par \textbullet~\emph{Slow} data inputs & \textbullet~Triggered by recent events; outcomes are updated in real-time \par \textbullet~\emph{Fast} data inputs & Indices suit long-term planning initiatives. Predictive algorithms are better for short-term, time-dependent actions.  \\
                \hline 
                Validation and evaluation 
                & \textbullet~Validity is lower priority, and is often based on social acceptance and deliberative process \par \textbullet~Sensitivity analyses can surface validity and reliability concerns
               & \textbullet~Evaluation, validity, and measuring impact are all critical \par
                \textbullet~Holdout sets and model retraining easily support evaluation & Indices lead to validity concerns, and justification for use may depend on social acceptance and historical practices. Predictive algorithms may also incur validity issues but are more easily tested and evaluated further.\\
                \hline
                Stakeholder capacity & 
                \textbullet~Lower technical barrier upfront \par
                \textbullet~May require sensitivity analyses to ensure validity & \textbullet~Involves more resources upfront \par
                \textbullet~Setting up technical infrastructure requires organizational buy-in & Both are resource-intensive: indices should involve sensitivity analyses; predictive algorithms may require up-front investment (e.g., developing a custom pipeline). \\
                \hline 
                Audience & \textbullet~Accessible to a public audience, but potential for misinterpretation \par 
                \textbullet~Better for communication and sharing, rather than decision-making &
               \textbullet~Primarily for internal decision-making \par
               \textbullet~May be less meaningful and even confusing to a public audience & Indices may work better for communicating policy issues to a public audience. Predictive algorithms better support decision-making with internal stakeholders. \\
                \hline
    \end{tabular}}
    \caption{This table describes characteristics of the different kinds of tasks that are better suited to either \textit{indices} or \textit{predictive algorithms}. We enumerate the trade-offs that practitioners should consider. More detail on each point is in Appendix~\ref{sec:appendix_trade-offs}.
    \vspace{-1.5em}}
    \label{tab:takeaways}
\end{table*}

There is growing interest in leveraging predictive algorithms to improve government decision-making. But will the use of these algorithms generate real improvements? Answering this question requires analyzing the status quo.
Though the algorithmic fairness community has highlighted numerous risks with using algorithms in high-stakes decisions,  indices -- the status quo for many decisions -- are similarly fallible. 
We believe there may be opportunities to make reasonable improvements.
At the same time, there are also decisions where indices may continue to be more appropriate.
As research has shown, predictive algorithms are not a panacea. There are high-profile cases of algorithm abandonment (decisions to stop using an algorithmic tool to mitigate harm)~\citep{johnson2024fall}, or where machine learning tools exacerbate social problems~\citep{lum2016predict}.
As we discuss in the next section, a path forward involves better understanding the alignment between decision-making needs and available tools.

\section{A Path Forward}
\label{sec:discussion}

Generalizing out from our case study into the instability of the NYC HVI and concerns related to validity, we now discuss how practitioners can move forward in other settings with selecting a method. 
We outline seven trade-offs and recommendations for practitioners to consider in Table~\ref{tab:takeaways}.
These trade-offs represent broader themes based on observations and reflections from working with a government agency to develop prototypical predictive algorithms for potential deployment. 
Predictive algorithms may be preferable when:
(1) the problem formulation aligns with a narrow decision task, (2) objectives are clearly defined, (3) values are not explicit in the project goals, (4) decisions are time-sensitive, (5)  validation of results is needed or desired,
 (6) stakeholders can invest even limited resources, and (7) the intended audience includes internal stakeholders and domain experts.
We briefly discuss each of the trade-offs below, focusing on indices. Appendix~\ref{sec:appendix_trade-offs} provides further detail.

\textbf{Problem Formulation:}
Prior scholarship has noted how problem formulation is often the product of a ``negotiated translation''~\citep{passi2019problem}, which can lead to imperfect formulations and validity concerns~\citep{jacobs2021measurement, coston2023validity}. Indices may appear to involve fewer upfront constraints relative to predictive algorithms. For example, terms like ``vulnerability'' can remain abstract. However, indices still involve many implicit, value-laden assumptions.

\textbf{Outcome Selection:}
Indices work well when there are no clear or measurable outcomes, or when the outcomes are sparse and challenging to measure (e.g., heat-related mortality or well-being). Various indices in the literature relate to abstract and contested notions like vulnerability or inequality -- a ``latent variable'' that can be measured only indirectly~\citep{spielman2020evaluating, cutter2024origin}. Thus, indices may be preferable when decision-makers' goals are broad.

\textbf{Value alignment:}
Policymakers and government officials may want to intentionally prioritize certain groups. Indices can make this prioritization explicit, though we caution that design decisions may still lead to unexpected results. For example, even simple decisions like equal weighting of inputs (e.g., the NYC HVI) versus hierarchical sub-indices for specific input types (e.g., the CDC HHI) can lead to differences in risk scores.

\textbf{Time horizon:}
Indices may be preferred for long-term planning. They often rely on static or intrinsic spatial characteristics that are updated less frequently. Typically, these are independent of specific historical events.

\textbf{Validation and evaluation:}
Validation is challenging with indices, as there is no clear ground truth. Prior work has often resorted to correlation as a validation approach~\citep{cutter2024origin, tate2012social}.
To some stakeholders, validation may not be necessary, particularly if the goals of the index (as described above) are sufficiently broad and abstract. Indices may also gain validity from widespread social acceptance~\citep{cutter2024origin} or the development process (e.g., deliberative or transparent processes, similar to the ``inductive rule selection'' that \citep{johnson2022bureaucratic} observe).

\textbf{Stakeholder capacity:}
On the surface, indices may appear to have fewer technical barriers and constraints due to simpler methodologies. They are also easier to share and reproduce. At the same time, we recommend careful reflection and the use of sensitivity analyses (similar to our approach in this paper). Organizations may find that applying such methods to indices may involve levels of investment comparable to predictive algorithms.

\textbf{Audience:}
Indices like the NYC HVI and the NRI are accessible to the general public: the inputs are easy to define and use public data. Both indices, for example, feature prominently in hazard mitigation plans, which are public documents. Still, public audiences may not understand the limitations of indices and the assumptions implicit in their broad scope. 

\subsubsection*{Examples} We now discuss several examples that reflect the nuances of navigating the trade-offs in Table~\ref{tab:takeaways}.

    $\blacktriangleright$ \textbf{Warnings and emergency notifications:} A common response action is to send emergency alerts to residents -- e.g., warnings to reduce electricity usage. The goals are clear and narrow (i.e., reduce power outages), and there is a single event trigger (e.g., National Weather Service advisories).  Ensuring relevant information reaches those most severely affected is critical (as opposed to prioritizing specific subpopulations). Furthermore, as opposed to a wide-ranging set of stakeholders, decision-makers are primarily domain experts who are experienced with emergency response.  \textit{Predictive algorithms can help with prioritizing areas for targeted and follow-up messaging.}
    
    $\blacktriangleright$ \textbf{Outreach}:  When the temperature passes a certain threshold (known as a ``Code Red''~\citep{heat_response_plan}), outreach workers increase their efforts to offer services to individuals who are unhoused. Unlike low-cost interventions like text message alerts, outreach can be resource-intensive and slow. \textit{Predictive algorithms related to emergency service utilization could help with prioritizing neighborhoods where services are needed most.}
    
    $\blacktriangleright$ \textbf{Advertising and public awareness:} 
    Advertisements in bus shelters and on social media  provide information like the locations of cooling centers or requirements for applying to the Home Energy Assistance Program (HEAP). 
    One  example is New York City Emergency Management's  ``Beat the heat'' campaign~\citep{extreme_heat_beat}.
Planning for such campaigns is long and starts well before the summer.
There are also competing priorities: it may be equally important to notify low-income households about the availability of HEAP funding and to warn elderly populations about heat stroke.
Government agencies may want to prioritize specific subpopulations out of equity concerns, e.g., neighborhoods hardest-hit by COVID-19~\citep{trie_nhood}.
\textit{Index tools -- with stable rankings over time and the flexibility to incorporate hand-selected inputs -- may better align with such decisions.}

    $\blacktriangleright$ \textbf{Funding and resource allocation:} Indices can inform funding decisions (e.g., CalEnviroScreen~\citep{calenviroscreen}, the Social Vulnerability Index for post-disaster flood assistance~\citep{blackwood2023application}), but this approach may obscure assumptions that are consequential. Relying on indices also hampers efforts to evaluate the funding's effectiveness without defining clear outcomes to measure and track over time. \textit{We recommend exploring whether funding decisions relate to more discrete priorities and outcomes, which can support further evaluation and predictive modeling.}
    
    $\blacktriangleright$ \textbf{Physical interventions:} Indices can influence decisions related to physical interventions (e.g., locating cooling centers, warming buses, or interim flood protection measures). \textit{However, we suggest that practitioners explore whether there are relevant outcomes (e.g., emergency calls) that might inform placement instead. Sensitivity analyses can further help with comparing allocation decisions under both approaches.}

\subsubsection*{Generalizability}
In this work, we use a case study to examine the implementation of existing indices in a real-world setting. We believe our analyses provide concrete recommendations more generally to index-based risk assessments in a wide range of decision-making settings, such as in environmental justice (e.g., CalEnviroScreen~\citep{huynh2024mitigating}) or food insecurity and nutrition (e.g., Global Hunger Index~\citep{global_hunger_index}). This is because indices can describe local phenomena (e.g., the NYC HVI~\citep{dohmh_interactive}, the NYC Flood Vulnerability Index \citep{nyc_fvi}, or the NYC Displacement Risk Index \citep{nyc_dri}) or enable country-level comparisons (e.g., the United Nations Human Development Index~\citep{un_hdi}); in fact, investigating the impact of spatial scale is an important step in evaluating indices.
Our work creates a foundation for future work to study the implications of using indices in relation to predictive algorithms, alongside other bureaucratic counterfactuals. 
Appendix~\ref{sec:other_indices}, which provides a table listing other index tools, can inform such follow-up work.

\subsubsection*{Limitations}

There are several limitations to our work that we address here. First, we focus on a single case study. Researchers should cautiously apply our findings to other indices, and not without similar empirical investigations and further input from stakeholders. Our recommendations can serve as a starting point for these discussions. At the same time, we believe our empirical results are robustly supported by literature in social science and public policy that has noted similar challenges with indices~\citep{huynh2024mitigating, schmidtlein2008sensitivity, dobbie2013robustness, jones2007vulnerability, conlon2020mapping}, and articulated their strengths and weaknesses~\citep{surminski2014policy, kaiser2021should}, though none has directly contrasted indices with predictive algorithms.
Second, we only compare indices to predictive algorithms. Similar trade-offs could be identified and discussed for other bureaucratic counterfactuals like threshold-based triggers or heuristics. Threshold-based triggers may allow stakeholders to prioritize specific values and require little stakeholder investment (similar to indices). At the same time, they are relatively inflexible and narrow in scope. Heuristics, in turn, may be difficult to validate and lack transparency, but can enable fast responses.
Third, our work is centered on the U.S., given the focus of our case study.
However, extreme heat is a global issue. Future work should expand on our analysis and study heat-related indices in other countries (e.g., recently developed indices in India~\citep{india_heat_vulnerability} and Australia~\citep{australia_heat_vulnerability}).

\section{Conclusion}
\label{sec:conclusion}
Much of the algorithmic fairness literature is concerned with studying the impacts of allocation decisions made by predictive algorithms -- typically involving machine learning methods. However, consequential government decisions are often made with much simpler methods that may also have flaws. In this paper, we highlight the importance of studying existing status quo methods for decision-making, in order to fully assess algorithmic tools.
We examine existing methods for decision-making related to extreme heat in NYC. We discuss how prevalent indices are in government decision-making, and empirically demonstrate how concepts like \textit{construct reliability}, \textit{convergent validity}, and \textit{predictive validity} can help provide a theoretical basis for critiquing them. 
Though indices aim to capture abstract, multi-dimensional concepts, they can also be inconsistent and arbitrary. Predictive algorithms do not necessarily solve these challenges, but they introduce a new set of trade-offs to consider. We call on decision-makers to incorporate sensitivity analyses of indices into their work and to carefully reflect on these trade-offs when deciding whether to use an index in the future.

\begin{acks}
We thank all members of the KLEAR lab for feedback on early drafts and extended abstracts (including Anna Choi, Camille Harris, Daniel Molitor, Emma Harvey, Isabel Corpus, and Samantha Gold). We are also grateful to stakeholders including government partners who provided helpful context and domain expertise alongside this project. Lastly, we would like to thank Cornell Tech's Siegel Family Endowment PiTech PhD Impact Fellowship and Rubinstein PiTech PhD Innovation Fellowship, both of which supported the first author’s work on extreme heat over the course of the past year.

\end{acks}

\section{Generative AI Statement}
No generative AI was used to generate the text in this paper. Claude Sonnet 4.5 was used to draft prototype JavaScript code, which supported the development of an interactive webmap. This webmap helped guide discussions with relevant stakeholders in emergency management, and these discussions informed the scope of this paper.
Claude Sonnet 4.5 was used to check the code for errors prior to publication. All suggested changes were human reviewed and manually implemented.

\section{Ethical Considerations Statement}
All of the data reported here involves only public data with some level of spatial aggregation (census tract, zip code, etc.).
We do not report any data that can identify a single individual.
We also preserve the privacy of the local government agencies mentioned in this paper. All of the information that we cite related to the emergency actions of any government agency relies on public data sources.
We do not anticipate any adverse effects of this work. Instead, we hope that it contributes to a more nuanced understanding of government decision-making, and the trade-offs associated with different methodologies, particularly in the context of extreme heat emergencies.

\bibliographystyle{ACM-Reference-Format}
\bibliography{references}

@misc{huynh2024mitigating,
  title={Mitigating allocative tradeoffs and harms in an environmental justice data tool. Nature Machine Intelligence 6, 2 (01 Feb 2024), 187--194},
  author={Huynh, Benjamin Q and Chin, Elizabeth T and Koenecke, Allison and Ouyang, Derek and Ho, Daniel E and Kiang, Mathew V and Rehkopf, David H},
  year={2024}
}

@misc{dohmh_mortality_2025,
    title={2025 NYC Heat-Related Mortality Report},
    author = {New York City Health Department},
    url={https://a816-dohbesp.nyc.gov/IndicatorPublic/data-features/heat-report/}
}

@misc{nasa,
    title={Temperatures Rising: NASA Confirms 2024 Warmest Year on Record},
    author={Bardan, Roxana},
    year={2025},
    url={https://www.nasa.gov/news-release/temperatures-rising-nasa-confirms-2024-warmest-year-on-record/}
}

@article{zhao2021global,
  title={Global, regional, and national burden of mortality associated with non-optimal ambient temperatures from 2000 to 2019: a three-stage modelling study},
  author={Zhao, Qi and Guo, Yuming and Ye, Tingting and Gasparrini, Antonio and Tong, Shilu and Overcenco, Ala and Urban, Ale{\v{s}} and Schneider, Alexandra and Entezari, Alireza and Vicedo-Cabrera, Ana Maria and others},
  journal={The Lancet Planetary Health},
  volume={5},
  number={7},
  pages={e415--e425},
  year={2021},
  publisher={Elsevier}
}

@misc{govtech,
    title={NYC’s Data-Driven Future: 46 Algorithms and Counting},
    author={Davidson, Nikki},
    year={2024},
    url={https://www.govtech.com/biz/data/nycs-data-driven-future-46-algorithms-and-counting}
}

@article{floridi2020artificial,
  title={Artificial intelligence as a public service: Learning from Amsterdam and Helsinki},
  author={Floridi, Luciano},
  journal={Philosophy \& Technology},
  volume={33},
  number={4},
  pages={541--546},
  year={2020},
  publisher={Springer}
}

@article{engstrom2020government,
  title={Government by algorithm: Artificial intelligence in federal administrative agencies},
  author={Engstrom, David Freeman and Ho, Daniel E and Sharkey, Catherine M and Cu{\'e}llar, Mariano-Florentino},
  journal={NYU School of Law, Public Law Research Paper},
  number={20-54},
  year={2020}
}

@article{pencheva2020big,
  title={Big Data and AI--A transformational shift for government: So, what next for research?},
  author={Pencheva, Irina and Esteve, Marc and Mikhaylov, Slava Jankin},
  journal={Public Policy and Administration},
  volume={35},
  number={1},
  pages={24--44},
  year={2020},
  publisher={SAGE Publications Sage UK: London, England}
}

@inproceedings{johnson2022bureaucratic,
  author = {Johnson, Rebecca Ann and Zhang, Simone},
title = {What is the Bureaucratic Counterfactual? Categorical versus Algorithmic Prioritization in U.S. Social Policy},
year = {2022},
isbn = {9781450393522},
publisher = {Association for Computing Machinery},
address = {New York, NY, USA},
url = {https://doi.org/10.1145/3531146.3533223},
doi = {10.1145/3531146.3533223},
booktitle = {Proceedings of the 2022 ACM Conference on Fairness, Accountability, and Transparency},
pages = {1671–1682},
numpages = {12},
location = {Seoul, Republic of Korea},
series = {FAccT '22}
}

@misc{dohmh_interactive,
    title={Interactive Heat Vulnerability Index},
    url={https://a816-dohbesp.nyc.gov/IndicatorPublic/data-features/hvi/#:~:text=In%20NYC%2C%20an%20extreme%20heat,days%20reaching%20100%20%C2%B0F.},
    author={{NYC Environment and Health Data Portal}}
}

@misc{nyc_fvi,
    title={Flood Vulnerability Index},
    url={https://a816-dohbesp.nyc.gov/IndicatorPublic/data-features/flood-vulnerability-index/},
    author={{Mayor's Office of Climate and Environmental Justice}}
}

@misc{nyc_dri,
    title={Displacement Risk},
    url={https://a816-dohbesp.nyc.gov/IndicatorPublic/data-features/displacement-risk/},
    author={{NYC Departments of City Planning (DCP)} and {Housing Preservation and Development (HPD)}}
}

@misc{calenviroscreen,
    title={CalEnviroScreen},
    url={https://oehha.ca.gov/calenviroscreen},
    author={{Office of Environmental Health Hazard Assessment (OEHHA)}}
}

@misc{ejscreen,
    title={EJSCREEN: Environmental Justice Screening and Mapping Tool},
    url={https://19january2021snapshot.epa.gov/ejscreen_.html},
    author={{U.S. Environmental Protection Agency}}
}

@misc{global_hsi,
    title={Global Health Security Index},
    url={https://ghsindex.org/},
    author={{Nuclear Threat Initiative} and {Brown University School of Public Health Pandemic Center} and {Economist Impact}}
}

@article{kyrkou2022machine,
  title={Machine learning for emergency management: A survey and future outlook},
  author={Kyrkou, Christos and Kolios, Panayiotis and Theocharides, Theocharis and Polycarpou, Marios},
  journal={Proceedings of the IEEE},
  volume={111},
  number={1},
  pages={19--41},
  year={2022},
  publisher={IEEE}
}

@article{kimutis2024emergency,
  title={Emergency management short term response to extreme heat in the 25 most populated US cities},
  author={Kimutis, Nicholas and Wall, Tamara and Darrow, Lyndsey},
  journal={International Journal of Disaster Risk Reduction},
  volume={100},
  pages={104097},
  year={2024},
  publisher={Elsevier}
}

@article{madrigano2015case,
  title={A case-only study of vulnerability to heat wave--related mortality in New York City (2000--2011)},
  author={Madrigano, Jaime and Ito, Kazuhiko and Johnson, Sarah and Kinney, Patrick L and Matte, Thomas},
  journal={Environmental health perspectives},
  volume={123},
  number={7},
  pages={672--678},
  year={2015},
  publisher={NLM-Export}
}

@article{nayak2018development,
  title={Development of a heat vulnerability index for New York State},
  author={Nayak, Seema G and Shrestha, Srishti and Kinney, PL and Ross, Zev and Sheridan, SC and Pantea, CI and Hsu, WH and Muscatiello, Nicola and Hwang, Syni-An},
  journal={Public health},
  volume={161},
  pages={127--137},
  year={2018},
  publisher={Elsevier}
}

@article{conlon2020mapping,
  title={Mapping human vulnerability to extreme heat: A critical assessment of heat vulnerability indices created using principal components analysis},
  author={Conlon, Kathryn C and Mallen, Evan and Gronlund, Carina J and Berrocal, Veronica J and Larsen, Larissa and O’Neill, Marie S},
  journal={Environmental health perspectives},
  volume={128},
  number={9},
  pages={097001},
  year={2020}
}

@article{niu2021systematic,
  title={A systematic review of the development and validation of the heat vulnerability index: major factors, methods, and spatial units},
  author={Niu, Yanlin and Li, Zhichao and Gao, Yuan and Liu, Xiaobo and Xu, Lei and Vardoulakis, Sotiris and Yue, Yujuan and Wang, Jun and Liu, Qiyong},
  journal={Current climate change reports},
  volume={7},
  number={3},
  pages={87--97},
  year={2021},
  publisher={Springer}
}

@misc{noaa_heatrisk,
    title={NWS HeatRisk},
    author={National Weather Service},
    url={https://www.wpc.ncep.noaa.gov/heatrisk/}
}

@misc{fema_nri,
    title={The National Risk Index},
    author={Federal Emergency Management Agency},
    url={https://hazards.fema.gov/nri/}
}

@misc{cdc_tracker,
    title={{Heat \& Health Tracker}},
    author={Center for Disease Control},
    url={https://ephtracking.cdc.gov/Applications/heatTracker/}
}

@inproceedings{benami2021distributive,
  author = {Benami, Elinor and Whitaker, Reid and La, Vincent and Lin, Hongjin and Anderson, Brandon R. and Ho, Daniel E.},
title = {The Distributive Effects of Risk Prediction in Environmental Compliance: Algorithmic Design, Environmental Justice, and Public Policy},
year = {2021},
isbn = {9781450383097},
publisher = {Association for Computing Machinery},
address = {New York, NY, USA},
url = {https://doi.org/10.1145/3442188.3445873},
doi = {10.1145/3442188.3445873},
booktitle = {Proceedings of the 2021 ACM Conference on Fairness, Accountability, and Transparency},
pages = {90–105},
numpages = {16},
location = {Virtual Event, Canada},
series = {FAccT '21}
}

@article{dolney2006relationship,
  title={The relationship between extreme heat and ambulance response calls for the city of Toronto, Ontario, Canada},
  author={Dolney, Timothy J and Sheridan, Scott C},
  journal={Environmental research},
  volume={101},
  number={1},
  pages={94--103},
  year={2006},
  publisher={Elsevier}
}

@article{seong2024spatio,
  title={Spatio-temporal patterns of heat index and heat-related Emergency Medical Services (EMS)},
  author={Seong, Kijin and Jiao, Junfeng and Mandalapu, Akhil and Niyogi, Dev},
  journal={Sustainable Cities and Society},
  volume={111},
  pages={105562},
  year={2024},
  publisher={Elsevier}
}

@article{calkins2016impacts,
  title={Impacts of extreme heat on emergency medical service calls in King County, Washington, 2007--2012: relative risk and time series analyses of basic and advanced life support},
  author={Calkins, Miriam M and Isaksen, Tania Busch and Stubbs, Benjamin A and Yost, Michael G and Fenske, Richard A},
  journal={Environmental health},
  volume={15},
  number={1},
  pages={13},
  year={2016},
  publisher={Springer}
}

@article{ke2023effects,
  title={Effects of heatwave features on machine-learning-based heat-related ambulance calls prediction models in Japan},
  author={Ke, Deng and Takahashi, Kiyoshi and Takakura, Jun'ya and Takara, Kaoru and Kamranzad, Bahareh},
  journal={Science of the total environment},
  volume={873},
  pages={162283},
  year={2023},
  publisher={Elsevier}
}

@misc{extreme_heat_beat,
    title={Extreme Heat: Beat the Heat!},
    url={https://www.nyc.gov/site/em/ready/extreme-heat.page},
    author={New York City Emergency Management}
}

@misc{cool_neighborhoods,
    title={Cool Neighborhoods NYC: A Comprehensive Approach to Keep Communities Safe in Extreme Heat},
    url={https://www.nyc.gov/assets/orr/pdf/Cool_Neighborhoods_NYC_Report.pdf},
    author={The City of New York}
}

@misc{hazard_mitigation_heat,
    title={Extreme Heat},
    author={NYC Hazard Mitigation Plan},
    url={https://nychazardmitigation.com/documentation/hazard-profiles/extreme-heat/}
}

@misc{nri_technical,
    author={FEMA},
    title={National Risk Index: Technical Documentation},
    year={2025},
    url={https://www.fema.gov/sites/default/files/documents/fema_national-risk-index_technical-documentation.pdf}
}

@misc{heataction_nyc,
    title={City Officials Urge New Yorkers to Stay Safe During Extreme Heat},
    author={New York City Office of the Mayor},
    year={2025},
    url={https://www.nyc.gov/mayors-office/news/2025/07/city-officials-urge-new-yorkers-to-stay-safe-during-extreme-heat}
}

@misc{chat,
    title={California Heat Assessment Tool},
    url={https://www.cal-heat.org/},
    author={{Four Twenty Seven, Argos Analytics, Habitat Seven, and the Public Health Institute (PHI)}},
    publisher={California Natural Resources Agency}
}

@misc{la_hmp,
    title={2024 Local Hazard Mitigation Plan},
    author={The City of Los Angeles},
    year={2024},
    url={https://drive.google.com/drive/folders/1HlmWGoSF6Gzhn4yXdw5lS3ew-dLK5fMY}

}

@misc{la_hrp,
    title={Emergency Operations Plan: Adverse Weather Hazard-Specific Annex},
    url={https://emergency.lacity.gov/sites/g/files/wph1791/files/2023-02/Adverse%20Weather%20Annex_%202022_MASTER%20FINAL%20.%20aa.pdf},
    author={The City of Los Angeles},
    year={2023}
}

@article{harlan2013neighborhood,
  title={Neighborhood effects on heat deaths: social and environmental determinants of vulnerable places},
  author={Harlan, Sharon L and Declet-Barreto, Juan H and Stefanov, William L and Petitti, DB},
  journal={Environmental Health Perspectives},
  volume={121},
  number={2},
  pages={197--204},
  year={2013}
}

@article{reid2009mapping,
  title={Mapping community determinants of heat vulnerability},
  author={Reid, Colleen E and O’neill, Marie S and Gronlund, Carina J and Brines, Shannon J and Brown, Daniel G and Diez-Roux, Ana V and Schwartz, Joel},
  journal={Environmental health perspectives},
  volume={117},
  number={11},
  pages={1730},
  year={2009}
}

@misc{maricopa_hmp,
    title={Maricopa County Multi-Jurisdictional Hazard Mitigation Plan},
    author={Maricopa County Emergency Management},
    year={2021},
    url={https://www.maricopa.gov/DocumentCenter/View/75143/Maricopa-County-MJMHMP-2021_MJPT-Final-2021-11-02}
}

@misc{phoenix_hrp,
    title={2025 Heat Response Plan},
    author={City of Phoenix},
    year={2025},
    url={https://www.phoenix.gov/content/dam/phoenix/heatsite/documents/FINAL%20-%202025%20HEAT%20RESPONSE%20PLAN.pdf}
}

@misc{forestry_plan,
    title={NYC Urban Forest Agenda:Toward a Healthy, Resilient, Equitable, and Just New York City},
    author={Forest for All NYC},
    year={2021},
    url={{https://forestforall.nyc/wp-content/uploads/2021/06/NYC-Urban-Forest-Agenda-.pdf}}
}

@article{deng2022independent,
  title={The independent and synergistic impacts of power outages and floods on hospital admissions for multiple diseases},
  author={Deng, Xinlei and Friedman, Samantha and Ryan, Ian and Zhang, Wangjian and Dong, Guanghui and Rodriguez, Havidan and Yu, Fangqun and Huang, Wenzhong and Nair, Arshad and Luo, Gan and others},
  journal={Science of the total environment},
  volume={828},
  pages={154305},
  year={2022},
  publisher={Elsevier}
}

@article{dominianni2018health,
  title={Health impacts of citywide and localized power outages in New York City},
  author={Dominianni, Christine and Lane, Kathryn and Johnson, Sarah and Ito, Kazuhiko and Matte, Thomas},
  journal={Environmental Health Perspectives},
  volume={126},
  number={6},
  pages={067003},
  year={2018}
}

@article{do2025impact,
  title={The Impact of Power Outages on Cardiovascular Hospitalizations Among Medicare Fee-for-service Enrollees in New York State, 2017--2018},
  author={Do, Vivian and McBrien, Heather Kathleen and Edmondson, Donald and Kioumourtzoglou, Marianthi-Anna and Casey, Joan Allison},
  journal={Epidemiology},
  volume={36},
  number={4},
  pages={458--466},
  year={2025},
  publisher={LWW}
}

@article{eugene2022using,
  title={Using NYC 311 call center data to assess short-and long-term needs following Hurricane Sandy},
  author={Eugene, Adriana and Alpert, Naomi and Lieberman-Cribbin, Wil and Taioli, Emanuela},
  journal={Disaster Medicine and Public Health Preparedness},
  volume={16},
  number={4},
  pages={1447--1451},
  year={2022},
  publisher={Cambridge University Press}
}

@article{vaidyanathan2019assessment,
  title={Assessment of extreme heat and hospitalizations to inform early warning systems},
  author={Vaidyanathan, Ambarish and Saha, Shubhayu and Vicedo-Cabrera, Ana M and Gasparrini, Antonio and Abdurehman, Nabill and Jordan, Richard and Hawkins, Michelle and Hess, Jeremy and Elixhauser, Anne},
  journal={Proceedings of the National Academy of Sciences},
  volume={116},
  number={12},
  pages={5420--5427},
  year={2019},
  publisher={National Academy of Sciences}
}

@article{alkire2021global,
  author = {UNDP (United Nations Development Programme)},
  title = {2025 Global Multidimensional Poverty Index (MPI)},
  journal = {UNDP (United Nations Development Programme)},
  year = {2025},
  location = {New York},
  URL = {}
}

@article{balica2012flood,
  title={A flood vulnerability index for coastal cities and its use in assessing climate change impacts},
  author={Balica, Stefania F and Wright, Nigel George and Van der Meulen, Frank},
  journal={Natural hazards},
  volume={64},
  number={1},
  pages={73--105},
  year={2012},
  publisher={Springer}
}

@article{dobbie2013robustness,
  title={Robustness and sensitivity of weighting and aggregation in constructing composite indices},
  author={Dobbie, Melissa J and Dail, David},
  journal={Ecological Indicators},
  volume={29},
  pages={270--277},
  year={2013},
  publisher={Elsevier}
}

@article{greco2019methodological,
  title={On the methodological framework of composite indices: A review of the issues of weighting, aggregation, and robustness},
  author={Greco, Salvatore and Ishizaka, Alessio and Tasiou, Menelaos and Torrisi, Gianpiero},
  journal={Social indicators research},
  volume={141},
  number={1},
  pages={61--94},
  year={2019},
  publisher={Springer}
}

@article{grupp2004indicators,
  title={Indicators for national science and technology policy: how robust are composite indicators?},
  author={Grupp, Hariolf and Mogee, Mary Ellen},
  journal={Research policy},
  volume={33},
  number={9},
  pages={1373--1384},
  year={2004},
  publisher={Elsevier}
}

@article{grupp2010review,
  title={Review and new evidence on composite innovation indicators for evaluating national performance},
  author={Grupp, Hariolf and Schubert, Torben},
  journal={Research Policy},
  volume={39},
  number={1},
  pages={67--78},
  year={2010},
  publisher={Elsevier}
}

@misc{heat_response_plan,
    title={NYC Emergency Management
Heat Emergency Plan: Local Law 85 of 2020},
    year={2020},
    url={https://www.nyc.gov/assets/em/downloads/pdf/local_law_reports/ll85_cooling_center_report_2023_b.pdf}
}

@article{heuton2025spatiotemporal,
  title={Spatiotemporal forecasting of opioid-related fatal overdoses: towards best practices for modeling and evaluation},
  author={Heuton, Kyle and Kapoor, Jyontika and Shrestha, Shikhar and Stopka, Thomas J and Hughes, Michael C},
  journal={American Journal of Epidemiology},
  volume={194},
  number={6},
  pages={1776--1782},
  year={2025},
  publisher={Oxford University Press}
}

@article{spielman2020evaluating,
  title={Evaluating social vulnerability indicators: criteria and their application to the Social Vulnerability Index},
  author={Spielman, Seth E and Tuccillo, Joseph and Folch, David C and Schweikert, Amy and Davies, Rebecca and Wood, Nathan and Tate, Eric},
  journal={Natural hazards},
  volume={100},
  number={1},
  pages={417--436},
  year={2020},
  publisher={Springer}
}

@article{cutter2024origin,
  title={The origin and diffusion of the social vulnerability index (SoVI)},
  author={Cutter, Susan L},
  journal={International Journal of Disaster Risk Reduction},
  volume={109},
  pages={104576},
  year={2024},
  publisher={Elsevier}
}

@misc{trie_nhood,
    title={Neighborhoods},
    author={The Taskforce on Racial Inclusion \& Equity},
    url={https://www.nyc.gov/site/trie/about/neighborhoods.page}
}

@misc{int_hvi,
    title={Interactive Heat Vulnerability Index},
    author={The City of New York, Environment \& Health Data Portal},
    url={https://a816-dohbesp.nyc.gov/IndicatorPublic/data-features/hvi/}
}

@article{beccari2016comparative,
  title={A comparative analysis of disaster risk, vulnerability and resilience composite indicators},
  author={Beccari, Benjamin},
  journal={PLoS currents},
  volume={8},
    doi={https://doi.org/10.1371/currents.dis.453df025e34b682e9737f95070f9b970},
  year={2016}
}

@book{joint2008handbook,
  title={Handbook on constructing composite indicators: methodology and user guide},
  author={Joint Research Centre},
  year={2008},
  publisher={OECD publishing}
}

@article{cutter2003social,
author = {Cutter, Susan L. and Boruff, Bryan J. and Shirley, W. Lynn},
title = {Social Vulnerability to Environmental Hazards},
journal = {Social Science Quarterly},
volume = {84},
number = {2},
pages = {242-261},
doi = {https://doi.org/10.1111/1540-6237.8402002},
url = {https://onlinelibrary.wiley.com/doi/abs/10.1111/1540-6237.8402002},
eprint = {https://onlinelibrary.wiley.com/doi/pdf/10.1111/1540-6237.8402002},
year = {2003}
}

@article{lewis2002disease,
  title={Disease outbreak detection system using syndromic data in the greater Washington DC area},
  author={Lewis, Michael D and Pavlin, Julie A and Mansfield, Jay L and O’Brien, Sheilah and Boomsma, Louis G and Elbert, Yevgeniy and Kelley, Patrick W},
  journal={American journal of preventive medicine},
  volume={23},
  number={3},
  pages={180--186},
  year={2002},
  publisher={Elsevier}
}

@article{henning2004syndromic,
  title={What is syndromic surveillance?},
  author={Henning, Kelly J},
  journal={MMWR: Morbidity \& Mortality Weekly Report},
  volume={53},
  year={2004}
}

@misc{svi_cdc_site,
    title={Social Vulnerability Index},
    author={Agency for Toxic Substances and Disease Registry},
    url={https://www.atsdr.cdc.gov/place-health/php/svi/index.html}
}

@misc{coolit,
    title={Cool It! NYC},
    author={New York City Department of Parks \& Recreation},
    url={https://www.nycgovparks.org/about/health-and-safety-guide/cool-it-nyc}

}

@article{wang2024against,
  author = {Wang, Angelina and Kapoor, Sayash and Barocas, Solon and Narayanan, Arvind},
title = {Against Predictive Optimization: On the Legitimacy of Decision-making Algorithms That Optimize Predictive Accuracy},
year = {2024},
issue_date = {March 2024},
publisher = {Association for Computing Machinery},
address = {New York, NY, USA},
volume = {1},
number = {1},
url = {https://doi.org/10.1145/3636509},
doi = {10.1145/3636509},
journal = {ACM J. Responsib. Comput.},
month = mar,
articleno = {9},
numpages = {45}
}

@inproceedings{kasy2021fairness,
  author = {Kasy, Maximilian and Abebe, Rediet},
title = {Fairness, Equality, and Power in Algorithmic Decision-Making},
year = {2021},
isbn = {9781450383097},
publisher = {Association for Computing Machinery},
address = {New York, NY, USA},
url = {https://doi.org/10.1145/3442188.3445919},
doi = {10.1145/3442188.3445919},
booktitle = {Proceedings of the 2021 ACM Conference on Fairness, Accountability, and Transparency},
pages = {576–586},
numpages = {11},
location = {Virtual Event, Canada},
series = {FAccT '21}
}

@article{liu2025bridging,
  title={Bridging prediction and intervention problems in social systems},
  author={Liu, Lydia T and Raji, Inioluwa Deborah and Zhou, Angela and Guerdan, Luke and Hullman, Jessica and Malinsky, Daniel and Wilder, Bryan and Zhang, Simone and Adam, Hammaad and Coston, Amanda and others},
  journal={arXiv preprint arXiv:2507.05216},
  year={2025}
}

@article{stevenson2024algorithmic,
  title={Algorithmic risk assessment in the hands of humans},
  author={Stevenson, Megan T and Doleac, Jennifer L},
  journal={American Economic Journal: Economic Policy},
  volume={16},
  number={4},
  pages={382--414},
  year={2024},
  publisher={American Economic Association 2014 Broadway, Suite 305, Nashville, TN 37203-2425}
}

@article{barocas2023automated,
  title={When is automated decision making legitimate},
  author={Barocas, Solon and Hardt, Moritz and Narayanan, Arvind},
  journal={Fairness and Machine Learning: Limitations and Opportunities. The MIT Press, Cambridge, MA, USA},
  year={2023}
}

@inproceedings{mashiat2024beyond,
  author = {Mashiat, Tasfia and DiChristofano, Alex and Fowler, Patrick J. and Das, Sanmay},
title = {Beyond Eviction Prediction: Leveraging Local Spatiotemporal Public Records to Inform Action},
year = {2024},
isbn = {9798400704505},
publisher = {Association for Computing Machinery},
address = {New York, NY, USA},
url = {https://doi.org/10.1145/3630106.3658978},
doi = {10.1145/3630106.3658978},
booktitle = {Proceedings of the 2024 ACM Conference on Fairness, Accountability, and Transparency},
pages = {1383–1394},
numpages = {12},
location = {Rio de Janeiro, Brazil},
series = {FAccT '24}
}

@inproceedings{mashiat2025pays,
  title={Who Pays the RENT? Implications of Spatial Inequality for Prediction-Based Allocation Policies}, volume={8}, url={https://ojs.aaai.org/index.php/AIES/article/view/36666}, DOI={10.1609/aies.v8i2.36666}, abstractNote={AI-powered scarce resource allocation policies rely on predictions to target either specific individuals (e.g., high-risk) or settings (e.g., neighborhoods). Recent research on individual-level targeting demonstrates conflicting results; some models show that targeting is not useful when inequality is high, while other work demonstrates potential benefits. To study and reconcile this apparent discrepancy, we develop a stylized framework based on the Mallows model to understand how the spatial distribution of inequality affects the effectiveness of door-to-door outreach policies. We introduce the RENT (Relative Efficiency of Non-Targeting) metric, which we use to assess the effectiveness of targeting approaches compared with neighborhood-based approaches in preventing tenant eviction when high-risk households are more versus less spatially concentrated. We then calibrate the model parameters to eviction court records collected in a medium-sized city in the USA. Results demonstrate considerable gains in the number of high-risk households canvassed through individually targeted policies, even in a highly segregated metro area with concentrated risks of eviction. We conclude that apparent discrepancies in the prior literature can be reconciled by considering 1) the source of deployment costs and 2) the observed versus modeled concentrations of risk. Our results inform the deployment of AI-based solutions in social service provision that account for particular applications and geographies.}, number={2}, journal={Proceedings of the AAAI/ACM Conference on AI, Ethics, and Society}, author={Mashiat, Tasfia and Fowler, Patrick J. and Das, Sanmay}, year={2025}, month={Oct.}, pages={1686–1697}
  }

@article{allen2026ethical,
  title={Ethical challenges and opportunities for integrating predictive analytics in community-based overdose prevention},
  author={Allen, Bennett and Urmanche, Adelya and Curtis, Brenda and Fisher, Celia},
  journal={The Lancet Regional Health-Americas},
  volume={55},
  pages={101345},
  year={2026},
  publisher={Elsevier}
}

@article{allen2024provident,
  title={PROVIDENT: development and validation of a machine learning model to predict neighborhood-level overdose risk in Rhode Island},
  author={Allen, Bennett and Schell, Robert C and Jent, Victoria A and Krieger, Maxwell and Pratty, Claire and Hallowell, Benjamin D and Goedel, William C and Basta, Melissa and Yedinak, Jesse L and Li, Yu and others},
  journal={Epidemiology},
  volume={35},
  number={2},
  pages={232--240},
  year={2024},
  publisher={LWW}
}

@article{karasaki2024machine,
  title={Machine learning for environmental justice: Dissecting an algorithmic approach to predict drinking water quality in California},
  author={Karasaki, Seigi and Morello-Frosch, Rachel and Callaway, Duncan},
  journal={Science of The Total Environment},
  volume={951},
  pages={175730},
  year={2024},
  publisher={Elsevier}
}

@article{rosenblatt2024fairlyuncertain,
  title={FairlyUncertain: A Comprehensive Benchmark of Uncertainty in Algorithmic Fairness},
  author={Rosenblatt, Lucas and Witter, R Teal},
  journal={arXiv preprint arXiv:2410.02005},
  year={2024}
}

@article{benesse2024fairness,
  title={Fairness seen as global sensitivity analysis},
  author={B{\'e}nesse, Cl{\'e}ment and Gamboa, Fabrice and Loubes, Jean-Michel and Boissin, Thibaut},
  journal={Machine Learning},
  volume={113},
  number={5},
  pages={3205--3232},
  year={2024},
  publisher={Springer}
}

@inproceedings{simson2024one,
  author = {Simson, Jan and Pfisterer, Florian and Kern, Christoph},
title = {One Model Many Scores: Using Multiverse Analysis to Prevent Fairness Hacking and Evaluate the Influence of Model Design Decisions},
year = {2024},
isbn = {9798400704505},
publisher = {Association for Computing Machinery},
address = {New York, NY, USA},
url = {https://doi.org/10.1145/3630106.3658974},
doi = {10.1145/3630106.3658974},
booktitle = {Proceedings of the 2024 ACM Conference on Fairness, Accountability, and Transparency},
pages = {1305–1320},
numpages = {16},
location = {Rio de Janeiro, Brazil},
series = {FAccT '24}
}

@article{kuzucu2024uncertainty,
  title={Uncertainty as a fairness measure},
  author={Kuzucu, Selim and Cheong, Jiaee and Gunes, Hatice and Kalkan, Sinan},
  journal={Journal of Artificial Intelligence Research},
  volume={81},
  pages={307--335},
  year={2024}
}

@article{fawkes2024fragility,
  author = {Fawkes, Jake and Fishman, Nic and Andrews, Mel and Lipton, Zachary},
 booktitle = {Advances in Neural Information Processing Systems},
 doi = {10.52202/079017-4356},
 editor = {A. Globerson and L. Mackey and D. Belgrave and A. Fan and U. Paquet and J. Tomczak and C. Zhang},
 pages = {137105--137134},
 publisher = {Curran Associates, Inc.},
 title = {The Fragility of Fairness: Causal Sensitivity Analysis for Fair Machine Learning},
 url = {https://proceedings.neurips.cc/paper_files/paper/2024/file/f7be3ebca4980b59fe3f665011115395-Paper-Datasets_and_Benchmarks_Track.pdf},
 volume = {37},
 year = {2024}
}

@misc{comptroller_report,
    url={https://comptroller.nyc.gov/reports/overheated-underserved/},
    author={NYC Comptroller},
    year={2022},
    title={Overheated, Underserved: Expanding Cooling Center Access}
}

@inproceedings{passi2019problem,
  author = {Passi, Samir and Barocas, Solon},
title = {Problem Formulation and Fairness},
year = {2019},
isbn = {9781450361255},
publisher = {Association for Computing Machinery},
address = {New York, NY, USA},
url = {https://doi.org/10.1145/3287560.3287567},
doi = {10.1145/3287560.3287567},
booktitle = {Proceedings of the Conference on Fairness, Accountability, and Transparency},
pages = {39–48},
numpages = {10},
location = {Atlanta, GA, USA},
series = {FAT* '19}
}

@misc{dssg_scoping,
    title={Data Science Project Scoping Guide},
    url={https://datasciencepublicpolicy.org/our-work/tools-guides/data-science-project-scoping-guide/},
    author={Data Science for Social Good}
}

@inproceedings{jacobs2021measurement,
  author = {Jacobs, Abigail Z. and Wallach, Hanna},
title = {Measurement and Fairness},
year = {2021},
isbn = {9781450383097},
publisher = {Association for Computing Machinery},
address = {New York, NY, USA},
url = {https://doi.org/10.1145/3442188.3445901},
doi = {10.1145/3442188.3445901},
booktitle = {Proceedings of the 2021 ACM Conference on Fairness, Accountability, and Transparency},
pages = {375–385},
numpages = {11},
location = {Virtual Event, Canada},
series = {FAccT '21}
}

@inproceedings{coston2023validity,
  author={Coston, Amanda and Kawakami, Anna and Zhu, Haiyi and Holstein, Ken and Heidari, Hoda},
  booktitle={2023 IEEE Conference on Secure and Trustworthy Machine Learning (SaTML)}, 
  title={A Validity Perspective on Evaluating the Justified Use of Data-driven Decision-making Algorithms}, 
  year={2023},
  volume={},
  number={},
  pages={690-704},
  doi={10.1109/SaTML54575.2023.00050}}

@article{hardt2016equality,
  author = {Hardt, Moritz and Price, Eric and Srebro, Nati},
 booktitle = {Advances in Neural Information Processing Systems},
 editor = {D. Lee and M. Sugiyama and U. Luxburg and I. Guyon and R. Garnett},
 pages = {},
 publisher = {Curran Associates, Inc.},
 title = {Equality of Opportunity in Supervised Learning},
 url = {https://proceedings.neurips.cc/paper_files/paper/2016/file/6a9659feb1216f14f7384ba499518b38-Paper.pdf},
 volume = {29},
 year = {2016}
}

@inproceedings{so2022beyond,
  author = {So, Wonyoung and Lohia, Pranay and Pimplikar, Rakesh and Hosoi, A.E. and D'Ignazio, Catherine},
title = {Beyond Fairness: Reparative Algorithms to Address Historical Injustices of Housing Discrimination in the US},
year = {2022},
isbn = {9781450393522},
publisher = {Association for Computing Machinery},
address = {New York, NY, USA},
url = {https://doi.org/10.1145/3531146.3533160},
doi = {10.1145/3531146.3533160},
booktitle = {Proceedings of the 2022 ACM Conference on Fairness, Accountability, and Transparency},
pages = {988–1004},
numpages = {17},
location = {Seoul, Republic of Korea},
series = {FAccT '22}
}

@article{rodolfa2021empirical,
  title={Empirical observation of negligible fairness--accuracy trade-offs in machine learning for public policy},
  author={Rodolfa, Kit T and Lamba, Hemank and Ghani, Rayid},
  journal={Nature Machine Intelligence},
  volume={3},
  number={10},
  pages={896--904},
  year={2021},
  publisher={Nature Publishing Group UK London}
}

@inproceedings{bell2022s,
  author = {Bell, Andrew and Solano-Kamaiko, Ian and Nov, Oded and Stoyanovich, Julia},
title = {It’s Just Not That Simple: An Empirical Study of the Accuracy-Explainability Trade-off in Machine Learning for Public Policy},
year = {2022},
isbn = {9781450393522},
publisher = {Association for Computing Machinery},
address = {New York, NY, USA},
url = {https://doi.org/10.1145/3531146.3533090},
doi = {10.1145/3531146.3533090},
booktitle = {Proceedings of the 2022 ACM Conference on Fairness, Accountability, and Transparency},
pages = {248–266},
numpages = {19},
location = {Seoul, Republic of Korea},
series = {FAccT '22}
}

@article{tate2013uncertainty,
  title={Uncertainty analysis for a social vulnerability index},
  author={Tate, Eric},
  journal={Annals of the association of American geographers},
  volume={103},
  number={3},
  pages={526--543},
  year={2013},
  publisher={Taylor \& Francis}
}

@article{tate2012social,
  title={Social vulnerability indices: a comparative assessment using uncertainty and sensitivity analysis},
  author={Tate, Eric},
  journal={Natural hazards},
  volume={63},
  number={2},
  pages={325--347},
  year={2012},
  publisher={Springer}
}

@article{schmidtlein2008sensitivity,
  title={A sensitivity analysis of the social vulnerability index},
  author={Schmidtlein, Mathew C and Deutsch, Roland C and Piegorsch, Walter W and Cutter, Susan L},
  journal={Risk Analysis: An International Journal},
  volume={28},
  number={4},
  pages={1099--1114},
  year={2008},
  publisher={Wiley Online Library}
}

@inproceedings{semenova2022existence,
  author = {Semenova, Lesia and Rudin, Cynthia and Parr, Ronald},
title = {On the Existence of Simpler Machine Learning Models},
year = {2022},
isbn = {9781450393522},
publisher = {Association for Computing Machinery},
address = {New York, NY, USA},
url = {https://doi.org/10.1145/3531146.3533232},
doi = {10.1145/3531146.3533232},
booktitle = {Proceedings of the 2022 ACM Conference on Fairness, Accountability, and Transparency},
pages = {1827–1858},
numpages = {32},
location = {Seoul, Republic of Korea},
series = {FAccT '22}
}

@book{lipsky2010street,
 ISBN = {9780871545244},
 URL = {http://www.jstor.org/stable/10.7758/9781610447713},
 author = {Michael LIpsky},
 publisher = {Russell Sage Foundation},
 title = {Street Level Bureaucracy: Dilemmas of the Individual in Public Services},
 urldate = {2026-01-25},
 year = {1980}
}

@article{roberts2019decision,
  title={Decision biases and heuristics among emergency managers: just like the public they manage for?},
  author={Roberts, Patrick S and Wernstedt, Kris},
  journal={The American Review of Public Administration},
  volume={49},
  number={3},
  pages={292--308},
  year={2019},
  publisher={SAGE Publications Sage CA: Los Angeles, CA}
}

@article{nearing2024global,
  title={Global prediction of extreme floods in ungauged watersheds},
  author={Nearing, Grey and Cohen, Deborah and Dube, Vusumuzi and Gauch, Martin and Gilon, Oren and Harrigan, Shaun and Hassidim, Avinatan and Klotz, Daniel and Kratzert, Frederik and Metzger, Asher and others},
  journal={Nature},
  volume={627},
  number={8004},
  pages={559--563},
  year={2024},
  publisher={Nature Publishing Group UK London}
}

@article{tran2025ai,
  title={AI improves the accuracy, reliability, and economic value of continental-scale flood predictions},
  author={Tran, Vinh Ngoc and Kim, Taeho and Xu, Donghui and Tran, Hoang and Le, Manh-Hung and Tran, Thanh-Nhan-Duc and Kim, Jongho and Tran, Trung Duc and Wright, Daniel B and Restrepo, Pedro and others},
  journal={AGU Advances},
  volume={6},
  number={3},
  pages={e2025AV001678},
  year={2025},
  publisher={Wiley Online Library}
}

@misc{natgeo_article,
    title={AI is helping seismologists find the next monster earthquake},
    author={Robin George Andrews},
    year={2024},
    publisher={National Geographic},
    url={https://www.nationalgeographic.com/science/article/ai-predict-earthquakes-seismology}
}

@article{lentz2022information,
  title={How do information problems constrain anticipating, mitigating, and responding to crises?},
  author={Lentz, Erin C and Maxwell, Daniel},
  journal={International Journal of Disaster Risk Reduction},
  volume={81},
  pages={103242},
  year={2022},
  publisher={Elsevier}
}

@misc{tufts_feinstein,
    title={Early Warning and Early Action for Increased Resilience of Livelihoods in the IGAD Region},
    year={2021},
    author={Maxwell, Daniel and Lentz, Erin and Simmons, Cori and Gottlieb, Gregory},
    url={https://fic.tufts.edu/wp-content/uploads/EW-EA-Executive-Summary-6-22.pdf},
    publisher={Tufts}
}

@inproceedings{johnson2025legacy,
    author = {Johnson, Nari and Silva, Elise and Leon, Harrison and Eslami, Motahhare and Schwanke, Beth and Dotan, Ravit and Heidari, Hoda},
    title = {Legacy Procurement Practices Shape How U.S. Cities Govern AI: Understanding Government Employees' Practices, Challenges, and Needs},
    year = {2025},
    isbn = {9798400714825},
    publisher = {Association for Computing Machinery},
    address = {New York, NY, USA},
    url = {https://doi.org/10.1145/3715275.3732049},
    doi = {10.1145/3715275.3732049},
    booktitle = {Proceedings of the 2025 ACM Conference on Fairness, Accountability, and Transparency},
    pages = {772–789},
    numpages = {18},
    location = {
    },
    series = {FAccT '25}
}

@misc{apwa,
    title={Artificial Intelligence and the Emergency Services Sector - Benefits and Challenges},
    year={2024},
    author={Cybersecurity and Infrastructure Security Agency},
    url={https://www.apwa.org/wp-content/uploads/Artificial-Intelligence-and-the-Emergency-Services-Sector-Case-Studies-Benefits-and-Challenges.pdf}
}

@article{frank2010development,
  title={The development of a walkability index: application to the Neighborhood Quality of Life Study},
  author={Frank, Lawrence D and Sallis, James F and Saelens, Brian E and Leary, Lauren and Cain, Kelli and Conway, Terry L and Hess, Paul M},
  journal={British journal of sports medicine},
  volume={44},
  number={13},
  pages={924--933},
  year={2010},
  publisher={British Association of Sport and Excercise Medicine}
}

@inproceedings{johnson2024fall,
  author = {Johnson, Nari and Moharana, Sanika and Harrington, Christina and Andalibi, Nazanin and Heidari, Hoda and Eslami, Motahhare},
title = {The Fall of an Algorithm: Characterizing the Dynamics Toward Abandonment},
year = {2024},
isbn = {9798400704505},
publisher = {Association for Computing Machinery},
address = {New York, NY, USA},
url = {https://doi.org/10.1145/3630106.3658910},
doi = {10.1145/3630106.3658910},
booktitle = {Proceedings of the 2024 ACM Conference on Fairness, Accountability, and Transparency},
pages = {337–358},
numpages = {22},
location = {Rio de Janeiro, Brazil},
series = {FAccT '24}
}

@article{lum2016predict,
  title={To predict and serve?},
  author={Lum, Kristian and Isaac, William},
  journal={Significance},
  volume={13},
  number={5},
  pages={14--19},
  year={2016},
  publisher={Oxford University Press}
}

@article{hamajima1999detection,
  title={Detection of Gene—Environment Interaction by Case-only Studies},
  author={Hamajima, Nobuyuki and Yuasa, Hidemichi and Matsuo, Keitaro and Kurobe, Yohko},
  journal={Japanese journal of clinical oncology},
  volume={29},
  number={10},
  pages={490--493},
  year={1999},
  publisher={Foundation for Promotion of Cancer Research}
}

@article{khoury1996nontraditional,
  title={Nontraditional epidemiologic approaches in the analysis of gene environment interaction: case-control studies with no controls!},
  author={Khoury, Muin J and Flanders, W Dana},
  journal={American journal of epidemiology},
  volume={144},
  number={3},
  pages={207--213},
  year={1996},
  publisher={Oxford University Press}
}

@article{messick1987validity,
  title={Validity},
  author={Messick, Samuel},
  journal={ETS research report series},
  volume={1987},
  number={2},
  pages={i--208},
  year={1987},
  publisher={Wiley Online Library}
}

@article{cronbach1955construct,
  title={Construct validity in psychological tests.},
  author={Cronbach, Lee J and Meehl, Paul E},
  journal={Psychological bulletin},
  volume={52},
  number={4},
  pages={281},
  year={1955},
  publisher={American Psychological Association}
}

@ArtifactDataset{cdc_places_data,
author = {Centers for Disease Control and Prevention},
title = {PLACES},
year=2023,
url = {https://www.cdc.gov/places},
lastaccessed = {December 1, 2025} }

@ArtifactDataset{acs_data,
author = {U.S. Census Bureau},
title = {American Community Survey 5-Year Estimates: Comparison Profiles 5-Year},
year = 2020,
url = {http://api.census.gov/data/2023/acs/acs5},
lastaccessed = {December 1, 2025} }

@ArtifactDataset{dec_data,
author = {U.S. Census Bureau},
title = {Decennial Census: Redistricting Data (PL 94-171)},
year = 2020,
url = {https://api.census.gov/data/2020/dec/pl.html},
lastaccessed = {December 1, 2025} }

@ArtifactDataset{ecostress,
author = {Hook, Simon and Hulley, Glynn},
title = {ECOSTRESS Land Surface Temperature and Emissivity},
year = 2019,
url = {https://doi.org/10.5067/ECOSTRESS/ECO2LSTE.001},
lastaccessed = {December 1, 2025} }

@ArtifactDataset{nri_data,
author = {Federal Emergency Management Agency},
title = {National Risk Index Census Tracts},
year = 2021,
url = {https://resilience.climate.gov/datasets/FEMA::national-risk-index-census-tracts/about},
lastaccessed = {December 1, 2025} }

@ArtifactDataset{cdc_hhi_data,
author = {United States Department of Health and Human Services},
title = {Heat and Health Index},
year = 2024,
url = {https://ephtracking.cdc.gov/Applications/heatTracker/},
lastaccessed = {December 1, 2025} }

@ArtifactDataset{nri_resilience_data,
author = {Hazards Vulnerability and Resilience Institute.},
title = { Baseline Resilience Indicators for Communities (BRIC) Index.},
year = 2025,
publisher={ Inter-university Consortium for Political and Social Research},
url = {https://doi.org/10.3886/E209669V2},
lastaccessed = {December 1, 2025} }

@ArtifactDataset{hvi_nta_level_dohmh,
    author = {NYC Department of Health and Mental Hygiene},
    url={https://a816-dohbesp.nyc.gov/IndicatorPublic/data-features/hvi/hvi-nta-2020.csv},
    title={Heat Vulnerability Index},
    lastaccessed = {December 1, 2025} }

@ArtifactDataset{hvi_zcta_level_dohmh,
    author = {NYC Department of Health and Mental Hygiene},
    url={https://data.cityofnewyork.us/Health/Heat-Vulnerability-Index-Rankings/4mhf-duep/about_data},
    publisher={NYC Open Data},
    year={2024},
    title={Heat Vulnerability Index Rankings},
    lastaccessed = {December 1, 2025} }

@ArtifactDataset{ems_data,
    author = {Fire Department of New York City (FDNY)},
    url={https://data.cityofnewyork.us/Public-Safety/EMS-Incident-Dispatch-Data/76xm-jjuj/about_data},
    publisher={NYC Open Data},
    year={2025},
    title={EMS Incident Dispatch Data},
    lastaccessed = {May 12, 2026} }

@ArtifactDataset{311_data,
    author = {Office of Technology and Innovation (OTI)},
    url={https://data.cityofnewyork.us/Social-Services/311-Service-Requests-from-2020-to-Present/erm2-nwe9/about_data},
    publisher={NYC Open Data},
    year={2026},
    title={311 Service Requests from 2020 to Present},
    lastaccessed = {May 12, 2026} }

@ArtifactDataset{landcover_data,
    title = {Land Cover Raster Data (2017) – 6in Resolution},
    url={https://data.cityofnewyork.us/Environment/Land-Cover-Raster-Data-2017-6in-Resolution/he6d-2qns/about_data},
    publisher={NYC Open Data},
    year={2017},
    author={NYC Office of Technology and Innovation (OTI)},
    lastaccessed = {December 1, 2025} }

@ArtifactDataset{uhp_lst_data,
    title = {Landsat 8-9 Operational Land Imager / Thermal Infrared Sensor Level-2, Collection 2 },
    url={https://urbanheat.nyc/#/download},
    publisher={Earth Resources Observation and Science (EROS) Center, NYC Urban Heat Portal},
    year={2020},
    author={Earth Resources Observation and Science (EROS) Center},
    lastaccessed = {December 1, 2025} }

@article{openshaw1979million,
  title={A million or so correlation coefficients: three experiments on the modifiable area unit problem.[In] Wrigley, H.(ed) Statistical applications in the spatial sciences},
  author={Openshaw, S and Taylor, PJ},
  journal={Pion: London. Pp127-144},
  year={1979}
}

@misc{311_website,
    title={About NYC311},
    url={https://portal.311.nyc.gov/about-nyc-311/},
    author={The City of New York},
    year={2025}
}

@ArtifactDataset{sheldus_data,
author={ASU Center for Emergency Management and Homeland Security},
year={2025},
title={The Spatial Hazard Events and Losses Database for the United States, Version 23.0},
url={https://sheldus.org},
    lastaccessed = {December 1, 2025} }

@ArtifactDataset{census_resilience,
    title={Community Resilience Estimates},
    author={U.S. Census Bureau},
    url={https://www.census.gov/programs-surveys/community-resilience-estimates.html},
    lastaccessed = {December 1, 2025}}

@article{jung2020simple,
  title={Simple rules to guide expert classifications},
  author={Jung, Jongbin and Concannon, Connor and Shroff, Ravi and Goel, Sharad and Goldstein, Daniel G},
  journal={Journal of the Royal Statistical Society Series A: Statistics in Society},
  volume={183},
  number={3},
  pages={771--800},
  year={2020},
  publisher={Oxford University Press}
}

@article{surminski2014policy,
  title={Policy indexes as tools for decision makers: the case of climate policy},
  author={Surminski, Swenja and Williamson, Andrew},
  journal={Global Policy},
  volume={5},
  number={3},
  pages={275--285},
  year={2014},
  publisher={Wiley Online Library}
}

@misc{un_hdi,
    title={Human Development Index (HDI)},
    publisher={United Nations Development Programme},
    year={2025},
    utl={https://hdr.undp.org/data-center/human-development-index#/indicies/HDI}
}

@misc{india_heat_vulnerability,
    title={How Extreme Heat is Impacting India},
    url={https://www.ceew.in/publications/mapping-climate-risks-and-impacts-of-extreme-heatwave-disaster-in-indian-districts},
    year={2025},
    authors={Prabhu, Shravan  and Suresh, Keerthana Anthikat  and Mandal, Srishti and Sharma, Divyanshu  andChitale,  Vishwas},
    publisher={{Council on Energy, Environment and Water}}
}

@misc{australia_heat_vulnerability,
    title={How vulnerable are Australia’s cities to extreme heat? Explore our maps},
    url={https://www.theguardian.com/environment/2025/sep/18/climate-crisis-how-vulnerable-australian-cities-sydney-melbourne-extreme-heat-explore-our-maps},
    author={Josh Nicholas},
    publisher={The Guardian},
    year={2025}
}

@article{jones2007vulnerability,
  title={Vulnerability index construction: methodological choices and their influence on identifying vulnerable neighbourhoods},
  author={Jones, Brenda and Andrey, Jean},
  journal={International journal of emergency management},
  volume={4},
  number={2},
  pages={269--295},
  year={2007},
  publisher={Inderscience Publishers}
}

@article{kaiser2021should,
  title={Should policy makers trust composite indices? A commentary on the pitfalls of inappropriate indices for policy formation},
  author={Kaiser, Matthias and Chen, Andrew Tzer-Yeu and Gluckman, Peter},
  journal={Health research policy and systems},
  volume={19},
  number={1},
  pages={40},
  year={2021},
  publisher={Springer}
}

@article{saltelli2020five,
  title={Five ways to ensure that models serve society: a manifesto},
  author={Saltelli, Andrea and Bammer, Gabriele and Bruno, Isabelle and Charters, Erica and Di Fiore, Monica and Didier, Emmanuel and Nelson Espeland, Wendy and Kay, John and Lo Piano, Samuele and Mayo, Deborah and others},
  journal={Nature},
  volume={582},
  number={7813},
  pages={482--484},
  year={2020},
  publisher={Nature Publishing Group UK London}
}

@article{razavi2020global,
  title={The Global Health Security Index: what value does it add?},
  author={Razavi, Ahmed and Erondu, Ngozi A and Okereke, Ebere},
  journal={BMJ global health},
  volume={5},
  number={4},
  year={2020},
  publisher={BMJ Publishing Group Ltd}
}

@article{abbey2020global,
  title={The Global Health Security Index is not predictive of coronavirus pandemic responses among Organization for Economic Cooperation and Development countries},
  author={Abbey, Enoch J and Khalifa, Banda AA and Oduwole, Modupe O and Ayeh, Samuel K and Nudotor, Richard D and Salia, Emmanuella L and Lasisi, Oluwatobi and Bennett, Seth and Yusuf, Hasiya E and Agwu, Allison L and others},
  journal={PloS one},
  volume={15},
  number={10},
  pages={e0239398},
  year={2020},
  publisher={Public Library of Science San Francisco, CA USA}
}

@article{aitken2020rethinking,
  title={Rethinking pandemic preparation: Global Health Security Index (GHSI) is predictive of COVID-19 burden, but in the opposite direction},
  author={Aitken, Tess and Chin, Ken Lee and Liew, Danny and Ofori-Asenso, Richard},
  journal={The Journal of infection},
  volume={81},
  number={2},
  pages={318},
  year={2020}
}

@misc{orgcode2015vulnerability,
  title={Vulnerability Index-Service Prioritization Decision Assistance Tool (VI-SPDAT) Prescreen Triage Tool for Single Adults},
  author={OrgCode Consulting},
  year={2015},
  publisher={Community Solutions}
}

@article{maxwell2003coping,
  title={The coping strategies index: A tool for rapidly measuring food security and the impact of food aid programs in emergencies},
  author={Maxwell, Daniel and Watkins, Ben and Wheeler, Robin and Collins, Greg and others},
  journal={Nairobi: CARE Eastern and Central Africa Regional Management Unit and the World Food Programme Vulnerability Assessment and Mapping Unit},
  year={2003},
  publisher={Cooperative for Assistance and Relief Everywhere, Inc.(CARE) Atlanta, GA, USA}
}

@misc{phil_hvi,
    title={Philadelphia Heat Vulnerability Index},
    author={{City of Philadelphia, Department of Health}},
    url={https://hip.phila.gov/emergency-response/philadelphia-heat-vulnerability-index/}
}

@misc{us_cvi,
    title={{U.S. Climate Vulnerability Index}},
    author={{Environmental Defense Fund} and {Texas A\&M University} and {Darkhorse Visualization}},
    url={https://climatevulnerabilityindex.org/}
}

@misc{global_hunger_index,
    url={https://www.globalhungerindex.org/},
    title={Global Hunger Index},
    author={{Concern Worldwide} and {Welthungerhilfe} and {the Institute for International Law of Peace and Armed Conflict}}
}

@misc{climate_risk_index,
    title={Climate Risk Index},
    url={https://www.germanwatch.org/en/cri},
    author={Germanwatch}
}

@misc{un_mpi,
    url={https://hdr.undp.org/content/2025-global-multidimensional-poverty-index-mpi#/indicies/MPI},
    title={2025 Global Multidimensional Poverty Index (MPI)},
    author={{United Nations Development Programme}},
    year={2025}
}

@inproceedings{agostini2024bayesian,
  title={A Bayesian Spatial Model to Correct Under-Reporting in Urban Crowdsourcing}, volume={38}, url={https://ojs.aaai.org/index.php/AAAI/article/view/30190}, DOI={10.1609/aaai.v38i20.30190}, abstractNote={Decision-makers often observe the occurrence of events through a reporting process. City governments, for example, rely on resident reports to find and then resolve urban infrastructural problems such as fallen street trees, flooded basements, or rat infestations. Without additional assumptions, there is no way to distinguish events that occur but are not reported from events that truly did not occur--a fundamental problem in settings with positive-unlabeled data. Because disparities in reporting rates correlate with resident demographics, addressing incidents only on the basis of reports leads to systematic neglect in neighborhoods that are less likely to report events. We show how to overcome this challenge by leveraging the fact that events are spatially correlated. Our framework uses a Bayesian spatial latent variable model to infer event occurrence probabilities and applies it to storm-induced flooding reports in New York City, further pooling results across multiple storms. We show that a model accounting for under-reporting and spatial correlation predicts future reports more accurately than other models, and further induces a more equitable set of inspections: its allocations better reflect the population and provide equitable service to non-white, less traditionally educated, and lower-income residents. This finding reflects heterogeneous reporting behavior learned by the model: reporting rates are higher in Census tracts with higher populations, proportions of white residents, and proportions of owner-occupied households. Our work lays the groundwork for more equitable proactive government services, even with disparate reporting behavior.}, number={20}, journal={Proceedings of the AAAI Conference on Artificial Intelligence}, author={Agostini, Gabriel and Pierson, Emma and Garg, Nikhil}, year={2024}, month={Mar.}, pages={21888–21896}
}

@inproceedings{10.1145/3490486.3538283,
author = {Liu, Zhi and Garg, Nikhil},
title = {Equity in Resident Crowdsourcing: Measuring Under-reporting without Ground Truth Data},
year = {2022},
isbn = {9781450391504},
publisher = {Association for Computing Machinery},
address = {New York, NY, USA},
url = {https://doi.org/10.1145/3490486.3538283},
doi = {10.1145/3490486.3538283},
booktitle = {Proceedings of the 23rd ACM Conference on Economics and Computation},
pages = {1016–1017},
numpages = {2},
location = {Boulder, CO, USA},
series = {EC '22}
}

@article{blackwood2023application,
  title={The application of the Social Vulnerability Index (SoVI) for geo-targeting of post-disaster recovery resources},
  author={Blackwood, Leah and Cutter, Susan L},
  journal={International Journal of Disaster Risk Reduction},
  volume={92},
  pages={103722},
  year={2023},
  publisher={Elsevier}
}

@misc{uri,
    title={Urban Risk Index},
    url={https://nychazardmitigation.com/documentation/resources/urban-risk-index/},
    author={{New York City Emergency Management}}
}

@misc{census_definition,
    title={Geography Program Glossary},
    url={https://www.census.gov/programs-surveys/geography/about/glossary.html},
    author={{United States Census Bureau}},
    lastaccessed = {May 12, 2026}
}

@misc{nta_definition,
    title={Neighborhood Tabulation Areas (NTAs)},
    url={https://www.nyc.gov/content/planning/pages/resources/datasets/neighborhood-tabulation},
    author={{NYC Department of City Planning}},
    lastaccessed = {May 12, 2026}
}

\newpage 
\appendix

\section{Discussion of Hazard Mitigation and Heat Actions}

We describe how emergency management agencies plan for and respond to extreme heat emergencies. Understanding the actions that outputs are meant to guide helps us compare indices and predictive algorithms. 
In our case study, we focus on NYC, which has high overlap with other major cities in the U.S., such as Los Angeles and Phoenix. 
\label{sec:comparison_other_cities}

\subsection{New York}

In addition to the hazard mitigation plan and the heat action plan, which are described in the main text, throughout the year, NYC advertises and conducts outreach for ``Beat the Heat''~\citep{extreme_heat_beat}. ``Beat the Heat'' is a citywide public health campaign that publicizes information on heat-related health risks and actions that individuals can take.
Heat vulnerability influences several other heat-related preparedness and response actions.
For example, during extreme heat events, there is an interactive map with information on both cooling centers and outdoor cool options (e.g., spray showers) shown alongside high heat vulnerable neighborhoods~\citep{coolit}.
Similarly, decisions about tree planting focus on heat vulnerable neighborhoods~\citep{forestry_plan}.

\subsection{Los Angeles} The 2024 Local Hazard Mitigation Plan discusses the history of extreme heat events in Los Angeles (LA) county, the potential for increased risk due to climate change, and several heat-related products that the county is exploring to combat extreme heat -- such as the experimental HeatRisk tool from the National Weather Service, which can be used to forecast an excessive heat
warning or heat advisory, and the California Heat Assessment Tool (CHAT)~\citep{la_hmp}. Through a combination of daily meteorological data and emergency department visits, CHAT estimates both historical and projected heat health events (i.e., events with negative public health effects regardless of the absolute temperature)~\citep{chat}. CHAT also includes a measure of heat vulnerability by weighting information on social, health, and environmental vulnerability. Response actions for extreme heat appear in the annex to the Los Angeles Emergency Operations Plan~\citep{la_hrp}. Actions that should occur in the event of a heat advisory or warning include: promoting cooling centers, tracking impacts to critical infrastructure (such as power outages), and distributing cooling supplies especially to vulnerable populations~\citep{la_hrp}.

\subsection{Phoenix} The hazard mitigation plan for the surrounding county includes detailed information on the National Weather Service's HeatRisk product, heat vulnerability (including a county-specific HVI~\citep{reid2009mapping, harlan2013neighborhood}), and an analysis on heat-associated death and illness. Long-term mitigation actions include identifying and communicating the locations of cooling centers, implementing ``cool roofs'' and ``cool pavements,'' and expanding green infrastructure~\citep{maricopa_hmp}. Similar to NYC and LA, the heat action plan provides guidance related to cooling centers, outreach, and public engagement. Some unique actions are to open a 24/7 ``Respite and Navigation Center,'' monitor the implementation of a cooling ordinance for rental buildings, and conduct outreach near trailheads and mobile homes~\citep{phoenix_hrp}.

\section{Replication of the NYC HVI and Other Datasets}
\label{sec:replication}
\subsection{The NYC HVI at the Neighborhood Level}
To produce the NYC HVI at the neighborhood level, we use data from NYC DOHMH~\citep{hvi_nta_level_dohmh}. This dataset includes the official 5-category HVI ranking along with all of the five inputs: surface temperature, median household income, greenspace, the percentage of households with air conditioning access, and the percentage of residents who are non-Latinx Black. Following the methodology in \citep{madrigano2015case} and the documentation on NYC Open Data~\citep{hvi_zcta_level_dohmh}, we recreate the 5-category HVI ranking with 98\% accuracy and a correlation coefficient of 0.99. Only 4 neighborhoods do not match, and the percentile rankings for these neighborhoods are all close to the threshold for defining quintiles (e.g., 20, 40, 60, or 80).

For ``Alt. 3'' in Table~\ref{tab:hvi_specification_formulas}, we include all comorbidities as separate inputs to the HVI formula. Estimates for comorbidities come from the 2024 CDC Places dataset, which uses model-based estimates from the 2021 and 2022 Behavioral Risk Factor Surveillance System (BRFSS) data and is the earliest version of the CDC Places data that aligns with data from the 2020 census. The full list of comorbidities includes: high blood pressure, current asthma, coronary heart disease, chronic obstructive pulmonary disease, high cholesterol, diabetes, no leisure-time physical activity, frequent mental distress, frequent physical distress, obesity, and stroke. 
Since the CDC Places data is provided at the census tract level, we estimate neighborhood-level percentages by reweighting tract-level percentages based on the total population over 18 from the 2020 census.

We selected these comorbidities based on potential adverse relationships with extreme heat. The CDC HHI, for example, uses a much smaller list of comorbidities from the CDC Places dataset: coronary heart disease, obesity, diabetes, chronic obstructive pulmonary disease, current asthma, and frequent mental distress.  Alternatively, practitioners might want to consider other health conditions such as self-care or mobility disabilities (e.g., populations that cannot easily evacuate in the case of heat waves or other hazards).

Across all possible subset combinations of the comorbidities selected in ``Alt. 3'', the Spearman correlation with the NYC HVI ranges from 0.856 - 0.987 for the percentile rankings and 0.833 - 0.967 for the quintile-based 5-point risk scores. The relationship to the NYC HVI does not necessarily depend on the number of comorbidities added. For example, while adding only a single comorbidity (e.g., stroke) might not lead to large deviations from the NYC HVI ($\rho = 0.987$), adding just six comorbidities (high blood pressure, coronary heart disease, chronic obstructive pulmonary distress, diabetes, high cholesterol, and stroke) leads to much larger differences ($\rho =0.858$).

A different approach to ``Alt. 3'' might instead involve a sub-index for comorbidities (e.g., first averaging across the percentages for all comorbidities or using the maximum percentage value). One could then add only the sub-index to the HVI formula, which would result in fewer changes to the number of inputs and increase the correlation with the original NYC HVI scores. For example, using the maximum percentage value across all comorbidities to produce a single sub-index results in a Spearman correlation coefficient of 0.95 for the percentile rankings and 0.91 for the quintile-based 5-point risk scores.

\subsection{The NYC HVI at the Census Tract Level}
We produce the NYC HVI at the census tract level using new data sources, as there is no widely accessible version of the NYC HVI at the census tract level to our knowledge. We produce census tract level estimates of all inputs except for households with air conditioning access. This variable comes from the NYC Housing and Vacancy Survey and is only publicly available at the Public Use Microdata Area (which is slightly larger than neighborhoods and roughly equivalent to zipcodes). To estimate air conditioning access, we use the corresponding values from the neighborhood-level NYC HVI.
Regarding the other inputs, we obtain census tract level estimates of median household income from the American Community Survey 5-year estimates (2016-2020)~\citep{acs_data} and census tract level estimates of the percentage of residents who are non-Latinx Black from the Decennial Census~\citep{dec_data}. We estimate the greenspace area percentage using the 2017 Land Cover Raster Data for NYC (available on the NYC Open Data portal~\citep{landcover_data}) and the average land surface temperature using NASA ECOSTRESS thermal imaging data from August 27, 2020~\citep{ecostress}. This date matches the date used in the NYC HVI from DOHMH. As we note in Figure~\ref{fig:lst_comparison}, there are striking differences in average land surface temperature for different satellite image capture dates. For the Land Cover Raster Data, we compute the greenspace area percentage using a zonal histogram; for the ECOSTRESS data, we compute a spatial average using zonal statistics. Both spatial operations were performed in QGIS.

To check whether the data sources we use for the census tract HVI replication introduce changes compared to the neighborhood-level version, we replicate our own version of the neighborhood-level HVI using the same data sources. We find that there are slight differences, potentially due to variation in software and rounding. The Spearman correlation coefficient between the neighborhood-level HVI with updated data inputs and the neighborhood-level HVI with original data inputs is 0.998; the quintile based 5-point risk scores match 94\% of the time. In Figure~\ref{fig:hvi_alt}(b) and Figure~\ref{fig:hvi_tract}, we use the neighborhood-level HVI with updated data inputs to ensure that the comparison is only due to differences in spatial scale, as opposed to variations in the data.

In updating the inputs of the NYC HVI to the census tract level, we find that the choice of satellite image capture date for computing average land surface temperature can lead to noticeable differences in the distribution of average land surface temperature throughout the city. Specifically, we compare estimates of the average land surface temperature from the NYC HVI (captured on August 27, 2020) to more recent estimates (July 30, 2025) as well as comparable Landsat data captured on July 9, 2020 from the NYC Urban Heat Portal~\citep{uhp_lst_data}. We find that even just across these three data sources, there is substantial variation in the distribution of average land surface temperature, which could have a meaningful effect on heat vulnerability rankings. As a result, we recommend that future versions of the NYC HVI incorporate multiple measurements for land surface temperature at different points in time.

\begin{figure}
    \centering
    \includegraphics[width=0.7\linewidth]{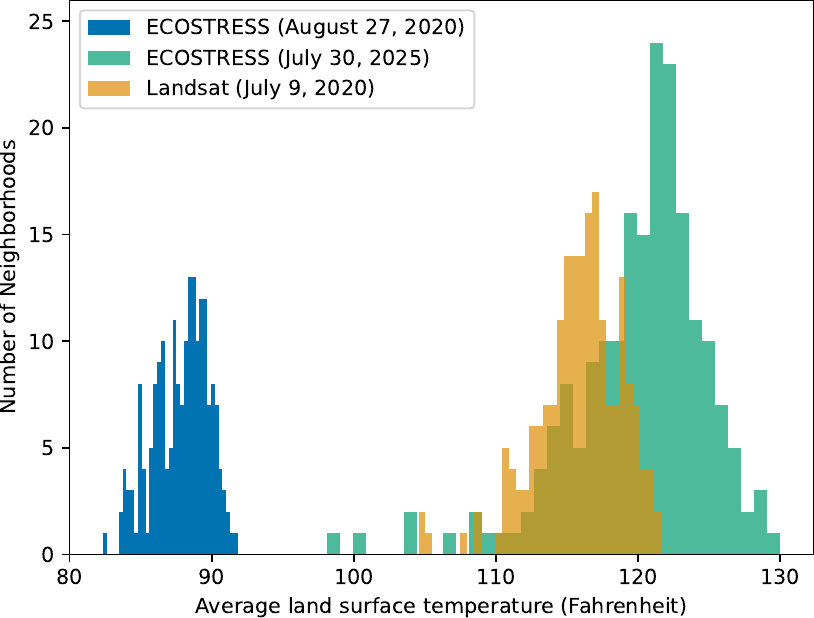}
    \caption{\textbf{Comparison of different average land surface measurements at the neighborhood level (n=197).} The average land surface temperature in the NYC HVI from ECOSTRESS thermal imaging on August 27, 2020 is compared to average land surface temperature estimates from ECOSTRESS thermal imaging (taken on July 30, 2025) and to Landsat data (taken on July 9, 2020). While the measurements are all positively correlated, the temperature estimates for the NYC HVI are notably lower. The Pearson correlation coefficient between the ECOSTRESS August 27, 2020 capture date and the Landsat estimates is 0.76; the Pearson correlation coefficient between the ECOSTRESS August 27, 2020 capture date and the ECOSTRESS July 30, 2025 capture date is 0.47.}
    \label{fig:lst_comparison}
    \Description[Three histograms are shown that compare the distribution of average land surface temperature for three different image capture dates.]{Three histograms (blue, teal, and gold) are shown that compare the distribution of average land surface temperature for three different image capture dates. The x-axis, which denotes land surface temperature, ranges from 80 to 130 degree Fahrenheit. The y-axis, which indicates the number of neighborhoods, ranges from 0 to 25. The average land surface temperature for August 27, 2020 is noticeably different (shifted left, toward lower land surface temperatures) from the other two capture dates.}
\end{figure}

\subsection{Other Indices}

To compare the NYC HVI to the CDC HHI, we use relevant public data available from the CDC~\citep{cdc_hhi_data}. The dataset for the HHI uses 2010 zipcode tabulation areas (ZCTAs). Many of the inputs to the index are included only in the form of national percentile rankings. Some of the inputs must be aggregated in this way, as they reflect highly sensitive health data. For example, data on heat-related illness comes from the National Emergency Medical Services Information Systems and is defined as the percentile rank of the three-year average heat-related EMS activation rate from 2020-2022. Given these restrictions, we are not able to recreate the HHI as fully as the NYC HVI.

We obtain census tract level data for the NRI from the Climate Mapping for Resilience \& Adaptation data portal~\citep{nri_data}. This dataset has many of the key inputs to the NRI data, including both the raw values, the state and national percentile rankings, and qualitative rating scales that are assigned via k-means. Key inputs include the annualized frequency of event-days, the expected annual loss, and the community risk factor (which combines social vulnerability with community resilience).

\section{Spatial Comparison of the NYC HVI to the CDC HHI}

In this section, we visually compare the NYC HVI at the neighborhood level to the CDC HHI. See Section~\ref{sec:cdc_hhi_comparison} for a full discussion of the HHI.
We observe that the NYC HVI and the HHI do not fully align. In particular, the NYC HVI identifies more areas as high risk (scores of 4 and 5) in central Brooklyn and Queens (the southeastern and southern areas of the map). The HHI, in contrast, identifies more areas as high risk in the Bronx and Staten Island (the northern and western areas of the map).
A limitation of the HHI is that it relies on national percentile rankings. In Figure~\ref{fig:hhi}, we show how coarser categories (e.g., bucketing the HHI into NYC-specific quintiles) can lead to distortions of risk.

\begin{figure}[htbp]
    \centering
    \begin{minipage}[t]{0.48\textwidth}
        \includegraphics[width=0.8\linewidth]{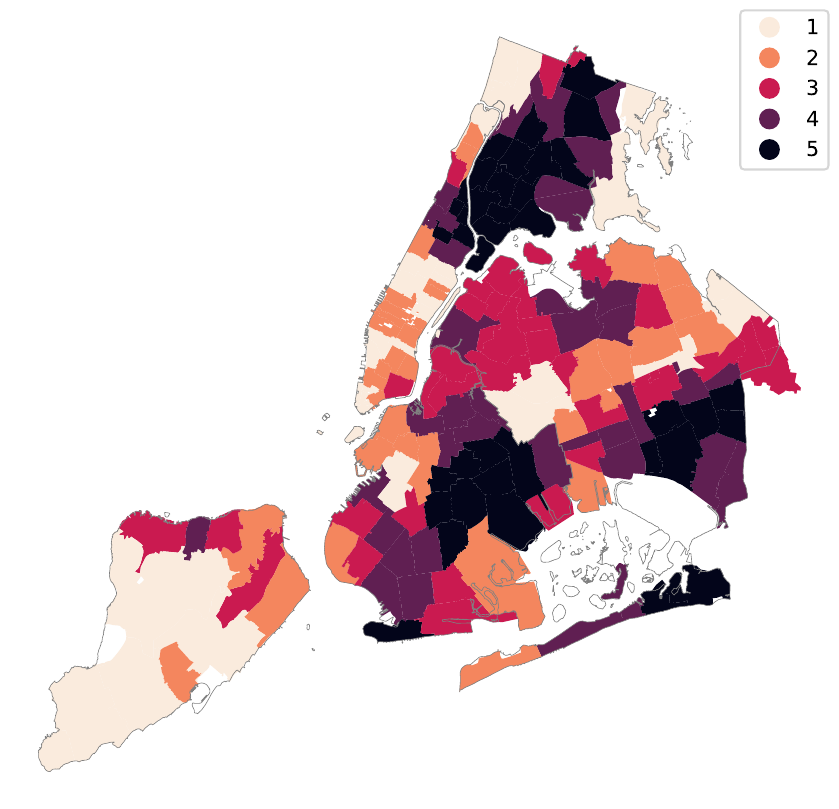}
        \caption{The NYC Heat Vulnerability Index (HVI)}
        \label{fig:hvi_appendix}
    \end{minipage}
        \hfill
            \begin{minipage}[t]{0.48\textwidth}
        \includegraphics[width=0.8\linewidth]{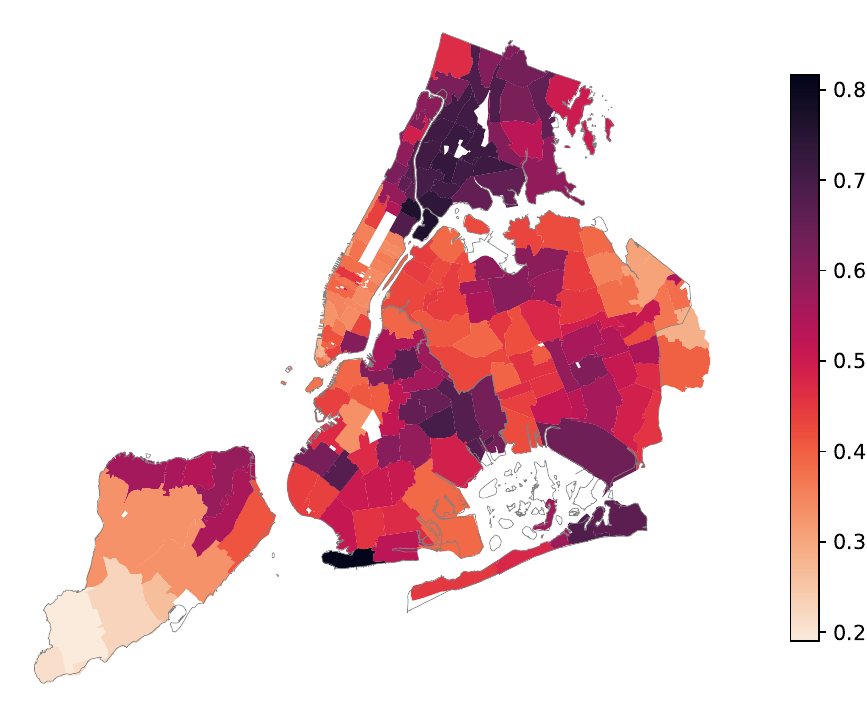}
        \caption{The CDC HHI with National Percentile Rankings}
        \label{fig:cdc_hhi_original_rankings}
    \end{minipage}
    
        \begin{minipage}[t]{0.48\textwidth}
        \includegraphics[width=0.8\linewidth]{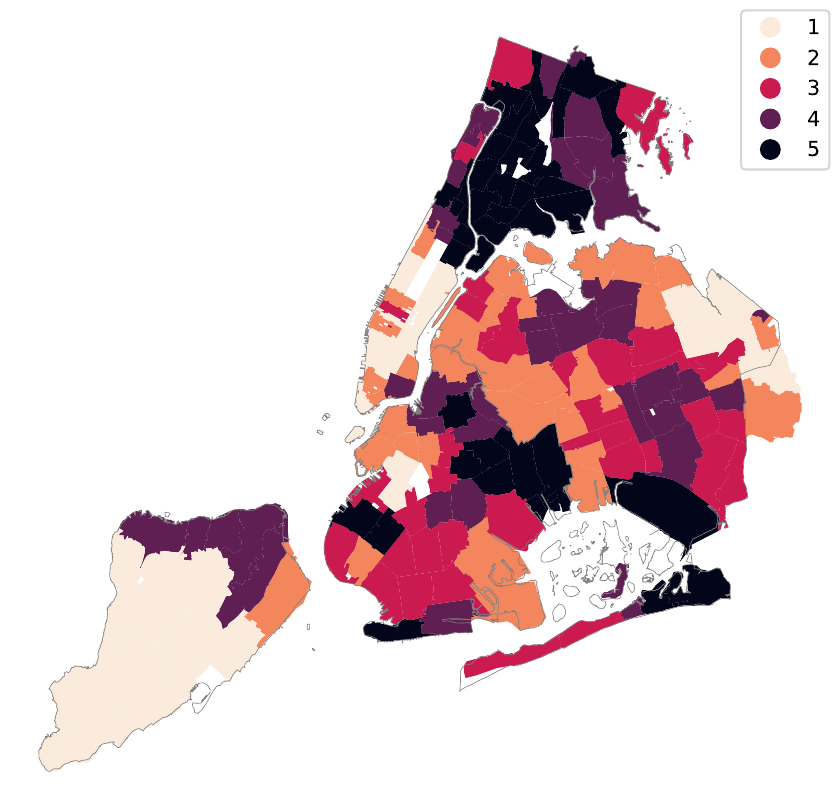}
        \caption{The CDC HHI with NYC-specific quintiles}
        \label{fig:hhi}
    \end{minipage}
    \hfill
    \begin{minipage}[t]{0.48\textwidth}
        \includegraphics[width=\linewidth]{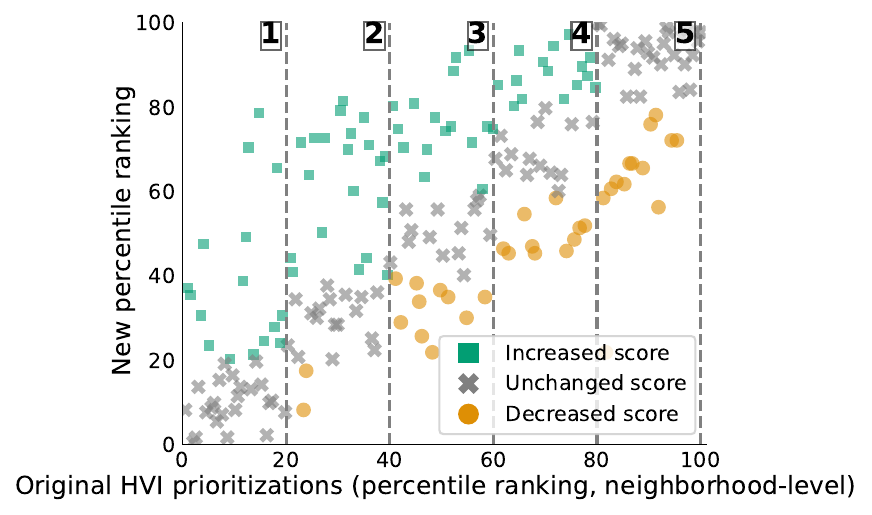}
        \caption{Comparison of the NYC HVI and the CDC HHI}
        \label{fig:hhi_scatter}
    \end{minipage}

    \caption*{\textbf{Comparing the NYC HVI to the CDC HHI.} The NYC HVI at the neighborhood level in Figure~\ref{fig:hvi_appendix} is compared to the CDC HHI at the 2010 zipcode tabulation area level with national percentile rankings (Figure~\ref{fig:cdc_hhi_original_rankings}) and NYC-specific quintiles (Figure~\ref{fig:hhi}). Regardless of the approach, there are notable differences between the two indices. In particular, while both identify areas in the northernmost parts of the city as high risk, the NYC HVI identifies more areas in the southeastern and southern areas as high risk compared to the HHI. Figure~\ref{fig:hhi_scatter} directly compares the two indices at the neighborhood level. We assign neighborhoods, which the NYC HVI uses, to 2010 zipcode tabulation areas, which the HHI uses, based on the highest spatial overlap.
    }
    \Description[Three maps and a scatter plot are shown to compare the NYC HVI and the CDC HHI.]{There are three maps. The first map (upper left) depicts the NYC HVI at the neighborhood level with its quintile-based 5-category risk scores. The second (upper right) and third (lower left) maps depict the CDC HHI. The graph in the lower right is a scatter plot that compares the neighborhood-level percentile rankings to the percentile rankings for the CDC HHI (originally computed at the zipcode tabulation area level). Green squares depict increases in the risk score, yellow dots depict decreases, and gray crosses indicate there is no change.}
\end{figure}

\FloatBarrier

\newpage
\section{Comparison of Heat-Related Measurements: the NYC HVI versus the NRI}

We present average land surface temperature (Figure~\ref{fig:lst_temperature}), taken from ECOSTRESS thermal imaging provided by NASA and the U.S. Geological Survey (USGS)~\citep{ecostress}. We produce a spatial average of land surface temperature at the census tract level using satellite imagery captured on August 27, 2020. Average land surface temperature from August 27, 2020 is one of the five inputs to the NYC HVI.
We then compare average land surface temperature to the annualized frequency of heat event-days (Figure~\ref{fig:nri_annualized_freq}), which FEMA uses in the NRI. The two measurements depict vastly different spatial distributions of heat risk.

\begin{figure}[htbp]
     \begin{minipage}[t]{0.48\textwidth}
        \centering
        \includegraphics[width=0.8\linewidth]{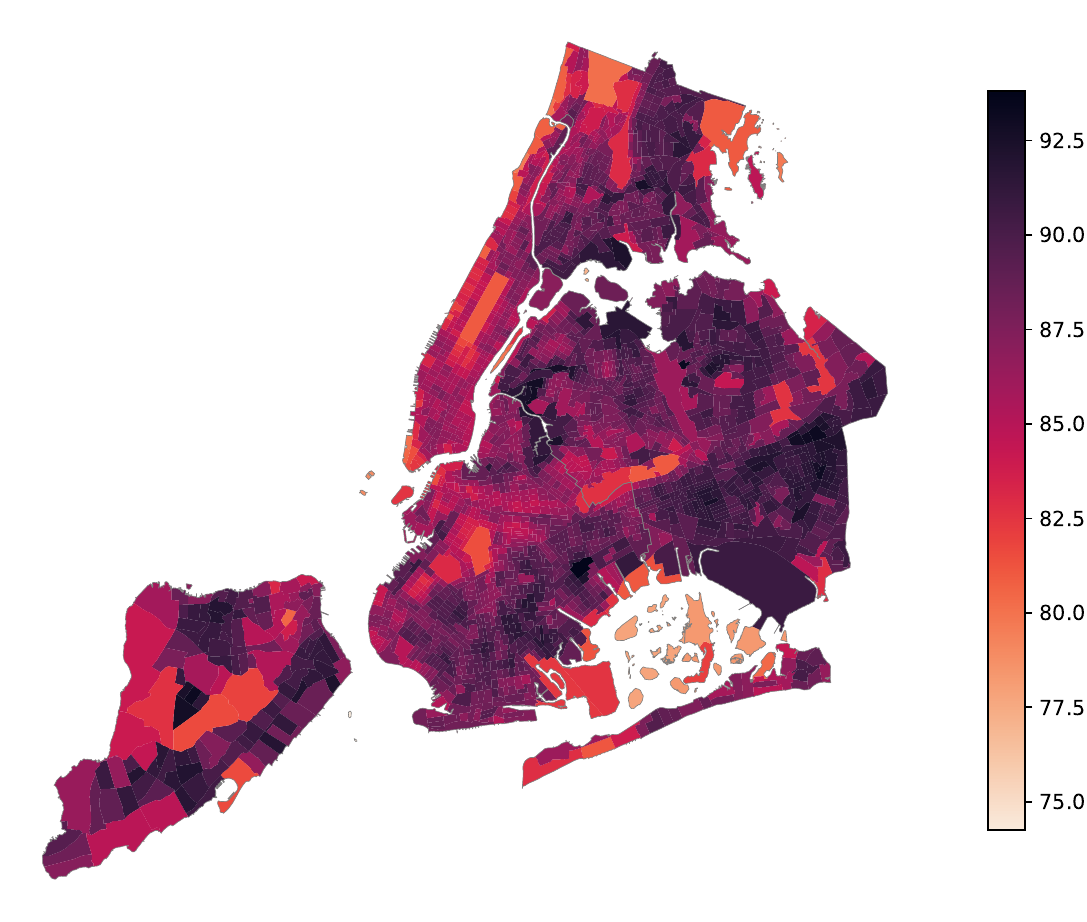}
    \caption{Average land surface temperature in Fahrenheit at the census tract level from NASA ECOSTRESS data (August 27, 2020). The pattern of average land surface temperature differs from the annualized frequency of heat event-days that the NRI uses (shown in Figure~\ref{fig:nri_annualized_freq}).}
    \label{fig:lst_temperature}
    \end{minipage}
    \hfill
   \begin{minipage}[t]{0.48\textwidth}
        \centering
        \includegraphics[width=0.8\linewidth]{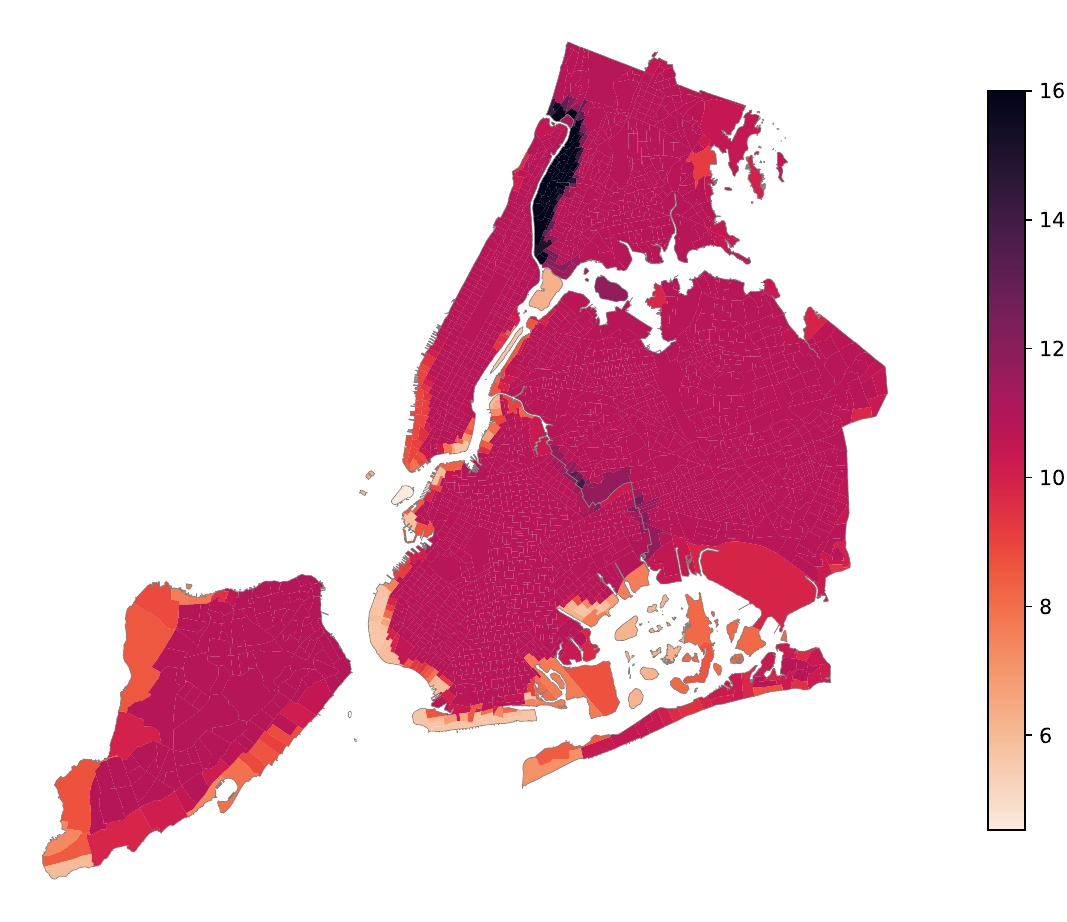}
    \caption{Annualized frequency of heat event-days. Adjacent census blocks can have drastically different annualized frequencies. Event-days are based on National Weather Service alerts and public forecast zones.}
    \label{fig:nri_annualized_freq}
    \end{minipage}
    \Description[Two maps are shown with an orange and red colorscale; each depicts different measurements related to extreme heat.]{The map on the left is a choropleth map of average land surface temperature; darker colors denote higher temperatures. The map on the right is a choropleth map that describes the annualized frequency of heat event days. The distribution of the colors look very different from the map on the left.}
\end{figure}
\FloatBarrier

\newpage 
\section{Spatial Analysis of the NYC HVI Sensitivity}
In this section, we expand on the analysis in Figure~\ref{fig:hvi_alt} by examining the spatial changes that result from either modifying the specification of the NYC HVI or constructing the HVI at the census tract level, as opposed to the neighborhood level. For each map, we highlight the corresponding neighborhoods with increased risk scores of either 4 or 5 under the new specification, relative to neighborhoods with risk scores of 1, 2, or 3 under the original NYC HVI.

\begin{figure}[htbp]
    \centering
    \includegraphics[width=0.9\linewidth]{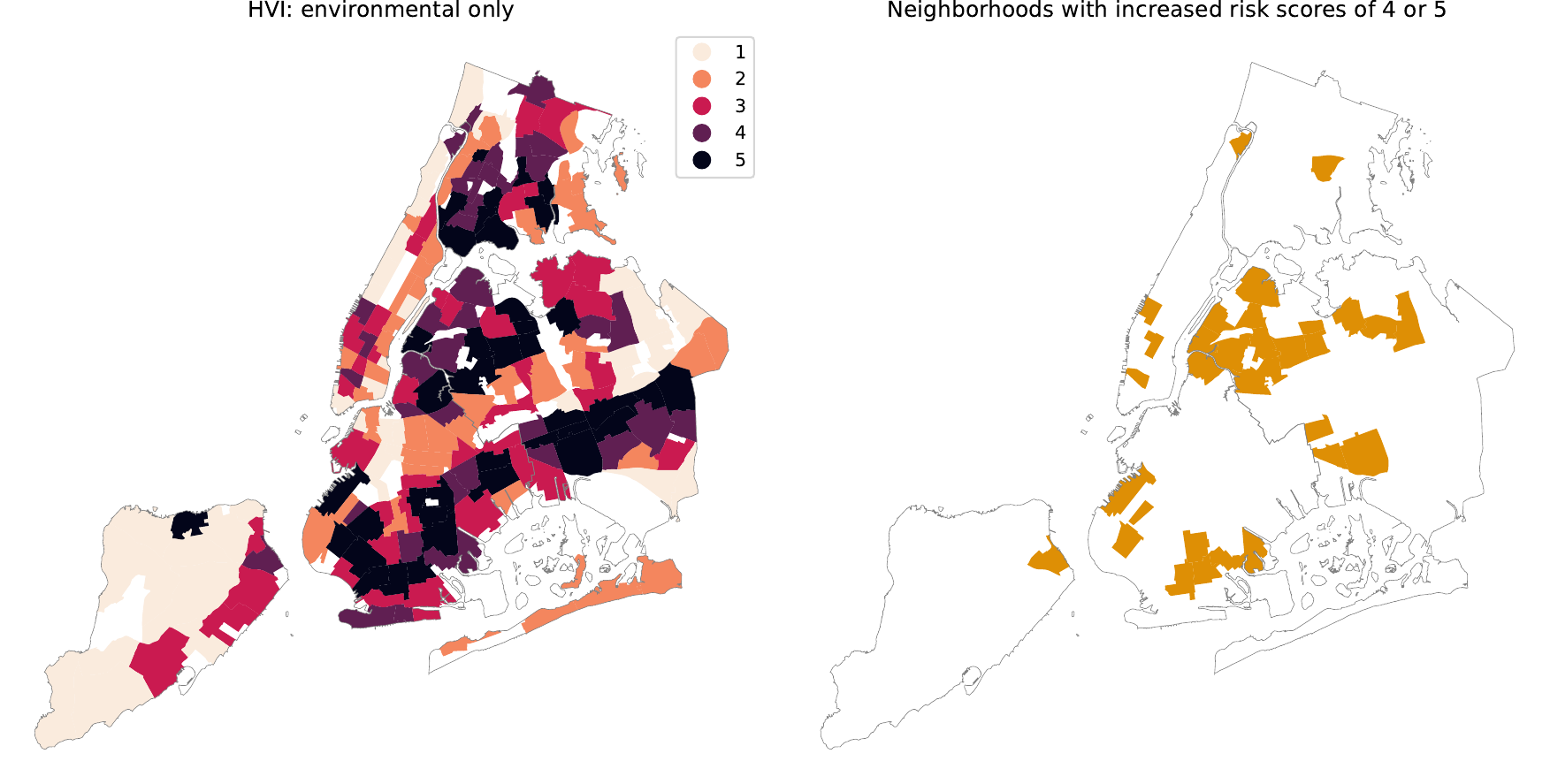}
    \caption{The NYC HVI if only environmental characteristics were used (based on the formula ``Alt. 1: Environmental'' in Table~\ref{tab:hvi_specification_formulas}).}
    \label{fig:hvi_sens}
\end{figure}

\begin{figure}[t!]
    \centering
    \includegraphics[width=0.9\linewidth]{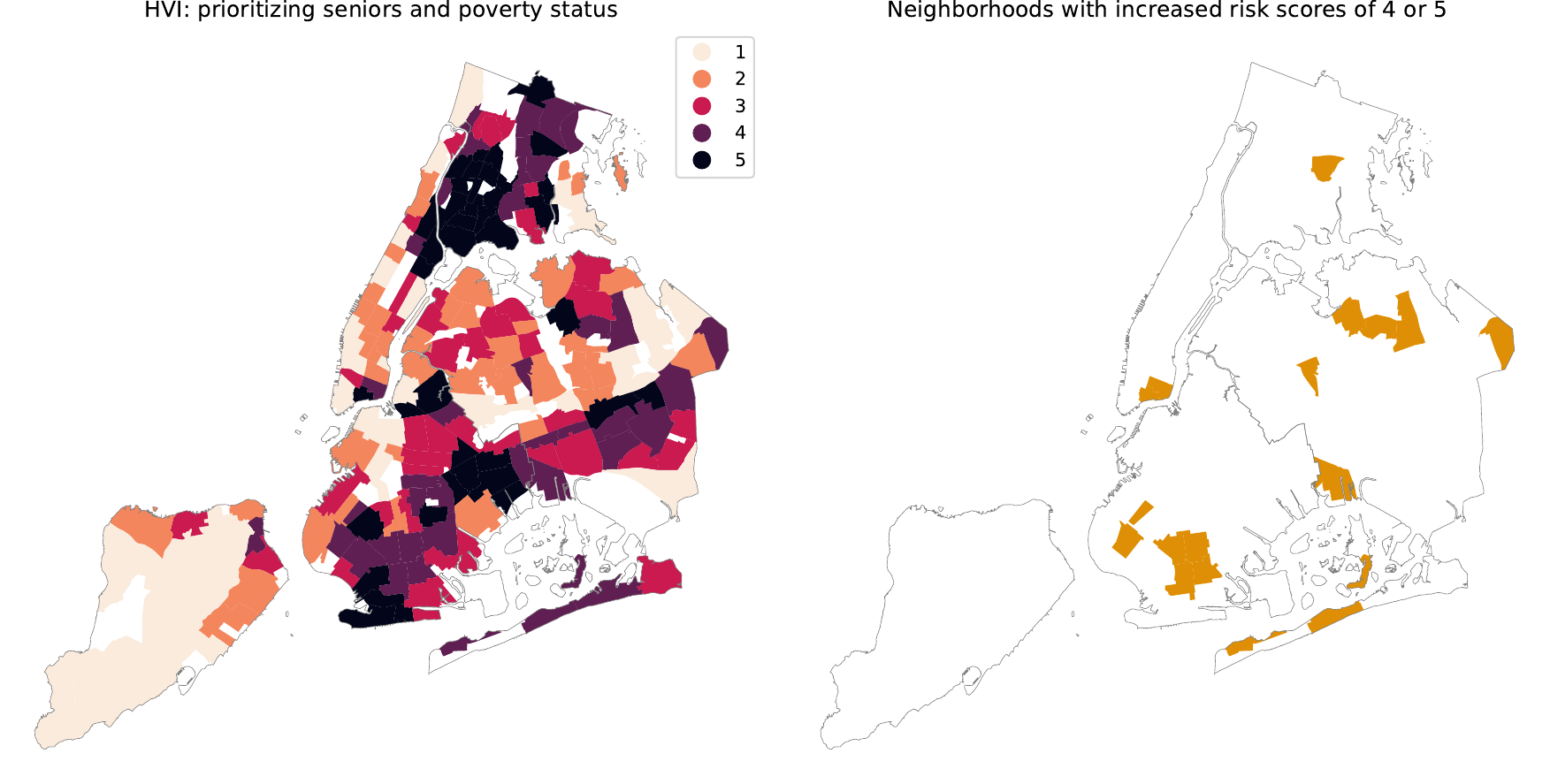}
        \caption{The NYC HVI if including information on seniors and poverty status (based on the formula ``Alt. 2: seniors and poverty status'' in Table~\ref{tab:hvi_specification_formulas}).}
    \label{fig:hvi_sens_enviro}
\end{figure}

\begin{figure}[htbp]
    \centering
    \centering
    \includegraphics[width=0.9\linewidth]{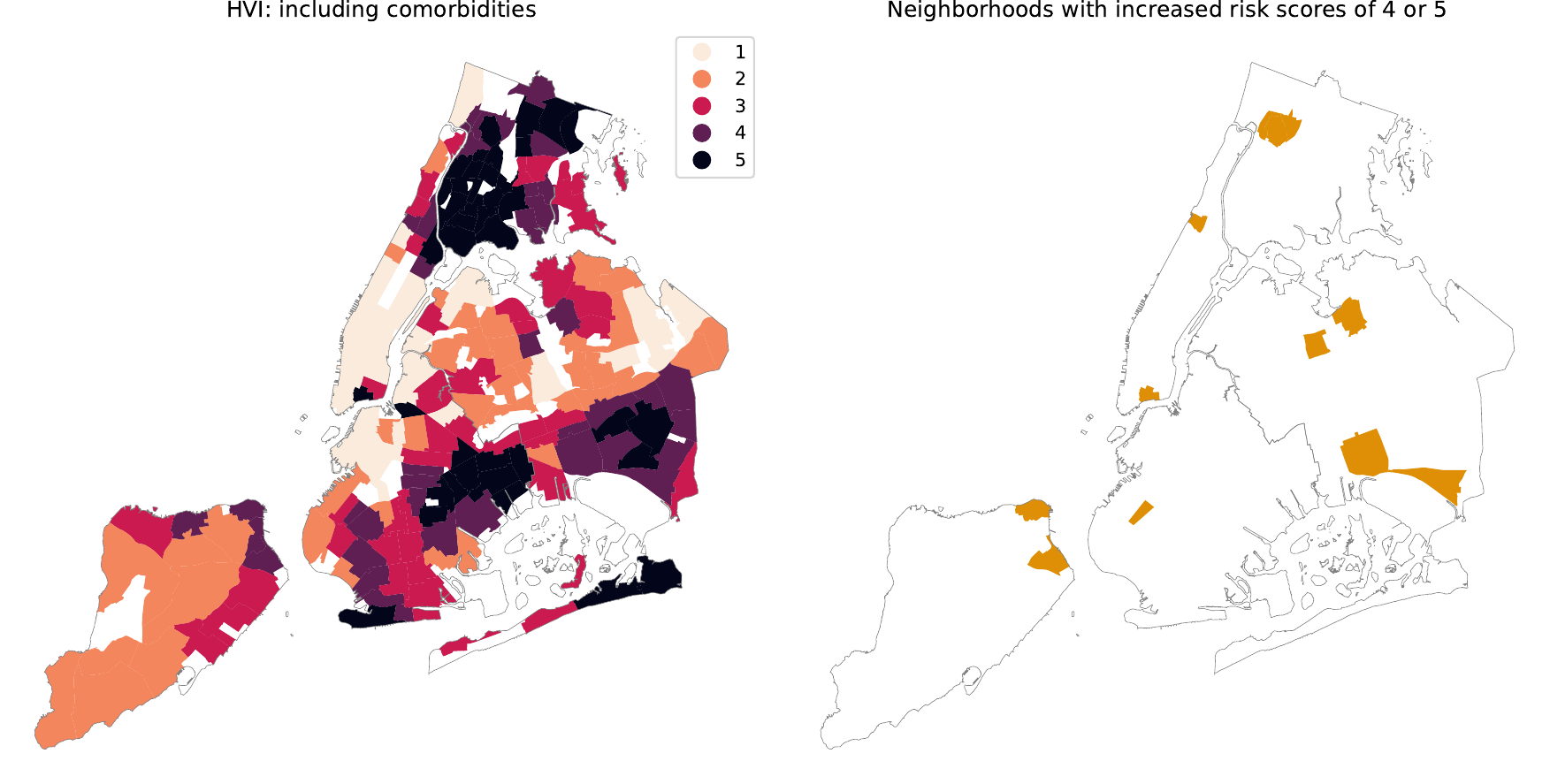}
        \caption{The NYC HVI if including individuals with comorbidities (based on the formula ``Alt. 3: Comorbidities'' in Table~\ref{tab:hvi_specification_formulas}).}
    \label{fig:hvi_sens_comorb}
    \end{figure}
\begin{figure}[htbp]
    \centering
    \includegraphics[width=0.9\linewidth]{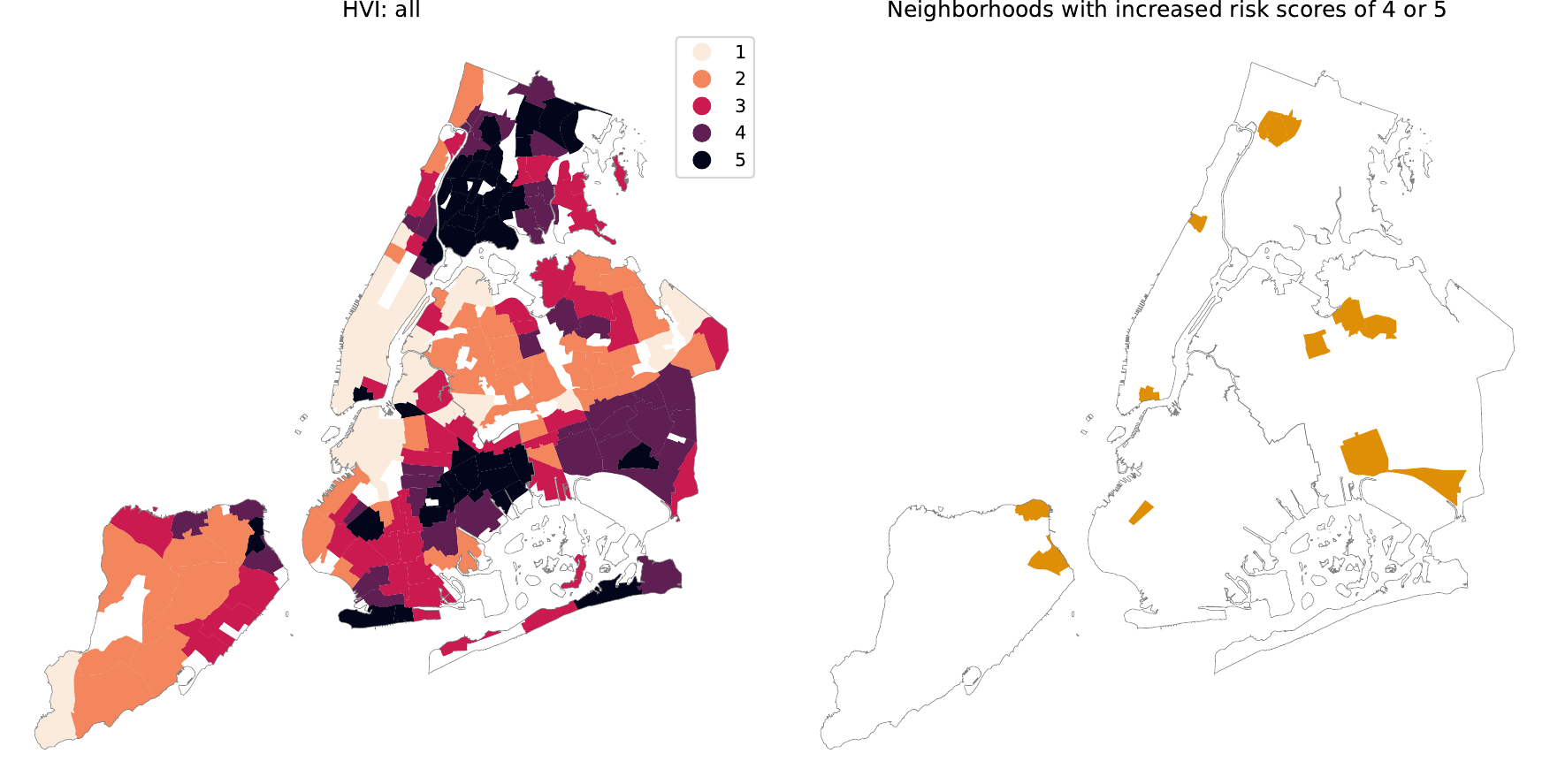}
        \caption{The NYC HVI if including all additional features (based on the formula ``Alt. 4: All'' in Table~\ref{tab:hvi_specification_formulas}).}
    \label{fig:hvi_sens_full}
\end{figure}

\begin{figure}[htbp]
    \centering
    \includegraphics[width=0.9\linewidth]{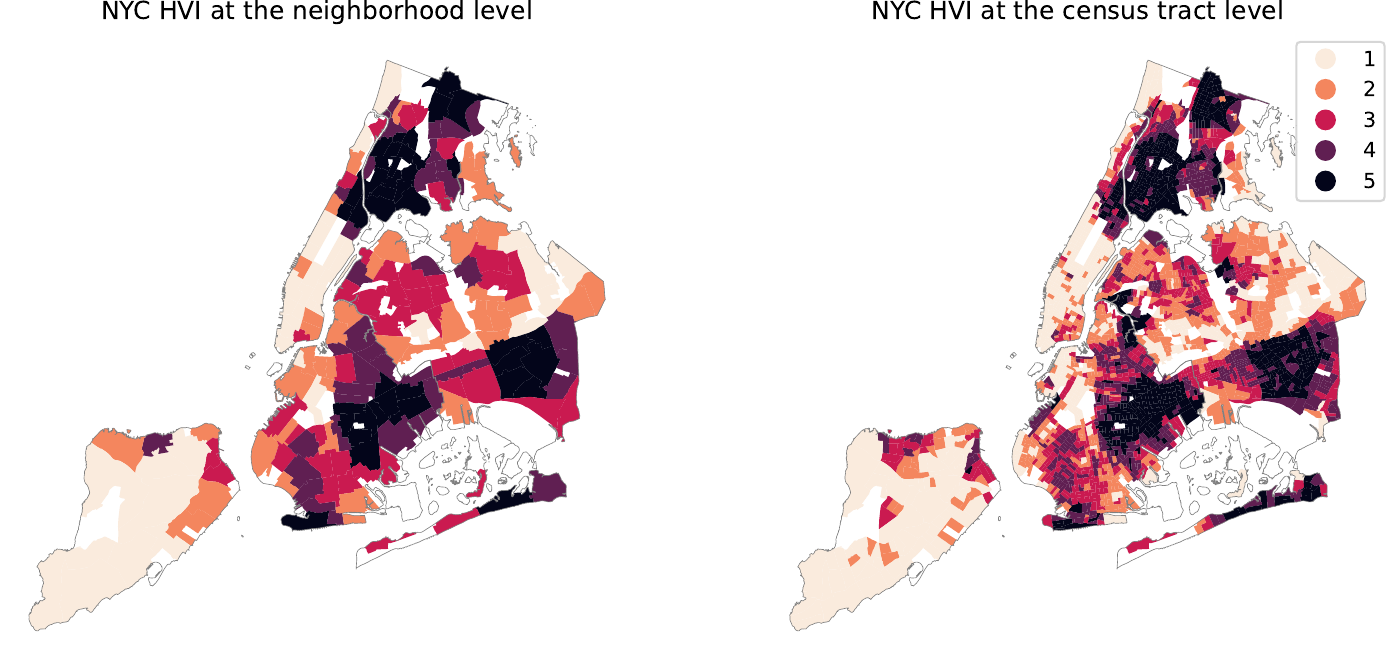}
    \caption{The NYC HVI if constructed at the neighborhood level (on the left) compared to the census tract level (on the right).}
    \label{fig:hvi_tract}
\end{figure}
\FloatBarrier

\section{Correlation Analysis and Alignment for Different NYC HVI Specifications}
\label{sec:correlation_alt_specifications}

In this section, we compute the correlations for different alternative specifications of the NYC HVI (enumerated in Table~\ref{tab:hvi_specification_formulas}) compared to the original NYC HVI. We estimate the correlations for both the underlying percentile rankings and the simpler 5-category risk scores (based on quintiles) using Spearman correlation and Kendall's Tau. We also estimate the overall alignment of the resulting 5-category risk scores. We define alignment as follows. Let $N$ equal the set of all neighborhoods. $HVI_i$ is the original 5-category HVI score for a neighborhood $i$ and  $HVI^{alt}_i$ is the 5-category risk score for an alternative specification: $$\text{Alignment} = \frac{\sum_{\forall i \in N} \mathbbm{1}_{\{HVI_i = HVI^{alt}_i\}}}{|N|}$$ We find that even when there is high correlation, alignment can still be low or moderate (between 32-55\%).

We additionally test the sensitivity of the HVI to a fifth alternative specification (``Alt. 5'') where we change percent Black in the original HVI formula to percent non-white. We find that the resulting HVI score is highly correlated with the original HVI score ($\rho=0.96$), though reduced alignment (only 72\% of neighborhoods match) indicates that even this small change could still lead to substantial differences in HVI risk scores across neighborhoods.

\begin{table*}[htbp!]
    \centering
    \adjustbox{max width=\textwidth}{
    \begin{tabular}{lcc|c}
\toprule
Alt specification & Correlation (percentile) & Correlation (quintiles) &  Alignment (\%) \\
\midrule
Alt 1. Environmental & 0.643 & 0.596 & 32.487 \\
Alt 2. Seniors and poverty status & 0.869 & 0.843 & 55.33 \\
Alt 3. Comorbidities & 0.885 & 0.858 & 53.299 \\
Alt 4. All & 0.87 & 0.848 & 49.239 \\
Alt 5. Percent non-white  &  0.957 & 0.929 & 71.574 \\
\bottomrule
\end{tabular}}
    \caption{\textbf{Comparison of the original NYC HVI formula to alternative specifications using Spearman correlation.} We estimate: (1) correlations of percentile rankings with the percentile rankings under the original NYC HVI formula, (2) correlations of the 5-category risk scores with the 5-category risk scores under the original NYC HVI formula, and (3) alignment between the 5-category risk scores and the 5-category risk scores under the original NYC HVI formula. Even when specifications have relatively high Spearman correlation coefficients, alignment across the 5-category risk scores can be low to moderate. This pattern illustrates how some neighborhoods may still experience substantial shifts in risk (even if correlation is high overall).}
    \label{tab:alt_specifications_correlation}
\end{table*}

\begin{table*}[htbp!]
    \centering
    \adjustbox{max width=\textwidth}{
    \begin{tabular}{lcc}
\toprule
Alt specification & Correlation (percentile) & Correlation (quintiles)  \\
\midrule
Alt 1. Environmental & 0.463 & 0.489 \\
Alt 2. Seniors and poverty status & 0.694 & 0.748\\
Alt 3. Comorbidities & 0.697 & 0.764\\
Alt 4. All & 0.674 & 0.749 \\
Alt 5. Percent non-white & 0.831 & 0.869 \\
\bottomrule
\end{tabular}}
    \caption{\textbf{Comparison of the original NYC HVI formula to alternative specifications using Kendall's Tau.} We estimate the same correlations as above (Table~\ref{tab:alt_specifications_correlation}). Since Kendall's Tau compares concordant and discordant pairings across the two rankings, it has a more intuitive meaning.  Kendall's Tau values tend to be lower than the Spearman correlations.}
    \label{tab:alt_specifications_correlation_kendalls}
\end{table*}
\FloatBarrier

\FloatBarrier

\section{Comparing the NYC HVI to Two Different NRI Specifications}
\label{sec:appendix_nri}

Here we briefly discuss the definition of the NRI and then compare it to the NYC HVI in Figure~\ref{fig:nri_comparison_full}.

The NRI has several components. The first is the \textbf{expected annual loss (EAL)}, which measures the expected amount of hazard-specific loss that may occur to people, buildings, or agriculture~\citep{nri_technical}. The EAL is measured in dollars to standardize comparisons. The EAL depends on other input data such as the expected annual frequency of hazard events and the historic loss ratio (information that typically comes from the SHELDUS database at Arizona State University~\citep{sheldus_data} and from the CDC). Separate EAL measures are produced for buildings, people, and agriculture. These measures are then summed together to produce an overall EAL.

The second component of the NRI is a \textbf{community's risk factor}, which combines social vulnerability (SV) and community resilience (CR). Social vulnerability is a percentile ranking based on 10 socioeconomic characteristics that comes from the Census
Community Resilience Estimates~\citep{census_resilience}.
Estimates of community resilience are based on 49 different indicators that capture six different types of resilience (such as social resilience or environmental resilience). Data for community resilience comes from the Hazards Vulnerability and Resilience Institute~\citep{nri_resilience_data}.

The two components are related by the following formula where $f$ denotes a mapping to a triangular distribution~\citep{nri_technical}: $$\text{RISK} =  \text{EAL} \cdot  f( \frac{\text{SV}}{\text{CR}})$$ 

Based on this RISK value, FEMA then produces percentile rankings for all census tracts and counties in the U.S. along with a qualitative 5-point risk rating, using k-means clustering~\citep{nri_technical}. For our comparisons to the NYC HVI, we opt to use quintile-based 5-point risk score categories to increase the similarity of the methods used across both indices (shown in Appendix Figure~\ref{fig:nri_comparison_appendix}). Recreating the qualitative 5-point risk rating for NYC based on k-means would be another approach. 

In our sensitivity analysis, we test for differences with the NYC HVI based on two specifications: (1) the EAL on its own and (2) the NRI risk score (combining EAL with SV and CR). As shown in Figure~\ref{fig:nri_comparison_full}, we do not observe any strong visual trends that differ between the comparisons using either specification. Both appear to have a relatively uniform distribution with respect to the NYC HVI, illustrating the overall weakness of the relationship. 

\begin{figure}[htbp]
    \centering
    \includegraphics[width=\linewidth]{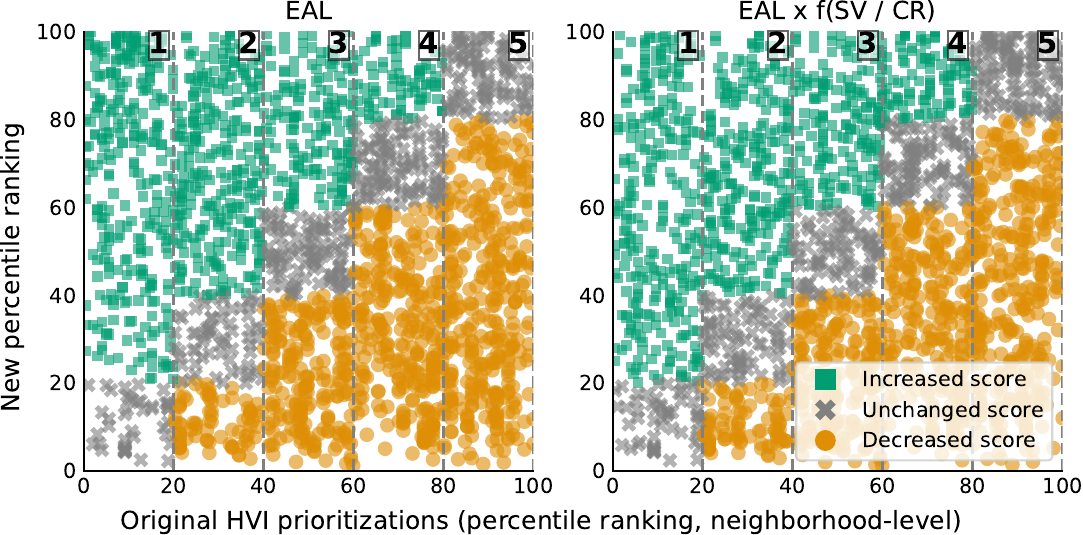}
    \caption{\textbf{Comparing the NYC HVI to two different specifications for the NRI: EAL and $\text{EAL} \cdot f( \frac{\text{SV}} {\text{CR}})$}. EAL = Expected Annual Loss, SV = Social Vulnerability, and CR = Community Resilience. Overall, we do not observe a strong relationship between the NYC HVI and the NRI using either specification.}
    \label{fig:nri_comparison_full}
\end{figure}

Similarly, as shown in Figure~\ref{fig:nri_comparison_appendix}, we can visualize the spatial distribution of the NRI with respect to the EAL on its own and the full NRI risk score (combining EAL with SV and CR). 

\begin{figure}[ht]
    \centering
    \includegraphics[width=\linewidth]{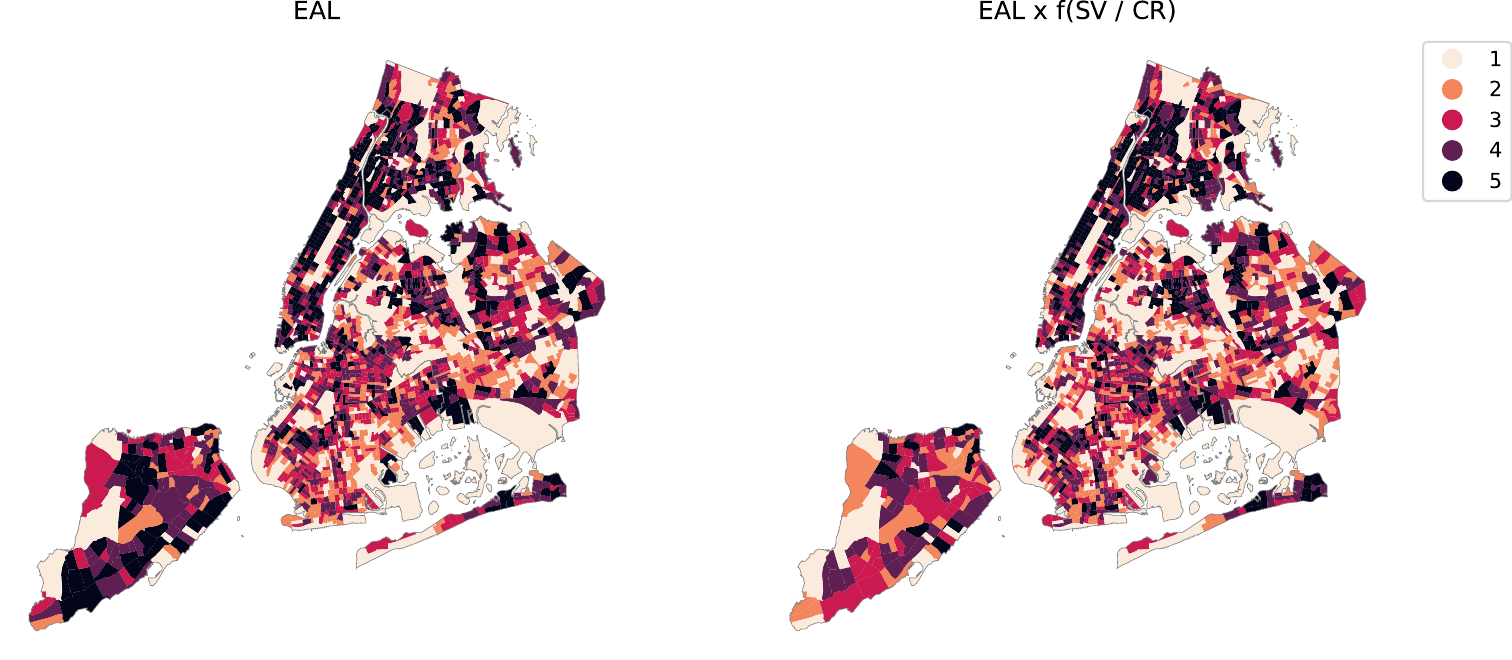}
    \caption{\textbf{Spatial distribution of the NRI according to the EAL alone and the full NRI risk score.} We observe only slight spatial variation  between the two methods.}
    \label{fig:nri_comparison_appendix}
\end{figure}

\section{Correlation Analysis between the NYC HVI, Indices, and Heat-Related Impacts}
\label{sec:correlation_analysis_methods}

We expand on the methodology used to facilitate the comparisons in  Tables~\ref{tab:correlation_matrix} and~\ref{tab:correlation_matrix_impacts}.
All correlation analyses are conducted at the census tract level (n=2,240).
We exclude census tracts for which there was no population count in the 2020 Decennial Census and that do not merge to the neighborhood-level NYC HVI.
Census tracts are subsets of neighborhood, defined by neighborhood tabulation areas (NTAs), and linking census tracts to neighborhood tabulation areas is straightforward.

To compare census tracts to other geographies like zip code tabulation areas (ZCTAs) and locality service areas (defined by the New York State Department of Public Service), we first estimate the overall spatial overlap between census tracts and the relevant polygons to be joined (e.g., ZCTAs). Many census tracts will intersect multiple polygons, and so we select the match with the highest spatial overlap. This process is similar to a point-to-polygon spatial join using census tract centroids, but results in fewer anomalous cases (e.g., the centroid location could be outside the census tract). One could perform the same calculations with the official U.S. Census relationship files (which indicate the amount of spatial overlap between polygons). Using the U.S. Census relationship files for ZCTAs produces the same result as manually calculating the spatial overlap.

In some cases, rankings based on historical heat-related impacts result in ties (e.g., many areas may not have any instances of hydrant-related 311 calls or heat-related EMS calls). We break ties by taking the average rank across all tied inputs. 
Furthermore, we define quintiles based on the percentile rank values that we compute (e.g., the first quintile corresponds to all percentile rank values $\leq 20$). This approach ensures that all inputs with the same underlying values are assigned to the same quintiles. Other functions, such as Python's default quantile function, may handle ties differently, leading to unexpected results.

\subsection{Data Processing of Heat-Related Impacts}

\textbf{Power outages:}
The power outage data we obtained from the Department of Public Service is available in 30-minute intervals for locality service areas (a type of geography that is slightly larger than a zipcode). For reference, there are 77 locality service areas that overlap with census tracts in NYC. Since it is not possible to identify individual customer-level cases of power outages in this dataset, we first compute the maximum daily power outage rate (i.e., the maximum of the  total number of customers experiencing outages divided by the total number of customers in the locality service area in any 30-minute period for each day in our dataset). We then compute the average maximum daily outage rate for each locality service area for all days occurring in May through September, 2021 - 2025. We obtain similar results for other kinds of calculations (e.g., ranking based on the cumulative outage rate for a locality service area across all 30-minute intervals from May through September, 2021 - 2025).

There are a few limitations with the power outage data. First, power outage data is commonly collected at the customer level. Customers are not equivalent to households; a single customer can refer to an entire apartment building. Second, data is provided for both network and radial distribution systems separately. Network distribution systems are underground while radial distribution systems typically refer to the above ground wires that are common in sparsely populated areas. While the counts for these distribution systems should be distinct, we noticed several instances of duplicate counts that inflated the total number of customer outages compared to news reports. As a result, we do not count customer outages for radial distribution systems when they exactly match the corresponding network distribution system and the number of customer outages is greater than 10.
Lastly, the power outage data, though public, is highly aggregated. Conducting similar analyses with more spatially granular power outage data (if available) might reveal important spatial patterns that this version of the data cannot capture.

\textbf{Heat-related EMS Calls:} For heat-related EMS calls, we pull EMS Incident Dispatch data from NYC Open Data for 2021 to 2025~\citep{ems_data}. We filter to all heat-related EMS calls where the final call type is equal to ``HEAT.'' An alternative approach would be to include initial call types equal to ``HEAT'' as well, but it is possible those were erroneously categorized. We exclude any incidents where either the date or zipcode is missing. Similar to the process for 311 complaints, we first compute the total count of heat-related EMS calls per zipcode. We do not normalize the counts of EMS calls by total population, as the counts are already quite small and, anecdotally, we noticed that many EMS incidents tend to occur in areas with more transient, high tourist populations. We estimate the total number of EMS incidents per zipcode from May through September, 2021 - 2025. We treat zipcodes as equivalent to 2020 census zipcode tabulation areas (which further enables linkages to other datasets).

\textbf{311 Complaints:} For 311 hydrant complaints, we pull data on all 311 service requests from NYC Open Data from 2021 to 2025~\citep{311_data}. We filter to any complaints with the following description: ``Hydrant Running Full (WA4)'', ``Hydrant Running (WC3)'', ``Illegal Use Of A Hydrant (CIN)'', ``Request To Open A Hydrant (WC4)''. There are other hydrant-related complaints that are less relevant to extreme heat (e.g., a car blocking a hydrant). We remove complaints where the resolution description indicates that the complaint is a duplicate. We count the number of complaints by census tract, standardizing by the total population for the census tract based on counts from the 2020 decennial census. To summarize, we estimate the total number of 311 complaints per census tract per 1,000 residents from May through September, 2021 - 2025. Standardizing by population may lead to large estimates in census tracts with small population counts, but ensures that our estimates of hydrant-related complaints do not simply reflect patterns in total population.

\subsection{Additional Correlation Results}
\begin{table*}[htbp!]
    \centering
\begin{minipage}[t]{0.44\textwidth}
    \centering
    \adjustbox{max width=\textwidth}{\begin{tabular}[t]
    {l S[table-format=3.2] S[table-format=3.2] S[table-format=3.2] S[table-format=3.2]}
    \toprule
        & \parbox{0.8in}{\centering HVI} & \parbox{0.4in}{\centering HVI} & \parbox{0.4in}{\centering NRI}  \\
    & \parbox{0.8in}{\centering (Neighborhood)} & \parbox{0.4in}{\centering (Tract)} & \parbox{0.4in}{\centering (Tract)} \\
    \midrule
    HVI (Tract) & 0.717 & & \\
    NRI (Tract) & 0.047 & 0.072 & \\
    HHI (ZCTA) & 0.512 & 0.468 & 0.197 \\
    \bottomrule
    \end{tabular}}
    \caption{Kendall's Tau correlations at the census tract level (n=2,240) comparing the percentile rankings from the NYC HVI (at the neighborhood and census tract levels) to percentile rankings based on the NRI at the census tract level and the HHI at the zipcode tabulation area level. While all four indices are positively correlated, the strength of the association varies and is lower with Kendall's Tau (as opposed to Spearman, shown in the main text). Directionally, correlation estimates using Kendall's Tau match those using Spearman.} 
    \label{tab:correlation_matrix_kendalls}
    \end{minipage}
    \hfill
    \begin{minipage}[t]{0.54\textwidth}
      \vspace{0pt}
    \adjustbox{max width=\textwidth}{\begin{tabular}[t]  {l S[table-format=3.2] S[table-format=3.2] S[table-format=3.2] S[table-format=3.2]}
    \toprule
    & \parbox{0.8in}{\centering HVI} & \parbox{0.4in}{\centering HVI} & \parbox{0.4in}{\centering NRI} & \parbox{0.4in}{\centering HHI} \\
    & \parbox{0.8in}{\centering (Neighborhood)} & \parbox{0.4in}{\centering (Tract)} & \parbox{0.4in}{\centering (Tract)} & \parbox{0.4in}{\centering (ZCTA)} \\
\toprule
Power outage & 0.121 & 0.096 & -0.128 & 0.028 \\
EMS & 0.129 & 0.133 & 0.174 & 0.266 \\
Hydrant & 0.206 & 0.237 & 0.058 & 0.271 \\
\bottomrule
\end{tabular}}
    \caption{Kendall's Tau correlations comparing census tracts (n=2,240) ranked by indices for heat and heat-related impacts. Similar to Table~\ref{tab:correlation_matrix_impacts}, we compare index-based rankings to percentile rankings based on (1) average daily maximum power outage rates at the locality level (2021-2025, May - September), (2) total counts of heat-related EMS calls at the zipcode level (2021-2025, May - September), and (3) total counts of 311 hydrant complaints (per 1,000 residents) at the census tract level. Correlations are weaker than in  Table~\ref{tab:correlation_matrix_kendalls}, and even negative.}
    \label{tab:correlation_matrix_impacts_kendalls}
    \end{minipage}
    \vspace{-1em}
\end{table*}

A limitation of this analysis is that comparing across different geographies can be misleading, as we show in Figure~\ref{fig:hvi_alt}. When possible, we have tried to independently verify the strength of the correlation relationships using similar geographies. For example, we can compare a version of the NYC HVI that uses 2020 zip code tabulation areas to the NYC HHI (which uses 2010 zip code tabulation areas) and the percentile rankings based on heat-related EMS calls (which use zipcodes). The correlation matrices using both Spearman and Kendall's Tau are below.

\begin{table*}[htbp!]
    \centering
\begin{minipage}[t]{\textwidth}
    \centering
    \adjustbox{max width=\textwidth}{\begin{tabular}
    {l S[table-format=3.2] S[table-format=3.2] S[table-format=3.2] | S[table-format=3.2] S[table-format=3.2] S[table-format=3.2]}
    \toprule
    & \multicolumn{3}{c}{Spearman} &  \multicolumn{3}{c}{Kendall's Tau} \\
    \midrule
        & \parbox{0.8in}{\centering HVI} & \parbox{0.4in}{\centering HHI} & \parbox{0.4in}{\centering EMS} & \parbox{0.8in}{\centering HVI} & \parbox{0.4in}{\centering HHI} & \parbox{0.4in}{\centering EMS}   \\
    \midrule
    HVI & 1.0 & & & 1.0 & & \\
    HHI & 0.744 & 1.0 & & 0.647 & 1.0 & \\
    EMS & 0.239 & 0.277 & 1.0 & 0.203 & 0.245 & 1.0 \\
    \bottomrule
    \end{tabular}}
    \caption{Spearman and Kendall's Tau correlations at the 2020 zip code tabulation area level for relevant indices and heat-related EMS calls. For these comparisons, we use the quintile-based 5-point risk scores, as the underlying data for the NYC HVI is not readily available. Correlations are higher than in Tables~\ref{tab:correlation_matrix} and~\ref{tab:correlation_matrix_impacts}, but the overall relationship is similar.} 
    \label{tab:correlation_zcta}
    \end{minipage}
    \vspace{-1em}
\end{table*}

\FloatBarrier 

\section{Discussion of the Experimental HeatRisk Product}
\label{sec:heatrisk}

A promising tool that many cities are beginning to explore is the experimental HeatRisk product from the National Weather Service (NWS)~\citep{noaa_heatrisk}. Unlike the indices discussed previously, the HeatRisk product produces an estimate of heat-related risk for an upcoming 24-hour period.
The trigger for heat activations in most cities usually depends on the heat index, which is a formula combining temperature information and humidity. The HeatRisk product offers an alternative that can be used for ``daily dynamic temperature thresholds''~\citep{vaidyanathan2019assessment, noaa_heatrisk}.
The fine-grained temporal and spatial granularity of the HeatRisk tool can inform heat early warnings systems, particularly as it is calibrated to heat-related health outcomes~\citep{vaidyanathan2019assessment}. 
However, the HeatRisk product is still primarily concerned with NWS forecasts and changing temperature thresholds; it does not directly measure the impact of extreme heat on health outcomes.

\section{Analysis of the Urban Risk Index}
\label{sec:uri}

In addition to the National Risk Index, we can also compare the NYC HVI to the Urban Risk Index, which is a local tool that was developed more recently for New York City (version 2.1 was released in February 2025)~\citep{uri}. The URI aims to replicate the NRI methodology for NYC with an expanded set of local datasets, particularly for estimating measures like community resilience. For example, the URI considers the placement of cooling centers, prior mitigation investments, and parks with water features. Instead of agricultural loss, it also focuses on power outages in its formulation of expected annual loss (EAL), along with heat-related mortality and morbidity.

In the same way that we compare the NYC HVI to the NRI (Section~\ref{sec:appendix_nri}), we can separately compare the NYC HVI to both the EAL estimates from the URI and the overall risk score, which combines EAL, social vulnerability, and community resilience, similar to the NRI. We present correlations with respect to both in Tables~\ref{tab:correlation_uri} and~\ref{tab:correlation_uri_kendalls}.

\begin{table*}[htbp!]
    \centering
    \centering
    \adjustbox{max width=\textwidth}{\begin{tabular}
    {l S[table-format=3.2] S[table-format=3.2] S[table-format=3.2] | S[table-format=3.2] S[table-format=3.2] S[table-format=3.2]}
    \toprule
    & \multicolumn{3}{c}{Percentile Ranking} &  \multicolumn{3}{c}{Risk Score (1-5)} \\
    \midrule
        & \parbox{0.8in}{\centering HVI} & \parbox{0.4in}{\centering URI} & \parbox{0.4in}{\centering URI (EAL)} & \parbox{0.8in}{\centering HVI} & \parbox{0.4in}{\centering URI} & \parbox{0.4in}{\centering URI (EAL)}   \\
    \midrule
    HVI & 1.0 &  &  & 1.0 &  &  \\
URI & 0.623 & 1.0 &  & 0.571 & 1.0 &\\
URI (EAL) & 0.155 & 0.338 & 1.0 & 0.122 & 0.293 & 1.0  \\
    \bottomrule
    \end{tabular}}
    \caption{Spearman correlations at the neighborhood level comparing the NYC HVI to the URI. For these comparisons, we compute Spearman correlations for both the underlying percentile rankings (left) and for the final assigned 1-5 risk scores (right). In contrast to previous analyses with the NRI, we use the official 5-point risk ratings, which are estimated using k-means rather than quintiles and are specific to NYC. Correlations are higher compared to the correlations with the NRI in Table~\ref{tab:correlation_matrix}. However, there is a noticeable difference between the correlations for the overall URI (which is based on EAL, social vulnerability, and community resilience) and the URI estimates for the EAL alone (i.e., the difference between the second and third rows). A reason for this difference is that the URI's estimate of community resilience accounts for a wide range of local characteristics that FEMA's NRI might overlook.} 
    \label{tab:correlation_uri}
    \vspace{-1em}
\end{table*}

\begin{table*}[htbp!]
    \centering
    \centering
    \adjustbox{max width=\textwidth}{\begin{tabular}
    {l S[table-format=3.2] S[table-format=3.2] S[table-format=3.2] | S[table-format=3.2] S[table-format=3.2] S[table-format=3.2]}
    \toprule
    & \multicolumn{3}{c}{Percentile Ranking} &  \multicolumn{3}{c}{Risk Score (1-5)} \\
    \midrule
        & \parbox{0.8in}{\centering HVI} & \parbox{0.4in}{\centering URI} & \parbox{0.4in}{\centering URI (EAL)} & \parbox{0.8in}{\centering HVI} & \parbox{0.4in}{\centering URI} & \parbox{0.4in}{\centering URI (EAL)}   \\
    \midrule
    HVI & 1.0 &  &  & 1.0 &  &  \\
URI & 0.438 & 1.0 &  & 0.487  & 1.0 &\\
URI (EAL) & 0.11 & 0.238 & 1.0 & 0.101 & 0.249 & 1.0  \\
    \bottomrule
    \end{tabular}}
    \caption{Kendall's Tau correlations at the neighborhood level comparing the NYC HVI to the URI. These correlations are the same as the table above (Table~\ref{tab:correlation_uri}) but use Kendall's Tau as opposed to Spearman correlation. The correlation values are smaller in magnitude, but substantively similar. The correlation between the EAL alone and the NYC HVI is much lower compared to the relationship between the URI's overall risk score and the NYC HVI.} 
    \label{tab:correlation_uri_kendalls}
    \vspace{-1em}
\end{table*}

\newpage 
\FloatBarrier
\section{Information on Heat-Related Impacts}
\label{sec:heat_impacts_appendix}

In this section, we present a table with detailed information on the three heat-related impacts introduced in Section~\ref{sec:heat-impacts}. We map each of these impacts to relevant decisions in emergency management.

\begin{table}[htbp!]
    \centering
    \begin{tabular}{|>{\raggedright\arraybackslash}p{0.75in}|>{\raggedright\arraybackslash}p{1in}|>{\raggedright\arraybackslash}p{0.75in}|>{\raggedright\arraybackslash}p{2.5in}|}
        \hline
        \textbf{Outcome} & \textbf{Data Source} & \textbf{Geography} & \textbf{Relevant Actions} \\
        \hline
        Hydrant-related 311 complaints & 311 Data \newline (NYC Open Data) & Census Tract & \textbullet~Indicates where cooling centers and outdoor cool options (e.g., spray showers) may be needed.
        \par \textbullet~Can support messaging and outreach for the FDNY's ``Spray cap'' program~\citep{extreme_heat_beat}.
       \\
        \hline
        Power outages & Department of \newline Public Service & Locality &
        \textbullet~Indicates zipcodes for additional messaging related to demand reduction. 
        \par \textbullet~Indicates areas where individuals who depend on critical infrastructure may be at risk~\citep{hazard_mitigation_heat}.
        \\
        \hline
        Heat-related EMS calls & FDNY \newline (NYC Open Data) & Zipcode & \textbullet~Indicates areas to prioritize for homeless outreach in the event of a Code Red~\citep{heat_response_plan}. \par
        \textbullet~Illustrates overall heat-related health risk.
        \newline 
        \\
        \hline
    \end{tabular}
    \caption{Selected impacts that are both related to extreme heat and relevant for downstream actions in emergency management.
    }
    \label{table:heat_outcomes}
\end{table}

\section{Extended Discussion on Trade-offs between Indices and Predictive Algorithms}
\label{sec:appendix_trade-offs}

Building on our takeaways in Table~\ref{tab:takeaways}, we discuss in more detail each of the seven trade-offs that practitioners should consider. For each trade-off, we provide specific considerations for  \textbf{indices} and \textbf{predictive algorithms}, as well as thoughts on potential areas where one method might excel.

These recommendations were motivated (1) by 
informal conversations with domain experts and government agency partners and (2) direct reflections from developing prototypical predictive models for extreme heat.
Specifically, a subset of the research team first compiled reflections and observations from using the NYC HVI in comparison to predictive models, grouped these observations into themes, and then identified relevant literature and examples to expand on these themes. We subsequently shared the main themes presented in Table~\ref{tab:takeaways} with domain experts, who independently reviewed and provided feedback.

\subsection{Problem Formulation}
Translating high-level goals into a narrow data science project is challenging and involves discretion.
This ``scoping'' process is lengthy and iterative. There are even tutorials on effective scoping in public policy~\citep{dssg_scoping}.
As discussed in the main text, \citet{passi2019problem} note how problem formulation in real-world settings is often the product of a ``negotiated translation.'' This process can lead to imperfect solutions in an effort to make the original problem tractable, which can exacerbate fairness-related concerns about measurement and validity~\citep{jacobs2021measurement, coston2023validity}.

\textbf{Indices} may be better suited to problems where it feels like compromising the integrity of the project to constrain it into a tractable data science task. On the surface, they appear to preserve ambiguity and allow for modeling abstract concepts like ``vulnerability.''
However, they can also make strong assumptions. 
Prior work has noted this trade-off in comparing indices to heuristic methods~\citep{kaiser2021should}.
Combining multiple models to improve robustness, conducting sensitivity analyses to understand the implications of different design decisions, and revisiting indices when applying them to new tasks can mitigate some of the limitations.

In contrast, the translation process for developing \textbf{predictive algorithms} may be imperfect, but it also forces decision-makers to confront their priorities and grapple with realistic trade-offs. In the context of extreme heat, decision-makers will need to decide what tasks or outcomes are most relevant: e.g. expanding cooling centers or measuring heat-related illness. These constraints may even help agencies identify previously unknown oversights. The need for clear and measurable outcome data may inspire an agency to initiate new data collection efforts or data sharing agreements, which in turn can help with measuring the effectiveness of their own interventions.

\subsection{Outcome Selection}
\label{sec:objectives}

Building on the previous section, predictive algorithms involve defining clear and measurable prediction targets. This process requires a shift from abstract concepts, like heat vulnerability, to measuring heat-related impacts.

As discussed in the main text, \textbf{indices} are useful in settings where outcomes are hard to define or data is challenging to obtain (e.g., sparse, hard to measure, or sensitive). 
 As an example, the NYC HVI is based on estimates of heat mortality; case counts for heat-related mortality are generally low, often due to undercounting, and there have instead been efforts to estimate heat-exacerbated deaths. In general, both heat-related and heat-exacerbated mortality estimates are low. 

In contrast, \textbf{predictive algorithms} require extensive data -- often from administrative datasets that may be public already or are managed by a government agency. These datasets require clear and labeled outcomes. Contested outcomes that are better answered by self-reported survey measures (e.g., well-being) may frustrate attempts to develop predictive algorithms.
These outcomes should additionally be relevant to the task at hand and meaningfully trigger actions. It is not useful to predict spurious outcomes that bear little relation to government decision-making, simply because they are available. Lastly, these outcomes should be reasonable to predict. Some outcomes may be highly unpredictable (e.g., many human-caused threats) or depend on information that is difficult to measure (e.g., street-level flooding).

A consideration is the potential for biased social and historical patterns to affect outcome validity (as in \citep{lum2016predict}). This can occur when outcomes are the products of human decisions as opposed to measurements from sensors or automatic data collection processes. In our work, outcomes like 311 hydrant complaints or heat-related EMS calls may capture variation in NYC resident behavior~\citep{10.1145/3490486.3538283, agostini2024bayesian}, as opposed to true differences in heat-related impacts. 
Outcomes that do not involve human reporting or decision-making (e.g., power outages or flood sensors) may be preferable.

\subsection{Value Alignment}
\citet{johnson2022bureaucratic} discuss the benefits and limitations of manually chosen decision rules in categorical prioritization. This discussion is similar to what we see in the design of \textbf{indices} where policymakers and government officials can decide whether to prioritize certain groups by including them in the index or not. For example, the CDC's social vulnerability index includes information on minority status (defined as the total population minus non-Hispanic white individuals) and the number of individuals who speak English less than well~\citep{svi_cdc_site}. The inclusion of this information reflects choices about what kinds of populations are more vulnerable. The NYC HVI tool is an interesting example because its inputs were selected based on a case-only study on heat-related mortality~\citep{madrigano2015case}. However, with the NYC HVI, including information on median household income and race still ensures that the index incorporates equity into its approach -- weighting sociodemographic characteristics equally alongside environmental ones.

In \textbf{predictive algorithms}, prior research efforts have focused on bias mitigation strategies, e.g. to reduce gaps in accuracy for sensitive groups or to equalize error rates by changing prediction thresholds~\citep{hardt2016equality}. These methods may ensure that some fairness criterion is met. Additionally, exploring consequential decisions in the model pipeline may reveal that some modeling approaches tend to prioritize socially disadvantaged groups (as \citet{benami2021distributive} discuss). However, predictive algorithms do not explicitly commit decisionmakers to uphold certain values or support specific sociodemographic groups. Perhaps it is possible to develop specific data science tasks that align with such explicit goals (similar to work on reparative algorithms in \citep{so2022beyond}), but such approaches are not typical.

\subsection{Time Horizon}

The timing of decisions is another important consideration.  This distinction maps onto broader concerns about slow versus fast data that can arise in emergency management and other government settings. ``Slow'' data inputs, which are infrequently or rarely updated, may be useful for long-term planning while ``fast'' data, which is collected and updated quickly, often in real-time, can inform more urgent decision-making and emergency response. \textbf{Indices} are best suited for long-term initiatives. They involve slow data inputs (e.g., updated annually) that may only describe the static or intrinsic spatial characteristics of a place. To use the example from the introduction and Section~\ref{sec:discussion}, public health advertising campaigns often last several months and are not meant to respond to time-sensitive events (e.g., a recent heat emergency). Relying on indices that summarize stable characteristics of a location or population makes sense in this context. In contrast, the targeting of text message notifications typically occurs in advance of an extreme heat event. \textbf{Predictive algorithms} can incorporate time series information to better support decision-making before or after an event. For example, anticipating the distribution of power outages just before an extreme heat emergency may lead to faster and more responsive actions.

\subsection{Validation and Evaluation}

Validation is challenging with \textbf{indices}. As we've discussed in this paper, reliability and validity concerns may arise frequently with additional testing or critical reflection. 
However, to some stakeholders the lack of validation may be acceptable, particularly if the goals, as described in Section~\ref{sec:objectives}, are sufficiently broad. The index may also gain validity from widespread social acceptance~\citep{cutter2024origin} or the design process, whether deliberative or transparent~\citep{johnson2022bureaucratic}.

A problem may arise, however, if practitioners start to rely on indices for decisions that are substantively different from the decisions for which they were initially designed. For example, given the ease of use, social vulnerability may start to become a proxy for a range of outcomes -- even when suitable, measurable outcomes are sufficient. Relying on indices in these settings makes it challenging to measure the effectiveness of policies and interventions. In contrast, validation is embedded in the process of developing \textbf{predictive algorithms}, whether through external test sets or model retraining to mitigate against distribution shift. The need for frequent data collection and monitoring of outcomes also allows researchers to assess the trade-offs of any given set of predictions, test whether an intervention is working, and determine if retraining is needed.

A question for practitioners to consider is whether predictive algorithms can obtain sufficient accuracy to warrant their use in decision-making.
Evaluating predictive algorithms, particularly in social sector settings, is non-trivial. Practitioners should leverage domain expertise to decide on an appropriate evaluation metric. As prior work demonstrates, domain-specific or task-specific evaluation metrics (such as ``top K'' precision, a metric that approximates real-world organizational constraints~\citep{rodolfa2021empirical}) may not align with conventional, \emph{technical} evaluation metrics (such as mean squared error for regression tasks) ~\citep{bell2022s, heuton2025spatiotemporal}.

\subsection{Stakeholder Capacity}
\textbf{Indices} may be preferred in government decision-making because they have fewer technical barriers and constraints. They are easier to share and reproduce. The values they embody may relate to a variety of decisions and goals -- making them easily adaptable to a range of settings and use cases. For example, heat vulnerability is a concept that appears in both NYC's Hazard Mitigation Plan and the \textit{NYC Urban Forest Agenda}~\citep{hazard_mitigation_heat, forestry_plan}. At the same time, over-reliance on these tools may lead to ineffective and slower progress on important policy goals, particularly if the index uses out of date information or misaligned outcomes. 
We recommend careful scrutiny and the use of sensitivity analyses (similar to our approach in Section~\ref{sec:case_study}). In applying these methods, organizations may find that indices require just as much upfront cost and investment as developing predictive algorithms.

Developing \textbf{predictive algorithms} can require technological capacity and expertise. Relevant stakeholders must be willing to collect and process data at regular intervals. Knowledgeable staff members should be able to maintain and monitor the system on their own. However, the level of investment may not always be significant. Prior work has demonstrated how even simple methods can be relatively effective for many kinds of prediction tasks ~\citep{heuton2025spatiotemporal, semenova2022existence, jung2020simple}. As a result, the process for developing predictive algorithms can sometimes be short and straightforward, especially compared to the process of designing an index from scratch, which might involve deliberation and iterative, expert review.

With both indices and predictive algorithms, there is the potential for over-reliance. Indices may appear as a panacea -- resolving all challenges associated with both defining and measuring high-level concepts. They allow practitioners to sidestep thorny questions related to validity (as in ~\citep{jacobs2021measurement}), and the legitimacy of these tools may gain traction through widespread social acceptance. Reliance on predictive algorithms can similarly lead to oversights, particularly if practitioners do not understand the implications of selected evaluation metrics or do not account for distribution shift.

\subsection{Intended Audience} 
A final consideration is the intended audience of a tool -- whether for the general public or domain experts who require specific information to make informed decisions. \textbf{Indices} like the NYC HVI and the NRI may be better-suited to a more general audience: the inputs are easy to define and often leverage public data while the concepts invoked are broad (e.g., vulnerability). However, the sensitivity of these tools to many design decisions and assumptions may lead to misleading takeaways and validity concerns, particularly if the uses stray from the original goals of the indices (e.g., using a social vulnerability index to allocate disaster assistance funding).
In contrast, \textbf{predictive algorithms}, which are tied to specific outcomes, may be better suited for internal uses. The outcomes in predictive algorithms are likely already tracked by the relevant agencies and therefore  translate well to the downstream operational decisions that teams in these agencies might make already.

\newpage
\section{Index Tools}
\label{sec:other_indices}

\small{
\begin{longtable}[c]{|>{\raggedright\arraybackslash}p{0.5in} |>{\raggedright\arraybackslash}p{0.6in} >{\raggedright\arraybackslash}p{1in} >{\raggedright\arraybackslash}p{1.5in} >{\raggedright\arraybackslash}p{0.6in} >{\raggedright\arraybackslash}p{0.6in} >{\raggedright\arraybackslash}p{0.5in} |}
        \hline
        \textbf{Type} & \textbf{Tool} & \textbf{Who Uses It} & \textbf{Input Variables} & \textbf{Outputs} & \textbf{Spatial Resolution} & \textbf{Last Updated} \\
        \hline
        \multirow{12}{*}{NYC} & Flood Vulnerability Index~\citep{nyc_fvi} & NYC City Agencies & Physical exposure to flooding (e.g., FEMA Special Flood Hazard Area data), Sociodemographics (e.g., disability status, language isolation, per capita income, and others) & Vulnerability score (1-5) & Census tract & 2024 \\
        \cline{2-7}
       & Displacement Risk Index~\citep{nyc_dri} & NYC Department of City Planning, NYC Department of Housing Preservation and Development  & Sociodemographics (e.g., race and ethnicity, income spent on rent), Housing characteristics (e.g., condition of housing stock, renter vs. owner status), Market Pressure (e.g., changes in the housing market) & Vulnerability score (1-5) & Public Use Microdata Area (PUMA) & 2025 \\
        \hline 
        \multirow{12}{*}{Heat} & New York State Heat Vulnerability Index~\citep{nayak2018development} & NY State Department of Health & Environmental and sociodemographic information & Score (1-6) & Census tract & 2017 \\
        \cline{2-7}
       & Philadelphia HVI~\citep{phil_hvi} & Philadelphia Department of Public Health & Environmental, temperature, sociodemographics, and community assets (e.g., community centers)  & Vulnerability score (1 - 6) & Census tract & 2024 \\
       \cline{2-7}
        & California Heat Assessment Tool (CHAT)~
        \citep{chat} & California Natural Resources Agency & Environmental, temperature and information on heat events, and sociodemographics  & Score (0 - 100) & Census tract & 2024 \\
        \hline 
        \multirow{7}{*}{Climate} & CalEnviro- Screen~\citep{calenviroscreen} & US Environmental Protection Agency, California Office of Environmental Health Hazard Assessment & Exposure indicators, Environmental effect indicators, Sensitive population indicators, and Socioeconomic factors & Score (0-100) & Census tract & 2026 \\
        \cline{2-7}
       & EJScreen~\citep{ejscreen} & U.S. Environmental Protection Agency  & Environmental and demographic information & Score (0-100) & Census block group & 2024 \\
        \cline{2-7}
       & U.S. Climate Vulnerability Index~\citep{us_cvi} & Environmental Defense Fund  & Environmental, social and economic, infrastructure, health, and climate change events & Score (0-1) & Census tract & 2023 \\
        \hline 
         \multirow{7}{*}{Global} & Global Hunger Index~\citep{global_hunger_index} & Concern Worldwide and Welthungerhilfe & Health and nutrition information (focused on children) & Score (0-100) & Country & 2025 \\
         \cline{2-7}          
         & Climate Risk Index~\citep{climate_risk_index} & Germanwatch & Climate-related impacts (such as fatalities and economic loss) & Score (0-100) & Country & 2026 \\
         \cline{2-7}
         & Global Health Security~\citep{global_hsi} & Nuclear Threat Initiative, Brown University School of Public Health, Economist Impact & Public data focused on COVID-19: prevention, detection and reporting, rapid response, robustness of health systems, and other factors & Score (0-100) & Country & 2026 \\
         \cline{2-7}
        & Human Development Index~\citep{un_hdi} & United Nations & Life expectancy, education, gross national income & Score (0-1) & Country & 2025 \\
        \cline{2-7}
        & Multidim- ensional Poverty Index~\citep{un_mpi} & United Nations & Information on health, education, standard of living & Score (0-1) & Country & 2025 \\
         \hline 
    \caption{\textbf{Different indices across relevant policy domains to which our findings and recommendations can be applied.} We highlight indices that are similar in scope to the indices for extreme heat that we discuss in depth in this paper. Many of these indices inform actions in climate and public health domains.}
    \label{tab:additional_indices}
\end{longtable}
}

\end{document}